\documentclass[12pt]{article}
\usepackage[utf8]{inputenc}
\usepackage[LGR, T1]{fontenc}
\usepackage{graphicx}
\usepackage{subcaption}
\usepackage{cancel}
\usepackage{float}
\usepackage[inline, shortlabels]{enumitem}
\usepackage[reqno]{amsmath} 
\usepackage{amsthm}
\usepackage{mathtools}
\usepackage{amssymb}
\usepackage{amsfonts}
\usepackage{eurosym}
\usepackage{geometry}
\usepackage{dsfont}
\usepackage{setspace}
\usepackage{natbib}
\usepackage{sectsty} 
\usepackage[colorlinks]{hyperref}
\hypersetup{citecolor=blue,linkcolor=blue, urlcolor=blue}
\usepackage{pgfplots}
\usepgfplotslibrary{colormaps,fillbetween}
\usepackage{tikz}
\usetikzlibrary{calc}
\usetikzlibrary{patterns}
\usetikzlibrary{datavisualization}
\usetikzlibrary{datavisualization.formats.functions}
\usetikzlibrary{positioning}
\usetikzlibrary{intersections, pgfplots.fillbetween}
\pgfplotsset{width=7cm,compat=1.15} 
\usepgfplotslibrary{patchplots}

\usepackage{afterpage}

\usepackage{adjustbox} 

\usepackage{cleveref}

\usepackage{xcolor} 

\definecolor{mtr0}{HTML}{bf8417} 
\definecolor{mtr1}{HTML}{292278} 
\definecolor{mte}{HTML}{c4395a} 

\usepackage{wrapfig} 
\usepackage{changepage}

\usepackage[flushleft]{threeparttable}
\usepackage[multiple]{footmisc} 
\usepackage{comment}

\usepackage{listings}
\usepackage{color}  

\definecolor{codegreen}{rgb}{0,0.6,0}
\definecolor{codegray}{rgb}{0.5,0.5,0.5}
\definecolor{codepurple}{rgb}{0.58,0,0.82}
\definecolor{backcolour}{rgb}{0.95,0.95,0.92}

\theoremstyle{plain} 
\newtheorem{assumption}{Assumption}

\newtheorem*{normalization}{Normalization}

\newcommand*{\myfont}{\fontfamily{qpl}\selectfont}
\DeclareTextFontCommand{\textmyfont}{\myfont}

\makeatletter
\newcommand{\leqnomode}{\tagsleft@true}
\newcommand{\reqnomode}{\tagsleft@false}
\makeatother

\DeclareSymbolFont{upgreek}{LGR}{cmr}{m}{n}
\SetSymbolFont{upgreek}{bold}{LGR}{cmr}{bx}{n}
\DeclareMathSymbol{\Epsilon}{\mathalpha}{letters}{"0F}
\DeclareMathSymbol{\Eta}{\mathalpha}{letters}{"11}
\DeclareMathSymbol{\epsilon}{\mathalpha}{letters}{`e}
\DeclareMathSymbol{\eta}{\mathalpha}{letters}{`h}

\newcommand{\indep}{\perp\!\!\!\perp}

\usepackage{changepage}

\usepackage{titlesec}
\titlespacing*{\section}{0pt}{19pt}{7pt}
\titlespacing*{\subsection}{0pt}{14pt}{5pt}
\titlespacing*{\subsubsection}{0pt}{12pt}{5pt}
\titlespacing*{\paragraph}{0pt}{9pt}{9pt}
\titleformat{\section}{\normalfont\fontsize{16}{15}\bfseries}{\thesection}{1em}{}
\titleformat{\subsection}{\normalfont\fontsize{14}{15}\bfseries}{\thesubsection}{1em}{}
\titleformat{\subsubsection}{\normalfont\fontsize{12}{15}\bfseries}{\thesubsubsection}{1em}{}

\allowdisplaybreaks

\definecolor{mtr0color}{RGB}{247, 186, 74}  
\definecolor{mtr1color}{RGB}{153, 147, 219}  
\definecolor{mtecolor}{RGB}{219, 147, 164}  
\definecolor{mtecolor2}{RGB}{222, 64, 170}  
\definecolor{mtr0color2}{RGB}{238, 112, 0}  
\definecolor{mtr1color2}{RGB}{102, 58, 150} 

\begin{document}


\title{{\fontsize{18}{20} \selectfont 
\textbf{\textmyfont{Don't (fully) exclude me, it's not necessary! \\
Causal inference with semi-IVs}}}}

\author{{\fontsize{14}{20} \textbf{Christophe Bruneel-Zupanc}}\footnote{E-mail address: \href{mailto:christophe.bruneel@gmail.com}{christophe.bruneel@gmail.com}. I am especially grateful to Jad Beyhum, Geert Dhaene, and Alexandre Gaillard for their tremendous help and to Gregory Veramendi for his help with the data of an earlier version. I also thank Jaap Abbring, Otilia Boldea, Christian Bontemps, Laurens Cherchye, Edoardo Ciscato, Ben Deaner, Sebastian Fleitas, Ferre de Graeve, Olivier de Groote, Vincent Han, Xavier d'Haultfoeuille, Yagan Hazard, Peter Hull, Toru Kitagawa, Tobias Klein, Denis Kojevnikov, Soonwoo Kwon, Louise Laage, Pascal Lavergne, Arthur Lewbel, Jan de Loecker, Iris Kesternich, Sebastian Maes, Thierry Magnac, Matt Masten, Arnaud Maurel, Costas Meghir, Geert Meesters, Bob Miller, Erwin Ooghe, Aureo de Paula, Krishna Pendakur, Christian Proebsting, Bram de Rock, Adam Rosen, Silvia Sarpietro, Sami Stouli, Alexandros Theloudis, Jo Van Biesebroeck, Frank Verboven, Frederic Vermeulen, Ao Wang, Martin Weidner, Frank Windmeijer, Nese Yildiz, Andrei Zeleneev, Mariana Zerpa, and participants in seminars and conferences in Antwerp, Duke, Leuven, Rotterdam, Bristol, Toulouse, Manchester, San Antonio, Tilburg, Oxford, Barcelona, and Thessaloniki for helpful discussions, suggestions, and comments. I acknowledge financial support from KU Leuven grant STG/21/040 and FWO grant G018725N. } \\
{\fontsize{14}{20} \textit{Department of Economics, KU Leuven}} } 
\date{{\fontsize{14}{20} \textit{\today}}} 

\maketitle

\vspace{-0.20in}

\begin{abstract}
\renewcommand{\baselinestretch}{1.2} 
{\normalsize 

\noindent This paper proposes \textit{semi-instrumental variables} (semi-IVs) as an alternative to instrumental variables (IVs) to identify the causal effect of a binary (or discrete) endogenous treatment.  
A semi-IV is a less restrictive form of instrument: it affects the selection into treatment but is excluded only from one, not necessarily both, potential outcomes. 
Having two continuously distributed semi-IVs, one excluded from the potential outcome under treatment and the other from the potential outcome under control, is sufficient to nonparametrically point identify marginal treatment effect (MTE) and local average treatment effect (LATE) parameters. In practice, semi-IVs provide a solution to the challenge of finding valid IVs because they are often easier to find: many selection-specific shocks, policies, prices, costs, or benefits are valid semi-IVs. As an application, I estimate the returns to working in the manufacturing sector on earnings using sector-specific characteristics as semi-IVs. \\

\noindent \textbf{Keywords:} instrumental variable, treatment effect, exclusion restriction, identification, selection, policy evaluation.} \\ 
\end{abstract}

\thispagestyle{empty}

\restoregeometry
\setstretch{1.25} 
\setlength{\abovedisplayskip}{6pt} 
\setlength{\belowdisplayskip}{6pt}

\setcounter{page}{1}

\pagebreak

\section{Introduction}

Identifying the causal effect of a treatment on an outcome is a central challenge in economics, particularly when the selection of treatment is endogenous. 
A common approach is to use instrumental variables (IVs), which must satisfy three 
conditions to be valid: they must affect treatment selection (relevance), be independent of unobserved factors (exogeneity), and have no direct effect on the potential outcomes (exclusion restriction). 
The IV approach is appealing because it can identify causal effects under relatively mild additional assumptions. 
For instance, when IVs influence treatment selection uniformly in the same direction for all individuals -- a condition known as monotonicity \citep{imbensangrist1994} -- they enable the nonparametric identification of average treatment effects for individuals whose treatment choice is influenced by the IVs. These include marginal treatment effect (MTE) \citep{heckmanvytlacil2005} and local average treatment effect (LATE) \citep{imbensangrist1994}   parameters, which are widely estimated in applied research. 
However, finding valid IVs that meet these conditions is challenging. In particular, the exclusion restriction is a strong and often controversial assumption, even for commonly used instruments, because it is difficult to find a variable that affects the selection without affecting any potential outcome. \\
%
%
%
\indent Meanwhile, in many settings with a discrete treatment, there are 
numerous variables that, while not excluded from \textit{all} potential outcomes, are credibly excluded from \textit{some}.
Think, for instance, of treatment-specific characteristics -- such as costs, prices, benefits, amenities, or policies -- that affect the outcomes of individuals only if they select that specific treatment but not if they select another one. 
I call such variables \textit{semi-instrumental variables} (semi-IVs): variables relevant for treatment selection and excluded from some, but not necessarily all, potential outcomes.  
This raises a critical question: is causal inference possible with semi-IVs alone? %
\noindent Remarkably, yes, this paper shows that semi-IVs can serve as substitutes for IVs in addressing endogeneity, even in general models with heterogenous treatment effects. \\
\indent When the treatment is binary, two continuously distributed semi-IVs -- one excluded from the potential outcome under treatment and the other from the potential outcome under control -- are sufficient to nonparametrically identify MTE and LATE parameters, just as an IV would. 
Crucially, this identification does not require any additional assumption beyond those in the standard IV framework (monotonicity of the selection, applied to both semi-IVs), nor any restriction on how the semi-IVs affect the potential outcomes from which they are not excluded. The identification is \textit{IV-like} and stems directly from the observable variations induced by the semi-IVs.
By expanding the variation available for causal inference, semi-IVs may offer a solution when valid IVs are difficult to find. Since semi-IVs are often more readily available in applied settings, they significantly broaden researchers' toolkit and open up new possibilities for identifying causal effects. \\
\indent To illustrate this, I present several examples of semi-IVs in occupation and location choice models, with various applications in labor, education, and industrial organization, among others (see Section \ref{subsec:example}).
As the main running example, which also corresponds to the empirical application in Section \ref{sec:application}, I focus on the identification of the earnings returns to working in the manufacturing sector \citep{heckmansedlacek1985, heckmansedlacek1990}. Workers select into sectors based on their unobservable attributes, including their sector-specific skills, leading to endogenous selection. As a result, the comparison of earnings between manufacturing and nonmanufacturing workers is confounded. Ideally, we would like to find an IV to address the selection on unobservables, but finding one that influences the sector choice without also influencing the earnings is particularly challenging. 
%
%
Fortunately, we have access to many sector-specific characteristics, such as sector size (e.g., GDP, number of firms and employees), productivity, and growth, with observable variations across time and markets (e.g., states), which can be leveraged for identification. 
These are promising semi-IV candidates. Indeed, they are likely to affect workers' sector choices: all else equal, the more attractive a sector's characteristics, the more likely a worker is to select it. Such aggregate (shocks to) sector characteristics are also independent of the workers' idiosyncratic sector-specific unobserved skills and preferences. 
Unfortunately, they are not valid IVs because a worker's earnings are typically determined as a function of the worker's unobserved skills and the characteristics of the selected sector \citep{heckmansedlacek1985}. The sector-specific characteristics are \textit{included} in the corresponding sector-specific potential outcomes. Crucially, however, conditional on the characteristics of the manufacturing sector, (shocks to) the characteristics of the nonmanufacturing sector should not directly affect the potential earnings of manufacturing workers, and vice versa. This partial exclusion is the key property that makes sector-specific characteristics -- and, more generally, many treatment-specific variables -- valid semi-IVs. Once characteristics specific to the treated are controlled for, similar characteristics for the untreated do not affect the potential outcomes under treatment, and vice versa.\footnote{Examples of models where sector-specific characteristics satisfy these semi-IV conditions include \citet{heckmansedlacek1985, heckmansedlacek1990} and, more recently, \citet{eckardt2024}. See Section \ref{subsec:example} for discussion.  }

%
%
\indent It is important to note what is \textit{not} required for semi-IVs to be valid. First, the two semi-IVs may be correlated, as is often the case. Their validity only requires that, conditional on the included semi-IV, the other one is relevant and rightfully excluded. In a sense, any effect of the excluded semi-IV on a given potential outcome must be subsumed by the effect of the included one.\footnote{
Thus, semi-IVs can be interdependent or jointly determined in equilibrium, as long as, once they are determined, the potential outcomes depend only on their treatment-specific semi-IV. This conditional exclusion property is particularly natural when semi-IVs are treatment-specific prices; see Section \ref{subsec:example} for discussion. }
Second, the exclusion restriction applies to individuals' (latent) potential outcomes, but each semi-IV may still indirectly influence the observable mean outcome of the group from which it is excluded via selection effects.
For instance, holding other factors constant, an increase in the size of the manufacturing sector encourages more workers (the compliers) to select into manufacturing, which, in turn, alters the observed average earnings of nonmanufacturing workers solely through changes in the workforce composition. \\
\indent In fact, the observable changes in the average nonmanufacturing earnings and the selection probability, both induced by changes in the excluded manufacturing sector characteristics (as semi-IV), are precisely the variations that identify the mean nonmanufacturing potential earnings for the compliers who select into manufacturing as a result of the change in the semi-IV. Similarly, shifts in the nonmanufacturing sector characteristics identify the mean manufacturing potential earnings for the compliers resulting from these shifts. 
This mirrors the standard IV nonparametric identification arguments but applied separately to the outcomes of the treated and untreated subpopulations instead of directly to all outcomes. \\
\indent Such a decomposition hints at why semi-IVs are sufficient to identify treatment effects. MTE and LATE parameters compare the potential outcomes under treatment (e.g., manufacturing) and under control (e.g., nonmanufacturing) for specific subpopulations of (marginal) compliers. 
An IV is excluded from both potential outcomes, so shifting it identifies the mean of both potential outcomes for the same set of compliers at once, and directly yields the treatment effect. While convenient, this is not necessary. 
Indeed, as previously described, two distinct semi-IVs can separately identify the means of the potential outcomes under treatment and control for specific subpopulations. 
Then, to reconstruct treatment effects, one needs to align the changes in each semi-IV that induce the same flow of compliers into treatment. 
This alignment is straightforward under the usual LATE monotonicity assumption. Indeed, as with multiple instruments \citep{carneiroheckmanvytlacil2011}, monotonicity ensures that the treatment probability summarizes the effect of all semi-IVs on selection into a single measure. 
At the margin, the marginal complier induced into treatment by an infinitesimal increase in the manufacturing sector size is the same as the one induced into treatment by an infinitesimal decrease in the nonmanufacturing sector size. More generally, any two semi-IV shifts causing the same observable shift in treatment probability induce the same set of compliers. \\
\indent Overall, the primary cost of the semi-IV approach is that researchers must use two semi-IVs, rather than a single IV, to identify similar treatment effects. This entails two distinct costs: a (typically modest) search cost and, in the binary treatment case, a stronger monotonicity requirement.\footnote{Throughout, I require continuously distributed semi-IVs so that complier flows can be aligned across treatment margins. For MTE, this adds no cost: standard IV-based MTE identification also require continuous IVs. For LATE, it is stricter than conventional IV identification, which works with discrete or even binary IVs. This extra cost is mild because most examples of semi-IVs are continuously distributed. Moreover, the support requirement can be relaxed: in fact, LATE can be identified if only one of the two semi-IVs is continuous. With two discrete semi-IVs, LATE may also be obtained by interpolation. } 
%
%
The search cost is usually small because valid semi-IVs often come in treatment-specific pairs (e.g., treatment-specific prices, size). So, researchers generally only need to find a single type of semi-IV that can be adapted to all potential outcomes. 
A related cost is that, by design, because we use at least two semi-IVs, monotonicity becomes a multi-instrument assumption, making it stronger than in the benchmark single IV case \citep{lee2018identifying, mogstadetal2021}. That said, since semi-IVs are typically representing the same type of incentive across alternatives (e.g., prices), assuming that all individuals respond in a uniform direction to changes in the semi-IVs (or in their relative comparison) may be more plausible than with unrelated incentives (e.g., college proximity and tuition fees in the standard multi-IV example). 
Note also that this additional monotonicity cost is specific to the binary treatment case. For discrete treatments with more than two distinct alternatives, standard IV identification also requires multiple instruments and thus a stronger monotonicity condition. 
On balance, the cost of the semi-IV approach -- especially in discrete-treatment settings -- is often modest relative to its main benefit: relaxing the stringent full exclusion restriction of IVs. Semi-IVs provide a convenient alternative to standard IVs for causal inference. \\
\indent Building on constructive identification arguments, this paper develops both nonparametric and semi-parametric methods for estimating MTEs and LATEs with semi-IVs. In particular, the semi-parametric method adapts the local IV MTE estimation \citep{heckmanvytlacil1999, carneiroheckmanvytlacil2011, andresen2018exploring} to the semi-IV framework. The estimator is fast, easy-to-implement, and performs well, as illustrated by simulations in Online Appendix \ref{app:montecarlo}. A user-friendly implementation is available in the companion R package \texttt{semiIVreg} \citep{semiivreg}.\footnote{The package is available for download on \href{https://github.com/cbruneelzupanc/semiIVreg}{Github}. See the associated vignette at \href{https://cbruneelzupanc.github.io/semiIVreg/}{https://cbruneelzupanc.github.io/semiIVreg/} for user guidelines and replicable simulated examples. } \\  
%
\indent To illustrate the method, I estimate the (marginal) earnings returns to working in the manufacturing sector in the US and their evolution from $1999$ to $2018$ using state-level sector-specific characteristics (size) as semi-IVs. 
The analysis focuses on young white men with only high school education, the group most affected by the decline of the manufacturing sector over the period considered \citep{pierce2016surprisingly}. 
I find significant heterogeneity in returns: some workers earn more in manufacturing, others more in nonmanufacturing jobs. 
Moreover, I find that the returns to working in manufacturing declined over the $20$-year period, closely mirroring the decline in manufacturing employment. This occurs despite an increasing observable earnings gap in favor of manufacturing over nonmanufacturing workers (as estimated by naive OLS), highlighting the importance of controlling for the endogenous change in sector composition. Perhaps most strikingly, 
I find that the potential earnings in both sectors declined over the period. This decline highlights a marked deterioration in the economic prospects of young, less-educated white males, regardless of their choice of sector. \\
\indent The main focus of this paper is on the effect of a binary treatment. This case provides the most transparent intuition for why semi-IVs can effectively substitute for IVs. However, 
the semi-IV approach naturally extends to the more general case of a discrete treatment with $J > 2$ distinct alternatives. In this setting, having one alternative-specific semi-IV -- which must be excluded from all but one potential outcome -- for each potential outcome (giving a total of $J$ distinct semi-IVs) is sufficient to nonparametrically identify discrete treatment effects, such as the effect of one treatment relative to the best alternative \citep{heckmanurzuavytlacil2006, heckmanvytlacil2007b} or margin-specific MTEs \citep{mountjoy2022community}. 
Notably, identifying these treatment effects with standard IV methods typically requires $J-1$ special "alternative-specific" IVs, which must affect the latent utility of only one alternative but must be excluded from all potential outcomes, unlike semi-IVs. Thus, semi-IVs provide a natural generalization by conveniently allowing the alternative-specific instruments to also impact their corresponding potential outcomes, hence enlarging the available variation to identify discrete treatment effects. This gain comes at the modest cost of requiring $J$ semi-IVs instead of $J-1$ IVs. Apart from this, and contrary to the binary treatment case, there is no additional cost of using semi-IVs in terms of monotonicity there.  In fact, margin-specific MTEs can be also be identified with semi-IVs under the same weaker form of monotonicity imposed on IVs by \cite{mountjoy2022community}, i.e., unordered partial monotonicity combined with a comparable complier assumption. 
The formal extension of the discrete treatment identification results with IVs to semi-IVs is provided in Online Appendix \ref{app:discrete}. The semi-IV approach does not extend to continuous treatments as it would require infinitely many semi-IVs. \\ 



\noindent \textbf{Related Literature.} Exploiting targeted exclusion restrictions from certain potential outcomes to achieve identification has been used in part of the literature on Roy models \citep{heckmansedlacek1985, heckmansedlacek1990, heckmanhonore1990, heckman1990, heckmanvytlacil2007b, bayeretal2011, frenchtaber2011, dhaultfoeuillemaurel2013, mourifie2020sharp}. For instance, the sector-specific variables in \cite{heckmansedlacek1985, heckmansedlacek1990} and \cite{dhaultfoeuillemaurel2013} can be interpreted as semi-IVs, excluded from the potential outcomes of workers selecting into other sectors.
Roy models can be \textit{parametrically} identified using distributional assumptions on the unobserved sector skills, possibly combined with sector-specific variables \citep{heckmanhonore1990}. \\
\indent However, \textit{nonparametric} identification results using only sector-specific variables (semi-IVs) are scarce and limited to models with specific selection rules.  
For \textit{Roy models} -- where the sector selection depends solely on the difference between the potential earnings, i.e., where workers select the sector giving them the highest potential earnings -- \cite{heckmanhonore1990} and \cite{frenchtaber2011} show that nonparametric identification of the treatment effects can be obtained with sector-specific variables if these variables have an additively separable effect with respect to the unobservables affecting the outcomes. The identification exploits the Roy model structure, which implies that the semi-IVs affect the first stage solely through their effect on the potential outcomes. Thus, by observing changes in sector choice probabilities induced by variations in semi-IVs, one can infer their effects on the corresponding outcomes and subsequently identify certain treatment effects. However, this reasoning is specific to Roy models.\footnote{Under the Roy model specification, the full model (i.e., the joint distribution of the potential outcomes) can be identified, but its identification relies on identification-at-infinity arguments, requiring sector-specific variables (semi-IVs) to shift the propensity score to zero or one, which is rarely satisfied in practice.} 
\noindent For \textit{extended Roy models} -- where the selection rule is a function of the differences between the potential earnings and a cost depending only on observable covariates, with no additional unobserved shocks beyond those affecting earnings -- \cite{dhaultfoeuillemaurel2013} show that nonparametric identification of the distribution of the treatment effects is also possible if the sector-specific variables have an additively separable effect on the outcomes.\footnote{\label{footnote:additive}In both, Roy and extended Roy models, additive separability in the outcome translates to the selection equation, yielding weak separability in the first stage and imposing a stronger form of LATE monotonicity.} 
Their identification results are more general than those in Roy models but still rely on the extended Roy model structure, in particular, the known link between the residuals in the outcome and the selection equations. \\ 
%
%
\indent In contrast, for the more widely applied \textit{generalized Roy models} -- where the selection rule is more general and includes additional unobserved shocks (e.g., individual unobserved preferences) that may correlate with those affecting earnings -- 
nonparametric identification of treatment effects has so far relied on IVs combined with the LATE monotonicity assumption \citep{heckmanvytlacil2005, heckmanvytlacil2007b, frenchtaber2011}.\footnote{The "LATE model" \citep{imbensangrist1994} is, implicitly, a generalized Roy model, in the sense that there is no restriction on the potential outcomes (allowing for heterogenous treatment effects) and no restriction on the selection, apart from monotonicity \citep[see][]{vytlacil2002}.  See \cite{heckmanvytlacil2007b} for a clear-cut distinction between Roy, extended Roy, and generalized Roy models.}\footnote{An exception is \cite{eckardt2024}, who estimates the effect of training on occupation-specific wages using occupation-specific vacancy rates as instruments to control for the selection, without invoking a Roy/extended Roy structure. Instead, because of the high-dimensional treatment space, the paper adapts \cite{lee1983generalized} and \cite{dahl2002} and imposes parametric assumption on the selection rule. Moreover, its main focus is to estimate homogenous `treatment effects' of training on occupation-specific wages (and not occupation premia). Although not a general nonparametric identification result in a generalized Roy model with semi-IVs, it provides a clear application in which occupation-specific variables help identifying some specific effects. } 
Even with IVs, the average treatment effect (ATE) for the entire population is generally not identified unless there exists a subpopulation for whom the probability of treatment is zero or one (identification-at-infinity). Fortunately, \cite{imbensangrist1994} showed that, even without this hard-to-satisfy identification-at-infinity requirement, an IV can still identify treatment effects of interest for subpopulations of compliers induced into treatment by varying the IV: the LATEs. \cite{heckmanvytlacil2005} further extended this result and showed that marginal variations of continuous IVs identify the MTEs for marginal compliers. \\
\indent This paper shows that, under the monotonicity assumption of \cite{imbensangrist1994}, semi-IVs can also nonparametrically identify treatment effects for subpopulations of compliers induced into treatment by varying the semi-IVs (MTEs and LATEs). The semi-IV identification argument parallels that for an IV, but applied separately to each potential outcome. It does not rely on a specific selection rule (apart from monotonicity), in contrast to existing identification results in Roy and extended Roy models.
Thus, semi-IVs provide an alternative to IVs for identifying MTE and LATE parameters in general models with monotonicity imposed on the selection rule. This is particularly advantageous in empirical applications, where we want to avoid structural assumptions and where semi-IVs -- such as occupation or location-specific characteristics -- are often readily available but have so far only been used as control variables, neglecting their identification power. \\
\indent This paper shows how semi-IVs can substitute for IVs in the LATE framework \citep{imbensangrist1994, angristimbensrubin1996}, i.e., under the LATE monotonicity assumption. A companion paper \citep{bruneelbeyhum2025semi} shows that semi-IVs can also substitute for IVs in the IV-quantile regression (IVQR) framework \citep{chernozhukovhansen2005} to identify quantile treatment effects (QTE) under the assumption of rank invariance on the potential outcomes instead of LATE monotonicity. 
The IVQR and LATE frameworks encompass most models used in applied research.  Since semi-IVs achieve similar identification results in both frameworks, they can substitute for IVs in many empirical applications.\footnote{This paper focuses on \textit{point} identification. Extensions of IV nonparametric \textit{set} identification results \citep[e.g.,][]{manskipepper2000, chesher2010, chesherrosen2017} to semi-IVs are for future research.} \\
%
%
\indent More broadly, this paper contributes to the literature on identifying treatment effects by relaxing key IV assumptions. 
A more detailed discussion of this literature is in Online Appendix \ref{app:literature}. For a recent survey, see \cite{lewbel2019identification} or \cite{mogstad2024handbook}. \\

\noindent \textbf{Outline.}  Section \ref{sec:framework} adapts the canonical binary treatment IV setup \citep{angristimbensrubin1996} to semi-IVs, outlines their properties and illustrates with detailed examples. Section \ref{sec:identification} establishes nonparametric identification of MTEs and LATEs using semi-IVs. Section \ref{sec:estimation} describes their nonparametric estimation, and shows how local IV estimation of MTEs can be extended to semi-IVs, offering a simple semi-parametric procedure. 
Finally, Section \ref{sec:application} applies this procedure to estimate the returns to working in the manufacturing sector in the US. \\ 




\section{Framework}\label{sec:framework}

First, this section shows how the canonical program evaluation problem of identifying the (possibly heterogenous) effect of an endogenous binary treatment on an outcome using an IV \citep{imbensangrist1994, angristimbensrubin1996, heckmanvytlacil2005} naturally extends to semi-IVs.  
Then, I discuss examples of semi-IVs in a variety of applications.

\subsection{The semi-IV model}\label{subsec:model}

\textbf{Potential outcomes and endogenous treatment.}  
We want to evaluate the effect of a binary treatment $D \in \mathcal{D}=\{0, 1\}$ on a scalar, real-valued outcome $Y$. Let $Y_d$ be the latent potential outcome under treatment state $d$ $\in \mathcal{D}$. 
The researcher only observes the realized outcome, $Y$, corresponding to the potential outcome of the selected alternative $D$. That is, 
\begin{align}\label{eq:split}
	Y = Y_D = D Y_1 + (1-D) Y_0,  
\end{align}
and the potential outcome corresponding to the non-selected treatment is latent. 
The challenge faced by economists is that $D$ may be endogenous with respect to the potential outcomes $Y_0$ and $Y_1$, so we cannot simply compare the observable distributions of $Y$ given $D=1$ and $D=0$ in order to identify the causal effect of $D$ on $Y$. It is well known that a valid IV, combined with a monotonicity assumption, allows us to nonparametrically recover some causal effects of interest: the LATE \citep{imbensangrist1994} and the MTE \citep{heckmanvytlacil1999, heckmanvytlacil2005}. I will show that valid semi-IVs allow us to recover these parameters as well. 
The intuition behind why semi-IVs can substitute for IVs can be gleaned from Equation \eqref{eq:split}. The IV approach requires a variable that is fully excluded from $Y$. In the semi-IV approach, it suffices to find two variables (the semi-IVs), one excluded from $Y_1$ to serve as an instrument for the effect of $D$ on $DY = DY_1$, and another one excluded from $Y_0$ to serve as an instrument for the effect of $D$ on $(1-D)Y = (1-D)Y_0$. \\  

\noindent \textbf{Semi-IVs.} 
%
%
%
\noindent Let $Z_0$ and $Z_1$ be the two (sets of) complementary semi-IVs, excluded from $Y_1$ and $Y_0$, respectively, with supports $\mathcal{Z}_0$ and $\mathcal{Z}_1$. Both semi-IVs are observed regardless of the selected treatment. Let $Z=(Z_0, Z_1) \in \mathcal{Z}$ be the set of all semi-IVs, with support $\mathcal{Z}$. Following \cite{imbensangrist1994}, denote by $D(z)$ the potential treatment choice if the semi-IVs are exogenously set to $Z=z$. The observed treatment is thus $D=D(Z)$.\footnote{The stable unit treatment value assumption (SUTVA) is implicitly made, meaning each individual's potential outcomes and treatment are unaffected by the others' treatment status or semi-IVs. }
The validity of the semi-IVs holds conditional on a set of control variables $X$, with support $\mathcal{X}$, which have no identifying power ($X$ may be neither excluded, exogenous, nor relevant). 

\begin{assumption}[Valid semi-IVs]\label{ass:semiiv} $(Z_0, Z_1)$ satisfies the following conditions: 
\begin{enumerate}[label=A\theassumption.\arabic*]
	\item (Exogeneity and Exclusion)\label{ass:exogeneity} For all $z \in \mathcal{Z}$, (i) $D(z) \indep Z | X$, and (ii) for $d=0,1$, $(Y_d, D(z)) \indep Z_{1-d} | (Z_d, X)$, where $\indep$ denotes conditional independence.  
	\item (Relevance)\label{ass:relevance}
	For $d=0,1$, the propensity score function, $P(z_0, z_1, x) = \mathbb{E}[D|Z_0=z_0, Z_1=z_1, X=x]$ is a nontrivial function of $z_d$ for all $(z_{1-d}, x) \in \mathcal{Z}_{1-d} \times \mathcal{X}$. 
\end{enumerate}
\end{assumption}

By definition, $(Z_0, Z_1)$ is a set of valid complementary semi-IVs if Assumption \ref{ass:semiiv} holds. A valid IV, $\tilde{Z}$, would be relevant for the selection into treatment and satisfy the stronger conditional independence assumption $(Y_0, Y_1, D(\tilde{z})) \indep \tilde{Z} | X$ \citep{imbensangrist1994}. In contrast, each semi-IV is relevant (\ref{ass:relevance}), but excluded from only one of the two potential outcomes, not both (\ref{ass:exogeneity}). Hence, we say that an IV is \textit{fully} excluded while a semi-IV is only \textit{partially} excluded from the outcome. 
\noindent Conditional on $Z_1$ and $X$, shifting $Z_0$ enables shifts in the selection probability without directly affecting $Y_1$: $Z_0$ acts as an IV for the effect of $D$ on $DY=DY_1$. Yet, $Z_0$ is not excluded from $Y_0$ and may affect it in an unrestricted manner, like any covariate $X$. The reverse holds for $Z_1$. 
Together, $Z_0$ and $Z_1$ provide an exclusion restriction for each potential outcome, making them \textit{complementary}. \\
%
%
%
\indent In the remainder of the paper, I refer to the conditional independence Assumption \ref{ass:exogeneity}(ii) as the \textit{(partial) exclusion restriction}, and call $Z_{1-d}$ \textit{excluded} from $Y_d$ given $(Z_d, X)$. This terminology is a simplification, as, similar to IVs \citep{angristimbensrubin1996}, satisfying $Y_d \indep Z_{1-d} | (Z_d, X)$ actually requires two conditions: exclusion \textit{and} exogeneity. To clarify the distinction between the two, denote by $Y_d(z, x)$ the potential outcome when $D$ is exogenously set to $d$, $Z$ to $z$, and $X$ to $x$. Thus, the potential outcomes defined earlier are $Y_d = Y_d(Z, X)$, and the observed outcome is $Y=Y_D=Y_D(Z,X)$. Then, the exclusion restriction is that, for all $x \in \mathcal{X}$, for all $(z_0, z_1)$, $(z_0, z_1')$, and $(z_0', z_1) \in \mathcal{Z}$, and for each $d=0, 1$, 
	\begin{align*} Y_d(z_d, z_{1-d}, x) = Y_d(z_d, z_{1-d}', x) =: Y_d(z_d,x). \end{align*} 
	That is, there is no direct effect on $Y_d$ of exogenously switching $Z_{1-d}$ from $z_{1-d}$ to $z_{1-d}'$, once we already control for $Z_d=z_d$ and $X=x$. Furthermore, exogeneity requires that $Z_{1-d}$ be independent of the potential outcomes and treatment, i.e., $(Y_d(z_d, x), D(z)) \indep Z_{1-d} | (Z_d, X)$.\footnote{\label{footnote:exogeneity}A stronger version of exogeneity in \ref{ass:exogeneity} is to assume that $(Y_0(z_0, x), Y_1(z_1, x), D(z)) \indep (Z_0, Z_1) | X$ for all $(z_0, z_1) \in \mathcal{Z}$ and $x \in \mathcal{X}$. This is still very general and only restricts how the semi-IVs affect their corresponding potential outcomes: through a direct effect rather than via their correlation with unobservables affecting the potential outcomes (see Online Appendix \ref{app:heckmanvyt}). This assumption enables a clearer causal interpretation of the effect of exogenous changes in semi-IVs on outcomes, though it is stronger than necessary for identification. } Notice that $Y_d(z_d, x)$ is not necessarily independent of  $Z_d$ nor $X$. This is because $Z_d$ and $X$ may be correlated with the unobservables affecting $Y_d(z_d, x)$. 
	The distinction between exclusion and exogeneity may be clearer when expressing the potential outcomes as latent variables in a generalized Roy model. 
	The reformulation of the semi-IV model as a generalized Roy model, analogous to the IV framework of \cite{heckmanvytlacil2005}, is detailed in Online Appendix \ref{app:heckmanvyt}. \\ 

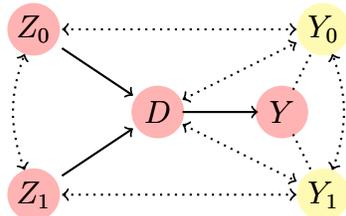
\begin{figure}[H]
\centering 
\caption{Causal graph of the semi-IV model (binary $D$)}\label{fig_graph_semiIV}
\begin{tikzpicture}[scale=1.1, node distance={100mm}, thick] 
\node (D) at (0,0) {$D$};
\node (W0) at (-1.5, 1) {$Z_0$};
\node (W1) at (-1.5, -1) {$Z_1$}; 
\node (Y) at (1.5, 0) {$Y$}; 
\node (Y0) at (2.0, 1) {$Y_0$};
\node (Y1) at (2.0, -1) {$Y_1$}; 
\path (D)[->] edge (Y); 
\path (W0)[->] edge (D);
\path (W1)[->] edge (D);
\path (W1)[<->, dotted] edge (Y1);
\path (W0)[<->, dotted] edge (Y0);
\path (Y)[-, dotted] edge (Y1);
\path (Y)[-, dotted] edge (Y0);
\path (W1)[<->, dotted] edge[bend left=20] (W0);
\node at (Y0) [shape=circle, fill=yellow,opacity=0.3, minimum size=0.7cm] {};
\node at (Y1) [shape=circle, fill=yellow,opacity=0.3, minimum size=0.7cm] {};
\path (Y1)[<->, dotted] edge[bend right=20] (Y0);
\node at (D) [shape=circle, fill=red,opacity=0.3, minimum size=0.7cm] {};
\path (D)[<->, dotted] edge (Y0);
\path (D)[<->, dotted] edge (Y1);
\node at (Y) [shape=circle, fill=red,opacity=0.3, minimum size=0.7cm] {};
\node at (W0) [shape=circle, fill=red,opacity=0.3, minimum size=0.7cm] {};
\node at (W1) [shape=circle, fill=red,opacity=0.3, minimum size=0.7cm] {};

\node at (Y1) {$Y_1$};
\node at (Y0) {$Y_0$}; 
\node at (D) {$D$};
\node at (W0) {$Z_0$};
\node at (W1) {$Z_1$}; 
\node at (Y) {$Y$};

\end{tikzpicture} 
\vspace{0pt} 
{\scriptsize{}}%
\noindent\begin{minipage}[t]{1\columnwidth}%
{\footnotesize{} \textit{ \tikz\draw[red,fill=red, opacity=0.3] (0,0) circle (1ex); $=$ observed, \tikz\draw[yellow,fill=yellow, opacity=0.3] (0,0) circle (1ex); $=$ potential outcome $Y_d$ is observed only if $D=d$. Solid arrows indicate causal effects. Dotted arrows indicate possible causal effects. Double-edged dotted arrows indicate a possible correlation.   }}%
\end{minipage}{\footnotesize\par}

\end{figure}

\noindent \textit{Causal graph representation.} 
\noindent Figure \ref{fig_graph_semiIV} provides a schematic causal graph representation of Assumption \ref{ass:semiiv} conditional on a specific $X=x$ (not displayed). The partial exclusions are visible by the absence of a direct link between $Z_{1-d}$ and $Y_d$ for each $d$.

\vspace{0.5\baselineskip}

\noindent \textbf{Selection into treatment.} As with standard IVs, identifying causal effects requires restricting how the semi-IVs affect treatment selection. I impose the same \textit{monotonicity}/no-defier condition that \cite{imbensangrist1994} impose on IVs on the semi-IVs. Beware that, since it is applied jointly to multiple semi-IVs, the assumption is stronger than with a single IV, as in the multi-IV case \citep{mogstadetal2021}. See discussion in Section \ref{subsec:discussion}.
\begin{assumption}[Monotonicity]\label{ass:monot} Conditional on $X$, for all $z, z' \in \mathcal{Z}$, either $D_i(z) \geq D_i(z')$ for all individuals $i$, or $D_i(z) \leq D_i(z')$ for all individuals $i$. 
\end{assumption}

%
%
Assumption \ref{ass:monot} states that, conditional on $X$, an exogenous change in the semi-IVs $Z$ from $z$ to $z'$ either weakly increases or weakly decreases selection into treatment for all individuals. This rules out non-uniform effects of $Z$ on treatment choice: if some individuals opt into treatment when $Z$ changes from $z$ to $z'$ ($D_i(z')=1 > D_i(z) = 0$ for some $i$), then no one opts out of the treatment because of this change ($D_i(z') = 0 < D_i(z) = 1$ for no $i$). Conversely, if some opt out, no one opts in. That is, there are only compliers and no defiers. \\
%


\noindent \textbf{Equivalent selection model with latent resistance to treatment.} For IVs, the monotonicity condition is equivalent to the existence of a weakly separable rule for the selection into treatment \citep{vytlacil2002}. The same holds for semi-IVs: monotonicity is equivalent to assuming that the treatment is determined by a weakly separable rule of the form
\begin{align}\label{eq:selection}
D = \mathds{1}\{ P(Z, X) - V \geq 0 \}	
\end{align}
for some function $P$ and a random variable $V$ that is continuously distributed given $X$. \\
\indent Under the selection rule \eqref{eq:selection}, Assumption \ref{ass:exogeneity} can be rewritten as follows.
{\theoremstyle{plain}
\newtheorem*{specialassumption}{Assumption \ref{ass:exogeneity}'}
\begin{specialassumption}
(i) $V \indep Z | X$, and (ii) for $d=0,1$, $(Y_d, V) \indep Z_{1-d} | (Z_d, X)$. 
\end{specialassumption}
}
\indent As is standard in the IV literature, since $V$ is continuously distributed given $X=x$ for all $x$, one may normalize $V|X \sim \text{Unif}(0, 1)$ without loss of generality.\footnote{This normalization is standard and innocuous given the model assumptions. Indeed, if the selection rule is $D = \mathds{1}\{ \nu(Z, X) - \Eta \geq 0\}$ where $\Eta$ is a general continuous random variable, the model can always be reparametrized with $P(Z, X) = F_{\Eta|X}(\nu(Z, X))$ and $V=F_{\Eta|X}(\Eta)$, ensuring $V|X \sim \text{Unif}(0, 1)$, which corresponds to Equation \eqref{eq:selection} under this normalization.}  Given the independence of the semi-IVs and $V$ (Assumption \ref{ass:exogeneity}'), it follows that $V | Z=z, X=x \sim \text{Unif}(0,1)$ for all $(z, x) \in \mathcal{Z}\times\mathcal{X}$. Under this normalization, the function $P(z, x)$ is the propensity score function, i.e., $P(z, x) \equiv \mathbb{E}[D | Z=z, X=x]$, and $P=P(Z,X)$ is the 
propensity score random variable with support $\mathcal{P}$. \\ 
\indent The normalized $V$ can be understood as the latent \textit{resistance to treatment} \citep{cornelissen2016late}, ranked from the lowest ($V=0$) to the highest (	$V=1$). Individuals with a low $V$ exhibit minimal resistance to treatment, opting in even when the propensity score $P$ is low, meaning they require fewer incentives to select into treatment. Conversely, those with a high $V$ have a strong resistance to treatment and only opt in if $P$ is high. \\

\noindent \textbf{Regularity conditions.} Following \cite{heckmanvytlacil2005}, identifying treatment effects also requires two additional mild regularity conditions: (i) $\mathbb{E}|Y_1|$ and $\mathbb{E}|Y_0|$ are finite (ensuring that the treatment parameters are well defined), and (ii) $0 < \textrm{Pr}(D=1|X=x) < 1$ for all $x \in \mathcal{X}$, ensuring the existence of a treated and an untreated population for every $x$. 

\subsection{Discussion about the model}\label{subsec:discussion}

\noindent \textbf{Discussion.} Assumptions \ref{ass:semiiv} and \ref{ass:monot}, along with the regularity conditions, define the semi-IV model. The flexibility of this model is best understood by considering what is \textit{not} imposed \citep{mogstad2018ecta}. First, similar to the IV model, it does not restrict the dependence between $D$ and the potential outcomes. This allows for rich forms of observed and unobserved heterogeneity, where the causal effect of $D$ on $Y$ can vary with observed covariates, semi-IVs, and unobservables. Second, except for the monotonicity assumption, the selection rule remains general.\footnote{Attention: when one assumes additive separability of the error term in the $Y_d$ equations \citep[e.g., ][]{frenchtaber2011, dhaultfoeuillemaurel2013}, monotonicity is weaker than the (extended) Roy selection rule. However, in general models with nonseparable unobservables in $Y_d$, monotonicity is distinct from -- rather than strictly weaker than -- the (extended) Roy selection rule with semi-IVs. For instance, interactions between semi-IVs and outcome-specific unobservables in $Y_d$ violate monotonicity under a Roy specification.  } The model does not specify the reasons behind the individuals' treatment choices, in contrast to Roy models, where $D = \mathds{1}\{ Y_1 > Y_0 \}$, or extended Roy models, where $D = \mathds{1}\{ Y_1 - Y_0 + \nu(Z, X) > 0 \}$ for some function $\nu$ of the observables $Z$ and $X$. This flexibility accommodates various decision-making processes, including partial selection on gains ($Y_1 - Y_0$). 
Third, the effect of the semi-IVs on the potential outcome from which they are not excluded is unrestricted and possibly heterogenous, varying with factors like latent resistance to treatment, $V$.  Finally, the semi-IVs can (and generally will) be correlated. This is not a problem since the exclusion restriction holds conditional on the included semi-IV (and covariates). The only cost of the correlation between the semi-IVs is that it may weaken their relevance by limiting their joint variation.\footnote{Assumption \ref{ass:relevance} rules out the degenerate case where $Z_0$ and $Z_1$  are perfectly correlated, as each semi-IV must be relevant for selection into treatment conditional on the other semi-IV (and covariates).} \\

\noindent \textbf{The role of monotonicity.} 
\noindent Conditional on $X=x$, if an individual selects $D=1$ when $Z=z$, one can conclude that $V \leq P(z, x)$, using \eqref{eq:selection} and the independence of $V$ and $Z$ given $X$. 
Conversely, $D=0$ when $V > P(z,x)$. Therefore, for any $z' \neq z$, if $P(z,x)=P(z',x)$, then $D_i(z', x) = D_i(z, x)$ for all individuals. The propensity score serves as a \textit{sufficient statistic} for the effect of the semi-IVs on selection: any two values of the vector of semi-IVs yielding the same propensity score select the same individuals into treatment. \\ 
\indent This property is key to recover treatment effects, as it allows us to \textit{match compliers} with respect to different semi-IV margins. Starting from a given environment, if a shift of $Z_0$ and another shift of $Z_1$ induce the same change in observable propensity score, under monotonicity we conclude that they induced the same compliers, i.e., the same set of $V$, into (or out of) the treatment. 
If each treatment-specific semi-IV represent a completely different type of incentive/disincentive (e.g., tuition fee versus distance from college), this assumption could be arguably strong \citep{mogstadetal2021}.\footnote{Although monotonicity is stronger with multiple IVs, many empirical paper still use multiple IVs because the practical benefits -- richer first stage variation -- can outweigh the cost \citep[e.g.,][]{carneiroheckmanvytlacil2011}.} 
Since the semi-IVs are typically the same type of incentive but treatment-specific (e.g., treatment-specific "prices"), assuming monotonicity, i.e., that all individuals uniformly value the semi-IVs (or their comparison) is more reasonable \citep{mogstad2024policy}.\footnote{It is possible to relax the monotonicity assumption. Indeed, to identify mean potential outcomes (marginal treatment responses) \textit{separately}, a weaker unordered partial monotonicity \citep{mogstadetal2021} assumption is sufficient. The issue is that these responses are identified for possibly different sets of compliers across semi-IVs. In order to align these sets of compliers and obtain marginal treatment effects, an additional assumption is needed. Standard monotonicity is a solution, but not the only one. For instance, complier alignment and point identification (of margin-specific marginal treatment effects) can also be achieved using \textit{comparable compliers} assumptions as in \cite{mountjoy2022community}. Online Appendix \ref{app:discrete} shows how partial monotonicity combined with comparable compliers identifies margin-specific MTE in the discrete-treatment case with semi-IVs; adapting it to the binary case is straightforward.} \\

\noindent \textbf{Implications of the partial exclusion.} 
\noindent Assumption \ref{ass:exogeneity}' (or \ref{ass:exogeneity}) implies that
\begin{align}\label{eq:mean_independence}
	\mathbb{E}[Y_d | V, Z_d, Z_{1-d}, X] = \mathbb{E}[Y_d | V, Z_d, X] \quad \text{ for } d = 0, 1. 
\end{align}
Intuitively, conditional on $X$, $Z_d$, and $V$, a shift in the excluded semi-IV $Z_{1-d}$ has no effect on the \textit{latent} average potential outcome $Y_d$. 
\noindent However, this shift affects the propensity score, thereby altering the composition of the treated population and, as a result, the \textit{observed} outcomes. 
Indeed, for the treated population, for all $z=(z_0, z_1) \in \mathcal{Z}$ and $x \in \mathcal{X}$, we have
\begin{align*}
	\mathbb{E}[DY|Z=z, X=x] &= \mathbb{E}[Y_1 | D=1, Z=z, X=x] \times \mathbb{E}[D|Z=z, X=x] \\
	&= \mathbb{E}[Y_1 | V \leq P(z, x), Z=z, X=x] \times P(z, x) \\
	&= \mathbb{E}[Y_1 | V \leq P(z, x), Z_1=z_1, X=x] \times P(z, x), 
\end{align*}
where the first equality follows from the law of total expectation, the second from $D=1$ being equivalent to $V\leq P(z,x)$, and the third from $(Y_1, V) \indep Z_0|Z_1,X$. 
Similarly, for the untreated population, we have
\begin{align*}
	\mathbb{E}[(1-D)Y|Z=z, X=x] = \mathbb{E}[Y_0 | V > P(z, x), Z_0=z_0, X=x] \times (1-P(z, x)). 
\end{align*}
This makes it clear that for each $d \in \{0,1\}$ the only effect of $Z_{1-d}$ on the conditional mean of $Y\mathds{1}\{D=d\}$ given $Z_d$ and $X$ is through the propensity score $P(Z,X)$. Since $P(Z,X)$ is identified from data on $D, Z,$ and $X$, we can condition directly on $P$ instead of $Z_{1-d}$, yielding
\begin{align}\label{eq:observable_property}
	\mathbb{E}[Y \mathds{1}\{D=d\} | Z=z, X=x]  
	&= \mathbb{E}[Y \mathds{1}\{D=d\} | P=P(z,x), Z_d=z_d, X=x],  
\end{align}
for $d=0,1$ and for all $z=(z_0, z_1) \in \mathcal{Z}$ and $x \in \mathcal{X}$. 
Similar to an IV, $Z_{0}$ (resp. $Z_1$) affects $DY$ (resp. $(1-D)Y$) only through its effect on the selection propensity $P$. The excluded semi-IV ($Z_{1-d}$) can shift the selection probability and alter the composition of the treated and control groups while holding the included semi-IV ($Z_d$) and covariates ($X$) fixed. Therefore, conditional on $Z_d$ and $X$, $Z_{1-d}$ acts as an IV for the effect of $D$ on $Y\mathds{1}\{D=d\}$. \\


\noindent \textbf{Comparison with IVs.} Each semi-IV acts as an IV for the observable outcomes of either the treated or the untreated subpopulation. An IV is stronger, excluded from the outcome $Y$, meaning it is excluded from both $DY$ and $(1-D)Y$. 
Since the conditions for a valid semi-IV are strictly weaker than those for an IV, a valid semi-IV is easier to find in practice. Indeed, as shown in the examples below, there are many selection-specific variables that are invalid IVs because they directly affect their alternative-specific potential outcome; but they are valid semi-IVs because they are credibly excluded from the other potential outcomes. \\ 
\indent While a single valid semi-IV is strictly easier to find than a valid IV, it cannot serve as a substitute for a valid IV on its own. 
To replace an IV, at least two complementary semi-IVs are needed, with at least one excluded per potential outcome. 
Thus, there are two possible approaches to address endogeneity: finding (at least) one IV or two complementary semi-IVs. The full exclusion restriction is often the main obstacle to finding a valid IV. Moreover, as discussed in the examples below, finding two complementary semi-IVs is often not much harder than finding one, since treatment-specific versions of a similar variable can be used as semi-IVs. Therefore, complementary semi-IVs should often be more accessible than an IV. \\
\indent Then, in the binary treatment case, the main cost of the semi-IV approach is that the monotonicity assumption jointly imposed on both semi-IVs is stronger than when imposed on a single IV. Practitioneers therefore face a trade-off between assuming stronger monotonicity on the selection (semi-IV) or full exclusion on the outcome equation (IV). \\
\indent This main trade-off disappears in the discrete treatment case (with $J > 2$ distinct alternatives). There, identification with IVs already requires multiple ($J-1$) IVs and a correspondingly stronger montonicity. I show in Online Appendix \ref{app:discrete} how identification of discrete treatment effects obtained with $J-1$ IVs \citep{heckmanurzuavytlacil2006, mountjoy2022community} can be obtained under the same assumptions with $J$ semi-IVs instead. For intuition, the main text focuses on the binary case.  \\
\indent Another cost of the semi-IVs approach is that matching compliers ideally require at least one continuously distributed semi-IV. This is not an additional cost for MTE identification which also requires continuous IVs. However, this is costly for the LATE which standard IV methods can identify with a single binary instrument while binary semi-IVs generally cannot. Fortunately, as visible in the examples described in the next subsection, most examples of semi-IVs are, in fact, continuously distributed.  \\
\indent Finally, and perhaps surprisingly, the possible non-exclusion of the semi-IVs from some potential outcomes has an unexpected upside: if individuals select into treatment partly based on their gains, $Y_1 - Y_0$, a variable affecting the potential outcomes should be relevant for the selection by construction. This should make it easier to find "non-weak" semi-IVs.  

\subsection{Examples and discussion}\label{subsec:example}

\noindent \textbf{Example 1 (Occupation choice).} 
Consider the question of identifying the effects of occupation choice ($D$) on earnings ($Y$). Examples include how earnings are impacted by working in the manufacturing sector \citep{heckmansedlacek1985, heckmansedlacek1990}, agriculture \citep{lagakoswaugh2013}, or a white-collar or blue-collar occupation \citep{keanewolpin1997}, as well as by joining a union \citep{lee1978unionism}, attending college \citep{angristkrueger1991, card1995, card2001, carneiroheckmanvytlacil2011}, or selecting a specific college major or training \citep{arcidiacono2004, arcidiacono2020ex, eckardt2024}.
To answer these, one needs to address endogeneity due to the self-selection of individuals into occupations based on their unobserved occupation-specific skills. To do so, occupation-specific characteristics such as the occupation-specific wage rate, unemployment rate, task content, firm size, revenue, productivity, or subsidies can serve as valid semi-IVs. 
More precisely, identification can be achieved by leveraging observable variations of these semi-IVs across time and geographic markets, and the corresponding variations in observed outcomes ($Y$) and occupation choices ($D$). \\ 
%
%
\indent \textit{Returns to working in manufacturing on earnings.} The main application in this paper (Section \ref{sec:application}) focuses on the effect of the manufacturing ($D=1$) versus the nonmanufacturing ($D=0$) sector choice on earnings ($Y$). I focus on young (aged $18$ to $30$) white men with only high school education, and all the analysis is conducted conditional on being in this subpopulation ($X$). I use the characteristics of the manufacturing ($Z_1$) and nonmanufacturing ($Z_0$) sector of the market (state of residence) of the workers as semi-IVs.  
First, these are likely to be relevant variables for sectoral choice because, all else equal, a larger or more active sector is more attractive to new (young) workers. Monotonicity assumes that workers have uniform valuation of the sector-specific characteristics (i.e., not interacted with latent $V$). 
Second, the aggregate sector-specific characteristics are likely to be exogenous with respect to atomistic workers' unobserved sector-specific skills and latent resistance to treatment ($V$). Exogeneity requires that, aside from differences in the semi-IVs, the various markets represent otherwise comparable environments. To ensure this, I control for time trends and pre-existing differences across markets by including year and market (state) fixed effects in $X$. This addresses concerns about the self-selection of high-skilled workers into more favorable markets and accounts for the global evolution of worker skills. Consequently, the identifying variation consists of deviations of sector-specific characteristics from these fixed effects, which I will occasionally refer to as "shocks to" the variables.\footnote{Instead of fixed effects, one could control for the permanent levels of the semi-IVs in each market in $X$, as done by \cite{carneiroheckmanvytlacil2011} for some of their IVs, or use shocks to these variables directly as semi-IVs. } 
Third, (shocks to) the sector characteristics are not valid IVs because a worker's earnings are determined as a function of the worker's unobserved skills and the characteristics of the selected sector \citep{heckmansedlacek1985}: the sector-specific semi-IVs are \textit{included} in the corresponding sector-specific potential earnings, i.e., $Z_d$ influences $Y_d$. 
However, once the included manufacturing characteristics are controlled for, similar characteristics of the other sectors do not affect the manufacturing potential earnings, and vice versa: $Y_d$ is independent of $Z_{1-d}$ conditional on $Z_d$ and $X$. \\
\indent Note that, in a given market, (the shocks to) the manufacturing and nonmanufacturing characteristics are likely to be correlated. As long as they are not perfectly correlated, so that some markets experience relatively larger shocks to their manufacturing sectors compared to their nonmanufacturing sectors than other markets, this is not a problem for identification. \\
\indent As with standard IVs, I work in partial equilibrium and assume that general equilibrium (GE) effects that could threaten partial exclusion restrictions are negligible. For instance, an increase in manufacturing attractivity might alter worker composition, thereby affecting equilibrium wage rates in both sectors and indirectly influencing nonmanufacturing workers' earnings. 
Fortunately, and unlike the IV case, GE concerns are partly mitigated with semi-IVs. 
Indeed, both sectors characteristics can be correlated, or jointly determined in equilibrium, the semi-IVs' validity only requires that, once determined, conditional on $Z_d$, $Z_{1-d}$ is excluded from $Y_d$. This property often holds, even in some GE models \citep{heckmansedlacek1985, heckmansedlacek1990} where the effect of $Z_{1-d}$ on $Y_d$ are captured by the "effect" of $Z_{1-d}$ on $Z_d$. Controlling for the appropriate sector characteristics $Z_d$ (e.g., productivity, size) absorbs a large part of the GE concerns. This property is most salient when $Z_0$ and $Z_1$ are the sector-specific wage rates: once we control for the equilibrium wage rate in sector $d$, the wage rate in the other sector has no effect on $Y_d$, which only depends on the workers' unobserved skills and on the sector wage rate/price of skills.\footnote{Note that the GE threats are further mitigated by the fact that I focus on a small subpopulation of young workers, which may be too small to influence the aggregate wage rates.} \\ 
\indent Which sector-specific characteristics to use? Ideally, one would like to observe sector-specific price of skills (wage rates) across markets. Contrary to other examples below, these are not available in this application. Instead, I use sector-specific size, as measured by sector-specific GDP, as semi-IVs in the application of Section \ref{sec:application}. One could use more characteristics per sector as semi-IVs, but according to the GE model of \cite{heckmansedlacek1985}, the size should be sufficient, as it determines the sector-specific marginal productivity of labor, and thus the sector-specific wage rates.\footnote{In this example, since I do not use the sector-specific wage rates as semi-IVs, a threat to the semi-IVs' validity could be monopsony power with strategic interactions between sectors, such as manufacturing firms adjusting wages based on the size of the nonmanufacturing sector. However, evidence suggests within-sector strategic responses are small, with \cite{staiger2010there} estimating an elasticity of only $0.128$ in the case of VA (Veteran Affairs) and non-VA hospitals. Between-sector interactions are likely smaller and negligible, as sectors do not explicitly compete. Given that I focus on large markets (states) with many firms in each sector, individual firms are relatively small and primarily compete within their own sector, reducing the risk of cross-sector strategic interactions. Thus, I assume that firms are price takers without monopsony power. } 
For more details and a complete example of a sector choice model where the sector-specific sizes (measured by total employee compensation in their case) would be valid semi-IVs, refer to \cite{heckmansedlacek1985, heckmansedlacek1990}. More recently, \cite{eckardt2024} employs a similar model to justify the use of sector-specific vacancy rates as valid semi-IVs. 
In terms of timing, I use the previous year (one-year lagged) sector sizes as semi-IVs, as I assume it reflects the local market conditions when the workers made their sector choices, i.e., at the beginning of the period. In principle, additional lags or current-year sizes could also be considered. \\


\noindent \textbf{Example 2 (Location/migration choice).} Consider the related problem of identifying the marginal earnings returns to endogenous location/migration choices of workers \citep{borjas1987, dahl2002, chiquiar2005international, roca2017learning} 
or firms \citep{combes2012productivity}. Location-specific (variations in) expected earnings \citep{kennanwalker2011}, welfare benefits \citep{borjas1999welfare, abramitzky2009effect, agersnap2020welfare}
, tax rates \citep{akcigit2016inventors, kleven2020taxation}, 
or more generally, any place-based policy \citep{neumark2015place} or measures of local amenities, may serve as valid semi-IVs. Again, these are (i) relevant for the location choice, (ii) exogenous with respect to the individuals' unobserved characteristics, and (iii) the partial exclusion holds because, conditional on one's location-specific shocks, the shocks to the other locations are excluded from one's potential outcome. One can use temporal variation in these location-specific semi-IVs for identification. \\

\noindent \textbf{Example 3 (Demand estimation).} 
Semi-IVs can be used to estimate the demand for different products. For instance, consider the effect of car type -- gasoline ($D=0$) or diesel ($D=1$) -- on subsequent energy consumption, of gasoline ($Y_0$) or diesel ($Y_1$). Endogeneity arises because individuals with different unobserved preferences (e.g., for higher mileage) select different cars. Estimating demand for gasoline/diesel cars is crucial for designing optimal energy policies, such as taxation or subsidies \citep{verboven2002quality, grigolon2018consumer}.
Local market energy-specific prices serve as semi-IVs: they are relevant for the choice of type of car but do not affect the energy consumption of owners of the other car type (e.g., gas prices do not affect the energy consumption of diesel-car owners).\footnote{To address the potential endogeneity of energy prices with respect to unobserved individual characteristics, one can use supply side shocks to the prices (e.g., cost shifters) or exogenous energy-specific subsidies/tax changes as semi-IVs instead of the prices. Additionally, one can control for market fixed effects. }
More broadly, energy-specific (shocks to) prices or subsidies can serve as semi-IVs to estimate the demand for energy based on appliance choice, such as solar panel adoption \citep{degrooteverboven2019, feger2022solar} or electric versus non-electric home appliance purchases \citep{dubin1984}. \\

\noindent \textbf{Example 4 (Measuring effectiveness).} 
Semi-IVs can also be used to measure the effectiveness (or value-added) of specific types of institutions or products, net of selection effects. For instance, in education, consider the estimation of the effect of charter schools ($D=1$) \citep{abdulkadiroglu2011charter, angristpathakwalters2013, dobbiefryer2020}, 
boarding schools \citep{behaghel2017ready}, catholic schools \citep{altonjietal2005}, or private schools \citep{angrist2002vouchers} 
compared to public schools ($D=0$), on academic achievements or future labor market outcomes ($Y$). In health, consider the estimation of the effectiveness of different insurance plans (e.g., private vs public) \citep{finkelsteinhendrenluttmer2019, abalucketal2021}, type of hospitals \citep{gowrisankaran1999estimating, geweke2003hospital}, 
or nursing homes \citep{einav2022producing} on health outcomes. 
In all these examples, variations in institution-specific prices, subsidies \citep{finkelstein2019subsidizing}, size, funding, or measures of quality (number of teachers/practitioners per student/patient) can serve as semi-IVs. \\ 

\noindent \textbf{Randomized experiments with semi-IVs.} 
Even in randomized experiments, a randomly assigned intervention is not necessarily a valid IV. While randomization ensures exogeneity with respect to ex ante unobservables, the intervention itself may not be excluded from the outcome of interest \citep{angristimbensrubin1996}. For instance, conditional cash transfers, such as vouchers for private schooling in Colombia \citep{angrist2002vouchers}, likely affect students’ achievements directly through income effects for families who receive them if the student attends a private school. As a result, only the intention-to-treat effect of the policy can be estimated, not the treatment effect of attending private school. 
However, the voucher assignment has no impact on students attending public schools, who receive no money regardless of their assignment, making it a valid semi-IV ($Z_1$), excluded from outcomes associated with attending a public school ($Y_0$).\footnote{The same kind of reasoning applies to the Vietnam lottery draft \citep{angrist1990}. Non-veterans who were drafted probably had to behave differently or face the consequence of their non-compliance (e.g., jail for conscientious objectors) compared to non-veterans who were not drafted. Thus, the draft is only excluded from the outcomes (subsequent earnings) of veterans, but not from the outcomes of non-veterans. Identifying the effect of being a veteran would require a complementary semi-IV excluded from the outcome of non-veterans, for example differential local tax rates/benefits for veterans at the time of enrollment. } 
To identify the (local) treatment effect of private schools, the Colombian government could have designed the experiment differently by assigning two random monetary amounts: one for public schools ($Z_0$) and one for private schools ($Z_1$). Both could be nonzero since $Z_0$ is excluded from private school outcomes ($Y_1$) and $Z_1$ from public school outcomes ($Y_0$). By randomly assigning the monetary amounts, this design would facilitate manipulating treatment probabilities. 
Moreover, it would allow policymakers to study two effects simultaneously: the treatment effect of private school attendance and the policy/semi-IV effect of income increases on achievement.
This approach also addresses ethical concerns by ensuring all families receive some support, not just a select few. \\

\noindent \textbf{Discussion.} 
With the exception of a few education-related examples, credible IVs are rarely available to address endogeneity in the settings and questions discussed above. Identification has so far mainly relied on distributional assumptions, functional form restrictions, structural models, or on assuming selection only on observables. Semi-IVs expand the researchers' toolkit by enabling IV-like identification with readily available variables, allowing them to credibly tackle a broader range of questions, even when credible IVs are hard to find.


\section{Identification}\label{sec:identification}

\indent Similar to IVs, semi-IVs identify average treatment effects for \textit{compliers} induced into treatment by their shifts: the LATE and the MTE. 
More precisely, each semi-IV separately identifies mean potential outcomes under treatment and control at specific unobserved $V$, which are then combined to obtain treatment effects. 
In this section, I show the identification of the MTE, followed by the LATE. 
I also describe the identification of the \textit{direct effects} of the semi-IVs on the potential outcomes from which they are not excluded.

\subsection{Marginal treatment effects and responses}\label{subsec:mte}

\subsubsection{Treatment parameters definition} 

\noindent \textbf{Marginal treatment effect (MTE).} The marginal treatment effect (MTE) was introduced by \cite{bjorklund1987estimation} and generalized to the nonparametric case by \cite{heckmanvytlacil1999, heckmanvytlacil2005, heckmanvytlacil2007b}. With semi-IVs, the MTE is naturally defined by
\begin{align}
	\text{MTE}(v, z_0, z_1, x) = \mathbb{E}[ Y_1 - Y_0 | V=v, Z_0=z_0, Z_1=z_1, X=x ].
\end{align}
The marginal treatment effect MTE$(v, z_0, z_1, x)$ is the average causal effect of $D$ on $Y$ for individuals with unobserved resistance to treatment $V=v$ and observed characteristics $X=x, Z_0=z_0, Z_1=z_1$. It is called "marginal" treatment effect because it is the effect of treatment for individuals at the margin of taking up treatment when $P=v$. \\
\indent The main difference with the MTE using IVs is that the MTE also depends on the semi-IVs here. 
Since $Z_0$ may affect $Y_0$, and $Z_1$ may affect $Y_1$, we need to condition on $(Z_0, Z_1)$, just as we condition on $X$. Fortunately, this conditioning does not impede identification. \\

\noindent \textbf{Marginal treatment responses (MTR$_d$).} 
Rather than identifying the MTE directly, we focus on the two \textit{marginal treatment responses} (MTR) functions \citep[see for example][]{mogstad2018ecta, mogstad2018review} instead. With semi-IVs, these are defined by
\begin{align}
	m_0(v, z_0, x) &= \mathbb{E}[ Y_0 | V=v, Z_0 = z_0, X=x], \tag{MTR$_0$} \label{eq:mtr0} \\
	\text{ and } \quad m_1(v, z_1, x) &= \mathbb{E}[ Y_1 | V=v, Z_1 = z_1, X=x]. \tag{MTR$_1$} \label{eq:mtr1}
\end{align}
For each $d$, the MTR$_d$ is the \textit{mean potential outcome} $Y_d$ for individuals with unobserved resistance to treatment $V=v$ and observed characteristics $Z_d=z_d$ and $X=x$. With the semi-IV exclusion property, $m_d$ only depends on $Z_d$ but not on $Z_{1-d}$. One may refer to $m_1$ and $m_0$ as the \textit{mean potential outcomes under treatment} and \textit{under control}, respectively. \\ 
\indent The MTE is equal to the difference of MTRs, 
\begin{align}
	\text{MTE}(v, z_0, z_1, x) &= \ \mathbb{E}[ Y_1 | V=v, Z=(z_0,z_1), X=x] - \mathbb{E}[ Y_0 | V=v, Z=(z_0, z_1), X=x ] \nonumber \\
	&=\ m_1(v, z_1, x) - m_0(v, z_0, x). \label{eq:mte_mtrlink}
\end{align}
The second equality holds because we can remove conditioning on the excluded semi-IV using the mean independence equation \eqref{eq:mean_independence}, i.e., using the partial exclusion of the semi-IVs.

\subsubsection{Identification} 

For notational convenience, omit $X$ in the notation and proceed conditional on $X=x$. \\
\indent In the data, we observe $(Y, D, Z_0, Z_1)$. Since we observe $(Z_0, Z_1)$ and $D$, we also identify the propensity score $P=P(Z_0, Z_1)=\textrm{Pr}(D=1|Z_0, Z_1)$. Therefore, consider that we observe $Y, D, Z_0, Z_1,$ and $P$. 
From there, the identification of each MTR with semi-IVs is similar to the identification of MTRs with a single IV \citep{mogstad2018review}. The only difference is that it is not the same source of variation (not the same semi-IV) that identifies both MTRs at once. Instead, $m_1$ is identified by using $Z_0$ as an instrument for the effect of $D$ on $DY$ given $Z_1$ and $X$, while $m_0$ is identified by using $Z_1$ as an instrument for the effect of $D$ on $(1-D)Y$ given $Z_0$ and $X$. $Z_0$ and $Z_1$ play, in turn, the role of an instrument (for $m_1$ and $m_0$, respectively) and the role of a covariate (for $m_0$ and $m_1$, respectively) in the proof. Once both MTRs have been separately identified, one can recover the MTE using Equation \eqref{eq:mte_mtrlink}. Note that the monotonicity of the selection implicitly plays a key role: although each MTR is identified by shifting a different semi-IV, any change in a semi-IV can be mapped to its corresponding marginal compliers, characterized by $V$, through the associated propensity score. 
\noindent First, I show the identification of each MTR separately, then that of the MTE. \\ 

\noindent \textbf{Identification of MTR$_1$.}  Let us proceed conditional on a specific $Z_1=z_1 \in \mathcal{Z}_1$. To identify the mean potential outcome under treatment ($m_1$), use $Z_0$ as an IV excluded from $Y_1$. $Y_1$ is not observed for the untreated individuals, but recall that $Y=DY + (1-D)Y$ and focus on the observable treated population outcomes, $DY = DY_1$. 
Any shift in $Z_0$ affects the outcome of interest ($DY$ here) only through its effect on the propensity score, $P=P(Z_0, z_1)$. This enables us to identify the MTR$_1$ by marginally shifting $P$ at fixed $Z_1=z_1$ using $Z_0$. \\
\indent Assume that the semi-IV excluded from $Y_1$, $Z_0$, is continuous and relevant for all $Z_0=z_0 \in \mathcal{Z}_0$, conditional on $Z_1=z_1$, i.e., that $\partial P(z_0, z_1)/\partial z_0 \neq 0$. In this case, $P$ is continuously distributed conditional on $Z_1=z_1$. 
\noindent Using the selection equation \eqref{eq:selection}, we have that $D=1$ when $V \leq P$. So, for any $P=p$ in the interior of the support of $P$ given $Z_1=z_1$, we observe
\begin{align}\label{eq:yd1}
	\mathbb{E}[DY | Z_1=z_1, P=p] 
	&= \mathbb{E}[ Y_1 | D=1, Z_1=z_1, P=p] \times p \nonumber \\
	&= \mathbb{E}[ Y_1 | V \leq p, Z_1=z_1] \times p \nonumber \\
	&= \int_0^p \frac{\mathbb{E}[Y_1 | V=v, Z_1=z_1]}{p} dv \times p \quad \text{ since } V \sim \mathcal{U}(0,1)  \indep Z_1 \nonumber \\
	&= \int_0^p m_1(v, z_1) dv.  
\end{align}
Thus, taking the derivative with respect to the propensity score, we have
\begin{align}\label{eq:mtr1}
	m_1(v, z_1) = \frac{\partial}{\partial p} \mathbb{E}[ DY | P = p, Z_1=z_1] \Big|_{p=v}, 
\end{align}
where $\mathbb{E}[ DY | P(z_1, Z_0) = p, Z_1=z_1]$ is observed and so is its derivative if $v$ is in the interior of the support of $P$ given $Z_1=z_1$. Therefore, for any $z_1$ and any $v$ in the interior of the support of $P$ given $Z_1=z_1$, the MTR$_1$ given $Z_1=z_1$ and $V=v$ is identified by \eqref{eq:mtr1}. \\
\indent The intuition behind why the derivative gives the MTR can be explained as follows. Fix $Z_1=z_1$. When $P=p$, individuals with $V \leq p$ are treated, and those with $V=p$ are at the margin of the treatment, meaning they are indifferent between being treated or not. Now, consider a marginal increase in $P$ from $p$ to $p'=p+dp$ by shifting $Z_0$ while holding $Z_1=z_1$ fixed. This shift moves the marginal individuals with $V=p$ into treatment. Their expected $Y_1$ is $\mathbb{E}[Y_1 | V=p, Z_1=z_1] = m_1(p, z_1).$ Since these individuals are now treated, the average observable outcome $DY$ changes by the proportion of individuals entering treatment times their average $Y_1$, i.e., $d(\mathbb{E}[DY | Z_1=z_1, P=p]) = dp \times \mathbb{E}[Y_1 | V=p, Z_1=z_1]$. Dividing both sides by $dp$ normalizes this change in $DY$ and yields $m_1(p, z_1)$. 
Thus, the derivative of the average $DY$ with respect to the propensity score identifies the mean potential outcome under treatment for individuals who are indifferent at $V=p$ (and $Z_1=z_1$). \\
\indent The key to identification is having a source of variation that can shift the propensity score at fixed $Z_1$ without directly affecting $DY$ otherwise. This is precisely the role of $Z_0$, as visible in \eqref{eq:observable_property}. Implicitly, the derivative in Equation \eqref{eq:mtr1} is identified by the local IV regression of $YD$ on $D$, instrumented by $Z_0$, at a specific value $\tilde{z}_0$ of $Z_0$ such that $P(\tilde{z}_0, z_1) = v$, i.e., \begin{align*}
	\frac{\partial}{\partial p} \mathbb{E}[DY | P=p, Z_1=z_1] \Big|_{p=v} = \frac{\partial \mathbb{E}[DY | P=P(z_0, z_1), Z_1=z_1] / \partial z_0 }{\partial P(z_0, z_1)/\partial z_0} \Big|_{z_0=\tilde{z}_0}.    
\end{align*} 
In order to marginally shift $P$ by shifting $Z_0$, a relevant and continuously distributed $Z_0$ is needed, similar to the continuous IV requirement for MTE identification with IVs. \\ 

\noindent \textbf{Identification of MTR$_0$.} Similarly, for any $Z_0 = z_0 \in \mathcal{Z}_0$, $m_0(v, z_0)$ is identified for any $v$ in the interior of the support of the propensity score $P$ given $Z_0=z_0$ by\footnote{\label{footnote:minus}The minus sign comes from the fact that, $(1-D)=1$ when $V > P$, so $\mathbb{E}[(1-D)Y | Z_0 = z_0, P=p] = \int_p^1 m_0(v, z_0) dv,$ and the derivative with respect to $p$ at $p=v$ is applied to the lower bound of the integral.}
\begin{align}\label{eq:mtr0}
	m_0(v, z_0) = - \frac{\partial}{\partial p} \mathbb{E} [ (1-D)Y | P=p, Z_0=z_0 ] \Big|_{p=v}.
\end{align}
Holding $Z_0=z_0$ fixed, we shift the propensity score by shifting the other semi-IV, $Z_1$. Again, obtaining marginal shifts in $P$ requires $Z_1$ to be continuously distributed and relevant. \\

\noindent \textbf{Identification of the MTE.} 
Given $Z_1=z_1$, $m_1$ is identified for all $v$ in the interior of the support of the propensity score given $Z_1=z_1$. Given $Z_0=z_0$, $m_0$ is identified for all $v$ in the interior of the support of the propensity score given $Z_0=z_0$. As a consequence, for any $(Z_0, Z_1) = (z_0, z_1) \in \mathcal{Z}$, the MTE is identified by Equation \eqref{eq:mte_mtrlink}, as 
\begin{align*}
	MTE(v, z_0, z_1) = m_1(v, z_1) - m_0(v, z_0), 
\end{align*}
for all $v$ in the interior of both, the support of $P$ given $Z_1=z_1$ and the support of $P$ given $Z_0=z_0$, i.e., the interior of the intersection of these two conditional supports.   \\

\noindent \textbf{Link with local IV} \citep{heckmanvytlacil1999}\textbf{.} With an IV, the MTE would be identified directly by the local IV regression of $Y$ on $D$, i.e., using 
\begin{align}\label{eq:localiv}
	\frac{\partial }{\partial p} \mathbb{E} \big[  Y | P=p, Z_0=z_0, Z_1 = z_1\big] \Big|_{p=v} &= \text{MTE}(v, z_0, z_1). 
\end{align}
The problem with semi-IVs is that it is not possible to shift $P$ while fixing \textit{both} $Z_0$ and $Z_1$. Instead, we split $Y$ into two subsamples, $YD$ and $Y(1-D)$, and identify $m_1$ and $m_0$ \textit{separately} on these two subsamples using a different excluded semi-IV ($Z_0$ and $Z_1$, respectively) as an instrument to shift $P$, which is feasible thanks to Property \eqref{eq:observable_property}, implied by the partial exclusions. We recover the MTE afterwards by aligning the compliers with the same underlying $V=v$. Thus, the main difference with standard local IV is that the shifts in $P$ used to identify $m_1$ and $m_0$ stem from different underlying semi-IV shifts. \\

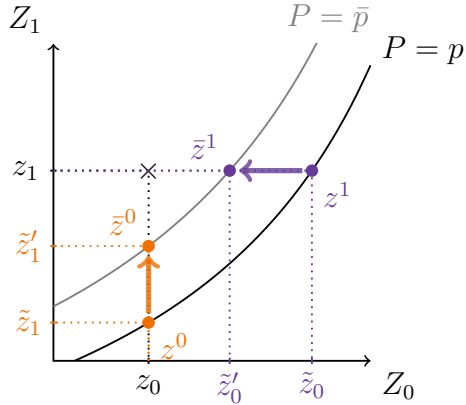
\begin{figure}[h!]
\begin{center}
\begin{tikzpicture}[x=120pt,y=120pt, line width=0.25mm] 

\def\deltac{0.3}
\def\deltawzero{-1.5}
\def\deltawone{3.5}
\def\deltatwo{-2}
\def\Probazero{0.20} 
\def\Probaone{0.90} 

\def\wone{0.6}
\def\wzero{0.3}

\draw[<->](0,1)--(0,0)--(1,0); 
\draw (0, 1) node[above left=0pt, black]{$Z_1$};
\draw (1, 0) node[below right=0pt, black]{$Z_0$};
\begin{scope} 
\clip (0, 0) rectangle (1, 1);
\draw[scale=1, domain=0:1, smooth, variable=\x, black] plot ({\x}, {(\Probazero-\deltac-\deltawzero*\x)/(\deltawone+\deltatwo*\x)});
\draw[scale=1, domain=0:1, smooth, variable=\x, gray] plot ({\x}, {(\Probaone-\deltac-\deltawzero*\x)/(\deltawone+\deltatwo*\x)});
\end{scope}
\draw (1, 0.9) node[above right=0pt, black]{$P=p$};
\draw (0.7, 1) node[above right=0pt, gray]{$P=\bar{p}$};

\draw (0.01, \wone) -- (-0.01, \wone) node[left=0pt]{$z_1$};
\draw (\wzero, 0.01) -- (\wzero, -0.01) node[below=0pt]{$z_0$};
\filldraw[black] (\wzero,\wone) node {$\times$};
\draw[dotted, black](0, \wone)--(\wzero, \wone)--(\wzero, 0);

\coordinate (z0) at (\wzero, {(\Probazero-\deltac-\deltawzero*\wzero)/(\deltawone+\deltatwo*\wzero)}); 
\coordinate (z0') at (\wzero, {(\Probaone-\deltac-\deltawzero*\wzero)/(\deltawone+\deltatwo*\wzero)}); 
\filldraw[mtr0color2] (z0) circle (2pt) node[below right=0pt] {$z^0$};
\filldraw[mtr0color2] (z0') circle (2pt) node[above left=0pt] {$\bar{z}^0$};
\draw[->, ultra thick, mtr0color2, line width=2pt, opacity=0.8] ($(z0) + (0, 0.03) $) -- ($(z0') - (0,0.03)$) node[midway, above] {}; 
\draw[dotted, mtr0color2](\wzero, 0)--(z0)--(0, {(\Probazero-\deltac-\deltawzero*\wzero)/(\deltawone+\deltatwo*\wzero)}) node[left=0pt, mtr0color2]{$\tilde{z}_1$};
\draw[dotted, mtr0color2](\wzero, 0)--(z0')--(0, {(\Probaone-\deltac-\deltawzero*\wzero)/(\deltawone+\deltatwo*\wzero)}) node[left=0pt, mtr0color2]{$\tilde{z}_1'$};
\draw[mtr0color2] (0.01, {(\Probaone-\deltac-\deltawzero*\wzero)/(\deltawone+\deltatwo*\wzero)}) -- (-0.01, {(\Probaone-\deltac-\deltawzero*\wzero)/(\deltawone+\deltatwo*\wzero)});
\draw[mtr0color2] (0.01, {(\Probazero-\deltac-\deltawzero*\wzero)/(\deltawone+\deltatwo*\wzero)}) -- (-0.01, {(\Probazero-\deltac-\deltawzero*\wzero)/(\deltawone+\deltatwo*\wzero)});

\coordinate (z1) at ({(\Probazero-\deltac-\deltawone*\wone)/(\deltawzero+\deltatwo*\wone)}, \wone); 
\coordinate (z1') at ({(\Probaone-\deltac-\deltawone*\wone)/(\deltawzero+\deltatwo*\wone)}, \wone); 
\filldraw[mtr1color2] (z1) circle (2pt) node[below right=0pt] {$z^1$};
\filldraw[mtr1color2] (z1') circle (2pt) node[above left=0pt] {$\bar{z}^1$};
\draw[->, ultra thick, mtr1color2, line width=2pt, opacity=0.8] ($(z1) - (0.03, 0) $) -- ($(z1') + (0.03, 0)$) node[midway, above] {}; 
\draw[dotted, mtr1color2] (0, \wone)--(z1)--({(\Probazero-\deltac-\deltawone*\wone)/(\deltawzero+\deltatwo*\wone)}, 0) node[below, mtr1color2]{$\tilde{z}_0$};
\draw[dotted, mtr1color2] (0, \wone)--(z1')--({(\Probaone-\deltac-\deltawone*\wone)/(\deltawzero+\deltatwo*\wone)}, 0) node[below, mtr1color2]{$\tilde{z}_0'$};
\draw[mtr1color2] ({(\Probazero-\deltac-\deltawone*\wone)/(\deltawzero+\deltatwo*\wone)}, 0.01) -- ({(\Probazero-\deltac-\deltawone*\wone)/(\deltawzero+\deltatwo*\wone)}, -0.01);
\draw[mtr1color2] ({(\Probaone-\deltac-\deltawone*\wone)/(\deltawzero+\deltatwo*\wone)}, 0.01) -- ({(\Probaone-\deltac-\deltawone*\wone)/(\deltawzero+\deltatwo*\wone)}, -0.01);
\end{tikzpicture} 
\centering
\captionsetup{justification=centering}
\caption{Semi-IV shifts identifying MTE$(p, z_0, z_1)$ (if $\bar{p}=p+dp$) and LATE$(p, \bar{p}, z_0, z_1)$} 
\label{fig:late_identification}
\end{center}
\end{figure}

\vspace{-1\baselineskip}

\noindent \textbf{Visual representation.} Figure \ref{fig:late_identification} illustrates the identification of the MTE (and LATE). It shows two distinct shifts in semi-IVs that separately identify $m_1$ and $m_0$. The black and grey curves represent all the combinations of semi-IVs $(Z_0, Z_1)$ yielding a propensity score of $P=p$ and $P=\bar{p}$, respectively. These are iso-probability curves, which are identified directly from the data.  Imagine $\bar{p}=p + dp$, where $dp$ is a marginal change in the propensity score. 
At a given unobserved resistance to treatment $V=p$, to identify $m_1(p, z_1)$, we use the shift represented by the \textcolor{mtr1color2}{purple arrow}, from $Z=z^1$ to $Z=\bar{z}^1$. This effectively shifts the selection probability from $p$ to $\bar{p}$ by exogenously changing  $Z_0$ from $\tilde{z}_0$ to $\tilde{z}_0'$ while holding $Z_1=z_1$ fixed. Similarly, to identify $m_0(p, z_0)$, we use the shift represented by the \textcolor{mtr0color2}{orange arrow}, from $Z=z^0$ to $Z=\bar{z}^0$, which shifts the selection probability from $p$ to $\bar{p}$ by exogenously changing  $Z_1$ from $\tilde{z}_1$ to $\tilde{z}_1'$ while holding $Z_0=z_0$ fixed. Since it is not possible to move the probability while holding both $Z_0$ and $Z_1$ fixed, the two shifts identifying the marginal treatment responses will always differ. Note that we generally do not use the point $(z_0, z_1)$ for identification, and it is not necessary that $P(z_0, z_1) = p$ to identify $m_0(p, z_0)$ or $m_1(p, z_1)$.\footnote{One could define the MTE at $Z=z=(z_0, z_1)$, with $P(z)=p$, in an alternative notation as $\widetilde{\text{MTE}}(z)$, equal to MTE$(P(z), z_0, z_1)$ in the general notation. Then for each $d$, one can identify the newly defined MTR,  $\tilde{m}_d(z)$ $= m_d(p, z_d)$ by marginally shifting $Z_{1-d}$ while holding $Z_d=z_d$ fixed. These treatment parameters are tied to specific values of the semi-IVs and are thus less general than the main definition. One advantage, however, is that these parameters can be identified while relaxing the monotonicity assumption. Indeed, since we focus on marginal shifts, unordered partial monotonicity combined with a comparable complier assumption \citep{mountjoy2022community} are sufficient to identify $m_d(z)$ and MTE$(z)$. See Online Appendix \ref{app:discrete}. } 

\indent Overall, the identification of both MTRs, and of the MTE, at a given $V=p$ and $(Z_0, Z_1)=(z_0, z_1)$, requires that there exists four points as illustrated in Figure \ref{fig:late_identification}: for each $d$, there must be two points, $Z=z^d$ and $\bar{z}^d$, which both have $Z_d=z_d$ and are such that $P(z^d) = p$ and $P(\bar{z}^d)=\bar{p}$ in order to identify $m_d(p, z_d)$. In other words, we need points with $Z_1=z_1$, and other points with $Z_0=z_0$, on both isocurves of probability $p$ and $\bar{p}$, allowing the propensity score to shift from $p$ to $\bar{p}$ while holding either $Z_1=z_1$ or $Z_0=z_0$ fixed. \\
\indent When we focus on the MTRs and MTE with continuous semi-IVs, if $z^1$ with $Z_1=z_1$ exists on the isocurve of probability $p$, the existence of $\bar{z}^1$ is guaranteed by the fact that $V=p$ is in the interior of the support of the propensity score given $Z_1=z_1$ (conversely for $z^0$ and $\bar{z}^0$), hence the previous conditions for the identification of the MTRs. \\
\indent In general, to identify the average treatment effects on a larger set of compliers with $V \in [p, \bar{p}]$ for any general $\bar{p} > p$ (i.e., not only marginal changes in $P$), the requirement can be directly stated in terms of the existence of these four points. This corresponds to the identification of the general LATE$(p, \bar{p}, z_0, z_1)$ that is developed in the next subsection. \\
\indent To contextualize this graph, in the application of earnings returns to manufacturing sector choice, identifying returns for workers with resistance $V=v$ in an environment with manufacturing size $Z_1=z_1$ and nonmanufacturing size $Z_0=z_0$ requires observing two environments with different relative sector sizes but the same propensity to work in manufacturing, $P=v$. The mean manufacturing potential earnings at $V=v$ and $Z_1=z_1$ are identified by marginally decreasing nonmanufacturing size (from $\tilde{z}_0$ to $\tilde{z}_0'$), which increases the propensity score from $p$ to $\bar{p}$ while holding $Z_1=z_1$ fixed. The resulting change in average $DY$ is due solely to the entry of the marginal compliers with $V=v$ into treatment, identifying their mean manufacturing earnings. 
A converse approach identifies the mean nonmanufacturing potential earnings. The difference between these two mean potential earnings gives the MTE, i.e., the returns to working in manufacturing at $V=v$ given $Z_1=z_1, Z_0=z_0$. 
Overall, identification relies on comparing four otherwise similar environments that differ in their semi-IV combinations, thus providing the required variation in treatment incentives.

\subsection{Local average treatment effect (and responses)} 
\subsubsection{Treatment parameters definition}
Following \cite{heckmanvytlacil2005}, define the generalized local average treatment effect (LATE) parameters as a function of the set of compliers they correspond to (in terms of $V$):
\begin{adjustwidth}{-1cm}{-0.5cm}
\vspace{-1\baselineskip}
	\begin{align}\label{eq:late}
	\text{LATE}(v, v', z_0, z_1, x) = \mathbb{E}[Y_1 - Y_0 | v \leq V < v', Z_0=z_0, Z_1=z_1, X=x ].
\end{align}
\end{adjustwidth}
The $\text{LATE}(v, v', z_0, z_1, x)$ is the average causal effects of $D$ on $Y$ for individuals with unobserved resistance to treatment $V \in [v, v')$ and observed characteristics $X=x, Z_0=z_0, Z_1=z_1$. It can also be interpreted as the mean gains for the individuals (compliers) who would be induced to switch into treatment if the probability of treatment changed from $v$ to $v'$, at fixed $X=x, Z_0=z_0, Z_1=z_1$.  This interpretation allows detaching the definition of the LATE parameter from the specific semi-IV shift that identifies it, which is especially convenient with semi-IVs. For a more standard definition and identification of the LATE$(z, z', x)$ tied to specific semi-IV shifts from $Z=z$ to $z'$, à la \cite{imbensangrist1994}, see Online Appendix \ref{app:late}. Since \cite{imbensangrist1994}'s LATE can always be mapped to the general LATE \eqref{eq:late}, I work with the latter for convenience due to its direct link with the MTE. \\
\indent Following the split of MTE in MTRs, define the \textit{local average treatment responses} (LATR): 
\begin{align}\label{eq:latr}
	\text{LATR}_d(v, v', z_d, x) = \mathbb{E}[Y_d | v \leq V < v', Z_d=z_d, X=x ] \text{ for } d=0,1.
\end{align}
For each $d$, the LATR$_d$ represents the mean potential outcome $Y_d$ for individuals with $V \in [v, v')$ and covariates $X=x, Z_d=z_d$. Because of the exclusion of the semi-IVs, the LATR$_d$ does not depend on $Z_{1-d}$, and this is what we exploit for identification. \\
\indent Similar to the MTE-MTR link in \eqref{eq:mte_mtrlink}, the LATE and LATRs are related by
\begin{align}\label{eq:late_latrlink}
	\text{LATE}(v, v', z_0, z_1, x) = \text{LATR}_1(v, v', z_1, x) - \text{LATR}_0(v, v', z_0, x).
\end{align}
Thus, the LATE is identified if one can identify both LATR$_d$ for compliers with $V \in [v, v')$.

\subsubsection{Identification}
Again, for notational convenience, omit $X$ and proceed conditional on $X=x$. \\ 

\noindent \textbf{Identification using the MTRs and MTE.} 
By definition, note that the LATR and the LATE are directly related to the MTR and MTE by
\begin{align}
\text{LATR}_d(v, v', z_d) = \int_v^{v'} \frac{m_d(\tilde{v}, z_d)}{v'-v} d\tilde{v}, \text{ and }  \text{LATE}(v, v', z_0, z_1) = \int_v^{v'} \frac{\text{MTE}(\tilde{v}, z_0, z_1)}{v'-v} d\tilde{v}.
\end{align}
Therefore, $\text{LATE}(v, v', z_0, z_1)$ is identified if MTE$(\tilde{v}, z_0, z_1)$ is identified for all $\tilde{v} \in [v, v')$, i.e., if for each $d=0, 1$, MTR$_d(\tilde{v}, z_d)$ is identified for all $\tilde{v} \in [v, v')$. \\

\noindent \textbf{Direct identification} (Figure \ref{fig:late_identification})\textbf{.} The LATE identification via the MTE is not possible if continuously distributed semi-IVs are unavailable to generate continuous variation in both the propensity score given $Z_1=z_1$ and given $Z_0=z_0$ over the entire range $P \in [v, v')$. 
Fortunately, following the intuition from Figure \ref{fig:late_identification}, the identification of this general LATE can be done directly using only four points (i.e., two semi-IV shifts) without identifying all the intermediary MTRs. 
\noindent The intuition for identifying the MTE extends directly to larger shifts in the propensity score, from $P=v$ to $P=v'$. Using \eqref{eq:yd1}, we have:
\begin{align*}
	\mathbb{E}[DY | Z_1=z_1, P=v'] - \mathbb{E}[DY | Z_1=z_1, P=v] = \int_v^{v'} m_1(\tilde{v}, z_1) d\tilde{v}. 
\end{align*} 
So, if there exists $z^1 = (\tilde{z}_0, z_1)$ with $P(z^1) = v$ and $\bar{z}^1 = (\tilde{z}_0', z_1)$ with $P(\bar{z}^1) = v'$, then LATR$_1(v, v', z_1)$ is identified by
\begin{align}\label{eq:latr1_identification}
	\text{LATR}_1(v, v', w_1) & = \frac{\mathbb{E}[DY|Z=\bar{z}^1] - \mathbb{E}[DY|Z=z^1]}{v' - v} \\
	&= \frac{\mathbb{E}[DY|Z_1=z_1, P=v'] - \mathbb{E}[DY|Z_1=z_1, P=v]}{v' - v}. \nonumber
\end{align}
This follows the standard IV interpretation, applied to the outcome $DY$: holding $Z_1=z_1$ fixed, shifts in $Z_0$ affect $DY$ only indirectly through their effects on selection probabilities (see Property \eqref{eq:observable_property}). At fixed $Z_1$, the semi-IV $Z_0$ serves as an IV for the effect of $D$ on $DY$. Thus, observable differences between $\mathbb{E}[DY|Z_1=z_1, Z_0=\tilde{z}_0']$ and $\mathbb{E}[DY|Z_1=z_1, Z_0=\tilde{z}_0]$ stem solely from the effect of shifting $Z_0$ from $\tilde{z}_0$ to $\tilde{z}_0'$ on selection and the corresponding set of compliers, $V \in [v, v')$. This shift corresponds to the \textcolor{mtr1color2}{purple arrow} in Figure \ref{fig:late_identification}. \\
\indent Similarly, $Z_1$ serves as an IV for the effect of $D$ on $(1-D)Y$ given $Z_0$. Thus, if $z^0 = (z_0, \tilde{z}_1)$ with $P(z^0) = v$ and $\bar{z}^0= (z_0, \tilde{z}_1')$ with $P(\bar{z}^0)=v'$ exist, LATR$_0(v, v', z_0)$ is identified by\footnote{The minus sign comes from the same reason as in \eqref{eq:mtr0}, see footnote \ref{footnote:minus}.} 
\begin{align}\label{eq:latr0_identification}
	\text{LATR}_0(v, v', w_0) = - \frac{\mathbb{E}[(1-D)Y|Z=\bar{z}^0] - \mathbb{E}[(1-D)Y|Z=z^0]}{v' - v}.
\end{align}
This identifying semi-IV shift corresponds to the \textcolor{mtr0color2}{orange arrow} in Figure \ref{fig:late_identification}.

\indent Therefore, for any $v, v' \in [0, 1]$, LATE$(v, v', z_0, z_1)$ is identified by Equation \eqref{eq:late_latrlink}, i.e., by 
\begin{align*}
	\text{LATE}(v, v', z_0, z_1) = \text{LATR}_1(v, v', z_1) - \text{LATR}_0(v, v', z_0), 
\end{align*}
provided that (at least) four semi-IVs combinations exist: (i) $z^1 = (\tilde{z}_0, z_1)$ with $P(z^1) = v$ and $\bar{z}^1 = (\tilde{z}_0', z_1)$ with $P(\bar{z}_1) = v'$, ensuring that LATR$_1(v, v', z_1)$ is identified by \eqref{eq:latr1_identification}, and (ii) $z^0 = (z_0, \tilde{z}_1)$ with $P(z^0) = v$ and $\bar{z}^0= (z_0, \tilde{z}_1')$ with $P(\bar{z})=v'$, ensuring that LATR$_0(v, v', z_0)$ is identified by \eqref{eq:latr0_identification}. 
Again, the monotonicity assumption plays a key role: it ensures that, although LATR$_1$ and LATR$_0$ are identified from two different semi-IV shifts, both shifts induce comparable sets of compliers, with $V \in [v, v')$, because they result in the same observed change in the propensity score, from $v$ to $v'$. Therefore, one can take the difference between these two LATRs to identify the LATE for $V \in [v, v')$. \\

\noindent \textbf{Discussion.} 
\noindent LATE parameters for compliers with $V \in [v, v')$ at $Z_1=z_1, Z_0=z_0$ can be identified if, for both $d$, there exist two semi-IV combinations with $Z_d=z_d$, one yielding $P=v$ and the other $P=v'$ (see Figure \ref{fig:late_identification}). 
With at least one continuously distributed semi-IV, a broad set of LATE is likely identifiable. 
With two discrete semi-IVs, exact complier alignment on the same propensity score values is unlikely. However, provided that the discrete supports are large enough, one can interpolate the LATRs in between the support points, and recover (or at least, bound) some LATE parameters accordingly. 
In the unfortunate case where the researcher only has two binary semi-IVs ($Z_d \in \{0, 1\}$), identifying any LATE requires specific conditions that the effect of the semi-IVs on the selection probabilities compensate each other exactly.\footnote{More precisely, if $P(0, 0) = P(1, 1)=p$, then LATE$(p, P(1,0), 0, 1)$ and LATE$(p, P(0,1), 1, 0)$ are identified, while if $P(0,1)=P(1,0)=p$ then LATE$(p, P(0,0), 1, 1)$ and LATE$(p, P(1,1), 0, 0)$ are identified. } 
This is very unlikely to hold, meaning that, contrary to binary IVs, the LATE can generally not be identified with two binary semi-IVs. \\ 
\indent Finally, note that the environment with $Z=(z_0, z_1)$ is generally not used to identify LATE$(v, v', z_0, z_1)$. Contrary to the standard LATE identification of \cite{imbensangrist1994} (see Online Appendix \ref{app:late}), the identification of the parameters is detached from the specific value of the semi-IVs. However, in the special case where $P(z_0, z_1)=v$ (resp. $v'$), then identification is simplified because it only requires two additional points belonging to the isocurve of probability $v'$ (resp. $v$): one with $Z_1=z_1$ and another one with $Z_0=z_0$.


\subsection{Other treatment effects}
\noindent \textbf{Direct effect of the semi-IVs (targeted-policy evaluation).} In some applications, especially when the semi-IVs are alternative-specific policies, one may be interested in identifying the \textit{direct effect} of these semi-IVs on their respective potential outcomes, net of the selection/compositional changes. 
Online Appendix \ref{app:directeffect} describes how to identify these effects. \\

\noindent \textbf{Policy relevant treatment effects (PRTE).} 
With MTRs identified with semi-IVs, any parameter expressed as a weighted function of the MTRs is also identified. The generalized LATE is already a specific (with unit weights) policy-relevant treatment effect (PRTE). More broadly, \cite{mogstad2018ecta} (Table 1) provide a list of parameters expressed as weighted functions of the MTRs, along with their corresponding weights with IVs. Many of these parameters and weights can be adapted to identify their counterparts using semi-IVs. \\
\indent Similarly, when the semi-IVs have limited support -- which restricts the observable propensity score and the range of unobserved $V$ for which MTRs are identified -- treatment effects can be extrapolated using approaches analogous to those in \cite{mogstad2018ecta} for IVs. 


\section{Estimation}\label{sec:estimation}

We observe a sample of $\{Y_i, D_i, Z_{0i}, Z_{1i}, X_i\}_{i=1}^N$. Each $Z_d$ may include multiple valid, continuously distributed semi-IVs.
Building on constructive identification arguments, we estimate the MTR$_d$ functions separately to obtain the MTE (and LATE).\footnote{This section focuses on estimating the MTE and MTR but can be adapted to the estimation of the LATE and LATR with only discrete semi-IVs using a similar control function approach for $P$.} The procedure involves estimating three key objects: the propensity score $P(Z_0, Z_1, X)$ and the two conditional expectations $\mathbb{E}[DY | P, Z_1, X]$ and $\mathbb{E}[(1-D)Y|P, Z_0, X]$, which are used to estimate $m_1$ and $m_0$ as in Equations \eqref{eq:mtr1} and \eqref{eq:mtr0}. The MTE is then obtained as their difference (Equation \eqref{eq:mte_mtrlink}) on the support where both MTRs are identified. \\
\indent I propose three estimation approaches. First, a fully nonparametric approach. 
Second, a more practical semi-parametric approach, which is the semi-IV counterpart to the typical MTE estimation with IVs \citep{carneiroheckmanvytlacil2011, andresen2018exploring}. 
Finally, I present a revisited $2$SLS for homogenous treatment effect models with semi-IVs.

\subsection{Nonparametric estimation} 
I estimate the three objects of interest using local linear regressions, similar to \cite{mountjoy2022community}, who studies a discrete treatment problem with multiple IVs. For practicality, assume that the functions of interest are globally linear in $X$ to avoid the curse of dimensionality. \\
\indent First, estimate the propensity score $P$ via local linear regression of $D$ on $Z_0, Z_1$ and $X$, using a bivariate kernel on $(Z_0, Z_1)$ for every values of $(z_0, z_1) \in \mathcal{Z}$.\footnote{Assume $Z_0$ and $Z_1$ each contain a single semi-IV; otherwise, estimation quickly becomes untractable. } \\
\indent Next, use the fitted $\hat{P}_i$ as a generated covariate in the second stage to estimate the conditional expectations of the outcome for $d=0$ and $d=1$ by 
\begin{adjustwidth}{-2cm}{-2cm}
\vspace{-1\baselineskip}
\begin{align*}
	&\begin{pmatrix}
		\hat{\beta}^{Y_d}_{Z_d}(z_d, p) \\
		\hat{\beta}^{Y_d}_{P}(z_d, p) \\
		\hat{\beta}^{Y_d}_{X}(z_d, p) 
	\end{pmatrix}  
	= \underset{\beta_{Z_d}, \beta_{P}, \beta_X}{\text{argmin}} \sum_{i=1}^N K\left( \frac{Z_{di}-z_d}{h_d}, \frac{\hat{P}_{i} - p}{h_P} \right) \left(Y_i\mathds{1}\{D_i=d\} - \beta_{Z_d} Z_{di} - \beta_{P} \hat{P}_{i} - X_i' \beta_X \right)^2. \nonumber 
\end{align*}
\end{adjustwidth}
For all values of $z_d$ and $p$, $X$ includes a constant, and $K(\cdot)$ is a bivariate kernel with bandwidths $h_d$ for each $Z_d$ and $h_P$ for the fitted $\hat{P}$. \\
\indent Local linear specifications are attractive because, or any observable $(z_d, p)$, the local coefficient on $P$ directly estimates MTR$_d$:
\begin{align*}
	\hat{m}_d(p, z_d) = \hat{\beta}_{P}^{Y_d}(z_d, p) = \partial \mathbb{E}[Y_d \mathds{1}\{D=d\} | Z_d=z_d, \hat{P}=p]/\partial \hat{P} \quad \text{ for } d =0, 1. 
\end{align*}
Then, given any $(z_0, z_1, p)$ for which both MTRs are identified and estimated, the MTE is estimated as their difference, using the empirical counterpart of Equation \eqref{eq:mte_mtrlink}:
\begin{align*}
	\text{MTE}(p, z_0, z_1) = \hat{m}_1(p, z_1) - \hat{m}_0(p, z_0). 
\end{align*}
\indent For inference, nonparametric bootstrap is used to obtain proper standard errors accounting for the estimation of $\hat{P}$ and its use as a generated regressor in the second stage.


\subsection{Semi-parametric estimation}\label{sec:estimation_semi}

I show how the local IV estimation \citep{heckmanvytlacil1999}, particularly the approach estimating both MTRs separately \citep{andresen2018exploring} with IVs, adapts to semi-IVs. \\


\noindent \textbf{The semi-parametric model.} The treatment is still determined by the weakly separable selection rule \eqref{eq:selection}. The potential outcomes are given by 
\begin{align}\label{eq:POsepa}
	Y_d = \mu_d(X, Z_d) + U_d = X\beta_d + Z_d \delta_d + U_d,  \quad \text{ for } d=0,1, 
\end{align}
where $U_d$ is a $d$-specific unobserved shocks affecting the potential outcomes, and $\mu_d$ is a $d$-specific parametric function, that is specified as linear for simplicity.\footnote{Linearity can be relaxed as long as $\mu_d$ follows a known parametric form. Since the dimensions of $X$ and $Z_d$ are unspecified, even a simple linear model can be made more flexible by adding interactions between $X$ and $Z_d$ or incorporating flexible functions of $Z_d$ (e.g., splines or polynomials).}
The joint distribution of $(U_0, U_1, V)$ is unrestricted, allowing for endogeneity between $Y_d$ and $D$. 
\noindent Under this model, the partial exclusion of the semi-IVs can be written as $(U_d, V) \indep Z_{1-d} | (Z_d, X)$ for each $d=0,1$. 
Property \eqref{eq:mean_independence} can be expressed in terms of $U_d$ by
\begin{align}\label{eq:mean_indep2}
	\mathbb{E}[U_d | V, Z_d, Z_{1-d}, X] = \mathbb{E}[U_d | V, Z_d, X] \quad \text{ for } d=0,1.
\end{align}

\noindent \textbf{Separability.} In addition, following the empirical literature estimating MTE \citep[e.g., ][]{carneiroheckmanvytlacil2011, maestas2013does, eisenhauer2015generalized, brinch2017beyond, cornelissen2018benefits}, and existing estimation packages \citep{brave2014estimating, andresen2018exploring}, I impose the following additional separability assumption to simplify the estimation.\footnote{A stronger version of this separability assumption imposing that $U_d \indep (Z_0, Z_1, X)$ for $d=0, 1$ is sometimes assumed instead. But it is stronger than necessary, especially for $X$, because Assumption \ref{ass:separability} does not restrict the dependence between $X$ and $V$. Since $Z_d \indep V$, the distinction is less important for the semi-IVs. See the Section $6.2$ of \cite{mogstad2018review} for more discussions. } 
\begin{assumption}[Separability]\label{ass:separability}
	For each $d \in \{0,1\}$, $\mathbb{E}[U_d | V, Z_0, Z_1, X] = \mathbb{E}[U_d | V].$
\end{assumption}
Separability is a stronger assumption than the implication \eqref{eq:mean_indep2} derived from the exclusion restriction in the general model. Importantly, separability does \textit{not} imply that the potential outcome $Y_d$ is independent of $Z_d$ and $X$. 
Rather, it restricts their dependence to be fully captured by the conditional mean function, $\mu_d(X, Z_d)$. Combined with the additive separability in \eqref{eq:POsepa}, it implies that the effect of $X$, and more importantly, of $Z_d$ on $Y_d$ are homogenous with respect to $U_d$, and thus, with respect to $V$. While this restriction is strong, it is commonly assumed (with respect to $X$) in the empirical literature estimating MTE. Since $Z_d$ behaves like a covariate $X$ for $Y_d$, it is natural to extend the separability assumption to $Z_d$. Treatment effects remain generally heterogenous under separability. \\
\indent Even under this stronger separability assumption, the exclusion of $Z_{1-d}$ from $Y_d$ still holds only conditional on $Z_d$ (and $X$), i.e., 
\begin{align*}
	\mathbb{E}[Y_d | V, Z_d, Z_{1-d}, X] = \mathbb{E}[Y_d | V, Z_d, X].
\end{align*} 
Thus if one does not control for $Z_d$, one cannot exclude $Z_{1-d}$ from the average $Y_d$, since $Z_{1-d}$ may indirectly affect the average $Y_d$ through its correlation with $Z_d$ in $\mu_d(X, Z_d)$. \\
%
%

\noindent \textbf{MTRs.} Under Assumption \ref{ass:separability} and model \eqref{eq:POsepa}, the MTRs are 
\vspace{-1\baselineskip}
\begin{align}\label{eq:kd}
	m_d(v, z_d, x) &= \mathbb{E}[Y_d | V=v, Z_d=z_d, X=x] = \mu_d(x, z_d) + \overbrace{\mathbb{E}[U_d | V=v]}^{:= \ k_d(v)} \nonumber \\
	&= x \beta_d + z_d \delta_d  + k_d(v) \quad \quad \text{ for } d=0,1. 
\end{align}
In other words, the mean potential outcomes, $m_d$, can be decomposed into two components: the effect of the observables, $\mu_d(x, z_d)$, and the mean unobservable $U_d$ given $V=v$, $k_d(v)$, which captures selection effects. While more restrictive than the general model, this structure has practical advantages. The direct effects of semi-IVs (and covariates) on $Y_d$ are homogenous and more easily interpretable as they do not vary with the unobserved resistance to treatment $V$. They are also easier to identify.\footnote{For any $z_d, z_d', x$, one can identify $\mu_d(x, z_d') - \mu_d(x, z_d)$ by controlling for $P$, for any $p \in \mathcal{P}$. Indeed, 
\vspace{-0.5\baselineskip}
\begin{align*}
	\mu_d(x, z_d') - \mu_d(x, z_d) = \mathbb{E}[Y | D=d, Z_d=z_d', P=p, X=x] - \mathbb{E}[Y | D=d, Z_d=z_d, P=p, X=x], 
\end{align*}
\noindent because $\mathbb{E}[U_d | D=d, Z_d, X, P] = \mathbb{E}[U_d | D=d, P]$ is independent of $Z_d$ and $X$. So the direct effect of changing $Z_d$ from $z_d$ to $z_d'$ (at $X=x$) is identified if one can shift $Z_d$ while holding $P$ fixed (for any $p$), which is feasible if $Z_{1-d}$ is relevant. With the linear specification, identification is even simpler: it suffices that there exist two $z_d \neq z_d'$ that yield the same $P=p$, for any $p \in \mathcal{P}$. In this case, $\delta_d$ is identified by: 
\vspace{-0.5\baselineskip}
\begin{align*}
	\delta_d = \Big(\mathbb{E}[Y | D=d, Z_d=z_d', P=p, X=x] - \mathbb{E}[Y | D=d, Z_d=z_d, P=p, X=x]\Big)\Big/\Big(z_d' - z_d\Big). \end{align*} 
	\vspace{-1\baselineskip}
 } 
Moreover, since $k_d(v)$ is independent of $(X, Z_d)$, the full support of the propensity score (given $D=d$) can be used for identification, rather than only its support conditional on specific $Z_d=z_d, X=x$ (and $D=d$). \\


\noindent \textbf{Partially linear model.} Under separability, the observed average outcomes in the treated and untreated subsamples of the semi-parametric model \eqref{eq:POsepa} follow a partially linear model: 
\begin{align}\label{eq:plm}
	\mathbb{E}[Y | D=d, Z_d, P, X] = \mathbb{E}[Y_d | D=d, Z_d, P, X] = X \beta_d + Z_d \delta_d + \kappa_d(P), \quad \text{ for } d =0,1,
\end{align}
where $\mathbb{E}[U_d | D=d, P, Z_d, X] = \mathbb{E}[U_d | D=d, P] =: \kappa_d(P)$, a nonparametric function of $P$ under separability.\footnote{The conditioning on $P=p$ remains because $D=1$ is equivalent to $V \leq p$ and $D=0$ to $V > p$.} Thus, average outcomes in each subsample follow a \textit{partially linear model} with a linear component in covariates and semi-IVs, and an additive nonparametric term in $P$. \\
\indent To estimate the MTRs \eqref{eq:kd}, it suffices to consistently estimate the partially linear model \eqref{eq:plm}, which directly provides an estimate for $\beta_d$ and $\delta_d$, and allows recovering $k_d(v)$ as a deterministic function of $\kappa_d(v)$. Indeed, $\kappa_1(p) = \mathbb{E}[U_1 | V \leq p] = \int_0^p \mathbb{E}[U_1|V=v]/p\ dv $ and $\kappa_0(p) = \mathbb{E}[U_0 | V > p] = \int_p^1\mathbb{E}[U_0 | V=v]/(1-p) dv$. Differentiating these gives
\begin{align*}
	k_1(p) = \frac{\partial \kappa_1(p)p}{\partial p} \quad \text{ and } \quad k_0(p) = - \frac{\partial \kappa_0(p)(1-p)}{\partial p}. 
\end{align*}


\noindent \textbf{Estimation.} Following the literature \citep{heckmanetal1998, heckmanurzuavytlacil2006}, I apply nonparametric \cite{robinson1988root}'s double residual regression to estimate the partially linear models separately for treated and untreated subsamples.\footnote{\textit{Alternative estimation.} \cite{robinson1988root}'s double residual regression is the most popular procedure to nonparametrically control for $P$, but alternatives exist. For instance, $\kappa_d(p)$ can be specified as flexible polynomials or splines, following a nonparametric sieve approach. In this case, after estimating $P$ in a first stage, the model can be estimated by running the regression of $Y$ on $X, Z_d,$ and a flexible function of $P$ in each subsample with $D=d$. If the chosen specification for $\kappa_d(p)$ is sufficiently flexible, this yields consistent estimates of $\beta_d$, $\delta_d$, and $\kappa_d(p)$ for values of $p$ in the interior of the support of $P$ given $D=d$.}
\noindent The estimation can be decomposed into three steps. In the first stage, the propensity score is estimated. Then, for each subsample of observations with $D=d$ separately, we estimate the effects of the semi-IVs and covariates on $Y$ (thus, on $Y_d$ in the subsample) by controlling for $P$ nonparametrically. More precisely, nonparametrically regress $Y, X, Z_d$ on $P$, and then run the linear regression of the residuals of $Y$ on the residuals of $X$ and $Z_d$ to estimate $\hat{\beta}_d$ and $\hat{\delta}_d$. Finally, we estimate the remaining selection effects, $k_d(v)$. To do so, nonparametrically regress $Y$ minus the effects of the covariates and semi-IVs, i.e., $Y-\hat{\beta}_d X + \hat{\delta}_d Z_d$, on $P$. The detailed estimation procedure with semi-IVs is described in Online Appendix \ref{app:estimation}. \\
\indent These three steps parallel those of MTE estimation with IVs \citep{andresen2018exploring}.\footnote{For more details, see also \cite{carneiroheckmanvytlacil2011}, Appendix A.2.1. of \cite{heckmanetal1998}, or the online supplement of \cite{heckmanurzuavytlacil2006}.} 
The key difference is the presence of the direct effects of the semi-IVs, $\delta_d$. Identifying these requires being able to fix the propensity score (i.e., to control for $P$) while shifting $Z_d$. This is achievable by also shifting the other excluded semi-IV, $Z_{1-d}$, to offset the effect of $Z_d$ on $P$. 
\noindent Note that covariates are already incorporated in the standard MTE estimation with IVs \citep{andresen2018exploring}. Thus, the effects of the semi-IVs are naturally incorporated as additional coefficients estimated in the second stage. Therefore, MTE estimation with semi-IVs is similar but also requires the relevance of the semi-IVs to provide identifying excluded variation. \\

\noindent \textbf{Estimated MTR and MTE.} For each $d$, once $\beta_d, \delta_d$ and $k_d(v)$ are estimated, the MTR$_d$ function is obtained by plugging these into \eqref{eq:kd} for any $v$ in the interior of the support of $\widehat{P}$ given $D=d$, and for any $(z_d, x)$ in the joint support of $Z_d$ and $X$ given $D=d$, i.e., 
\begin{align*}
	\widehat{m}_d(v, z_d, x) = x\widehat{\beta}_d + z_d \widehat{\delta}_d + \widehat{k}_d(v). 
\end{align*} 
\indent Similarly, the MTE is estimated for any $(z_0, z_1, x) \in \mathcal{Z}\times\mathcal{X}$ and for any $v$ in the \textit{common support} of $\widehat{P}$, i.e., at the intersection of the supports of $\widehat{P}$ given $D=1$ and given $D=0$:\footnote{In practice, there are very few observations at the extreme ends of the propensity score support, leading to poor estimation of the MTRs in those regions. To address this, it is common to trim the supports further to focus on well-estimated interior points \citep[e.g.,][]{brinch2017beyond}. For example, one might restrict $P$ in each subsample to values between the $1\%$ and $99\%$ percentiles of its conditional distribution given $D=d$ and define the common support as the intersection of these trimmed supports.}   
\begin{align*}
	\widehat{MTE}(v, z_0, z_1, x) = \quad \underbrace{x (\widehat{\beta}_1 - \widehat{\beta}_0) + z_1 \widehat{\delta}_1 - z_0 \widehat{\delta}_0}_{\text{effects of the observables}} \quad + \ \underbrace{\widehat{k}_1(v) - \widehat{k}_0(v).}_{\text{effects of the selection}} 
\end{align*}
Under separability, the MTE can be decomposed into two components: one driven by observables, including the semi-IVs, and another driven by the selection. 
\noindent Note that the MTRs and MTE still depend on the covariates and semi-IVs. Typically, we report MTE/MTR curves at average covariates and semi-IV values (i.e., at $\bar{x}, \bar{z}_0, \bar{z}_1$) across all identified values of $v$ (see Figure \ref{fig:est_effects} for example). 
\noindent Once the MTRs and MTE are estimated, they can be used to estimate more complex parameters such as the LATR and LATE, or other PRTEs. \\

\noindent \textbf{Inference.} As in its IV counterpart, raw analytical standard errors from the final-stage local polynomial estimation of the MTR and MTE are biased, as they ignore that the propensity score and the effect of the semi-IVs and covariates were estimated in previous stages. To account for this, nonparametrically bootstrapping the standard errors is recommended. \\

 \noindent \textbf{Performances.} The semi-parametric estimator performs well with reasonable sample sizes, as illustrated using Monte Carlo simulations in Online Appendix \ref{app:montecarlo}. \\

\noindent \textbf{Practical implementation.} The MTE and MTR estimation can be performed using the \texttt{semiivreg()} command from the \texttt{semiIVreg} package \citep{semiivreg}. The syntax of the function is designed to mimic the \texttt{ivreg()} command for IV estimation. 
\begin{lstlisting}[language=R] 
semiivreg(y ~ d | z0 + x | z1 + x, data=data)
\end{lstlisting}

\subsection{Two-Stage Least Squares with semi-IVs}\label{sec:estimation_2sls}
Consider a simpler model with homogenous treatment effects: 
\begin{align}\label{eq:tsls}
	Y = \alpha + D\delta + D Z_1 \delta_1 + (1-D) Z_0 \delta_0 + DX\beta_1 + (1-D)X\beta_0 + U,
\end{align}
where $\mathbb{E}[U] = 0$ and $\mathbb{E}[U | Z_0, Z_1, X] = 0$ serves as the counterpart to the separability Assumption \ref{ass:separability} in this model. The treatment is still determined by Equation \eqref{eq:selection}. 
This is a model with homogenous treatment effects with respect to the unobservables, meaning that MTE$(v, z_0, z_1, x)$ is independent of $v$. However, the average treatment effect remains heterogenous with respect to observables. Indeed 
\begin{align*}
	\text{ATE}(z_0, z_1, x) &= \mathbb{E}[Y_1 - Y_0 | X=x, Z_0=z_0, Z_1=z_1] = \delta + x (\beta_1 - \beta_0) + z_1\delta_1 - z_0 \delta_0.  \
\end{align*}
Even when covariates $X$ have the same effects on both potential outcomes ($\beta_1=\beta_0$), the ATE still depends on the semi-IVs. Thus, we typically report specific ATEs, such as the ATE at the mean, ATE$(\bar{z}_0, \bar{z}_1, \bar{x})$. \\
\indent Standard IV-GMM estimation does not directly apply to Model \eqref{eq:tsls} since the number of parameters exceeds the number of instruments. 
However, the model can be estimated using a modified two-stage least squares (2SLS) approach. In the first stage, estimate $P(Z_0, Z_1, X)$. Then run a 2SLS of Equation \eqref{eq:tsls} instrumented by the optimal instruments, $\widehat{P}, \widehat{P}Z_1, (1-\widehat{P})Z_0$ \citep[e.g., ][]{chamberlain1987asymptotic}. This yields consistent estimates of all parameters $\alpha, \delta, \delta_0, \delta_1, \beta_0$, and $\beta_1$.\footnote{If the true underlying model is a model with heterogenous treatment effects, e.g., Model \eqref{eq:POsepa}, the question of whether the estimates from this semi-IV-2SLS can be interpreted as convex combinations of positively weighted semi-IV LATE (like standard IV 2SLS) remains open.} 


\section{Application: returns to working in manufacturing}\label{sec:application}

\subsection{Context: the manufacturing decline}

\begin{figure}[h!]
    \centering
    \caption{Evolution of the manufacturing sector $1999-2018$}
    \label{fig:emp}
    
    \begin{adjustbox}{max width=1\textwidth,center}
        \begin{subfigure}[b]{0.5\textwidth}
            \centering
            \includegraphics[width=0.85\textwidth]{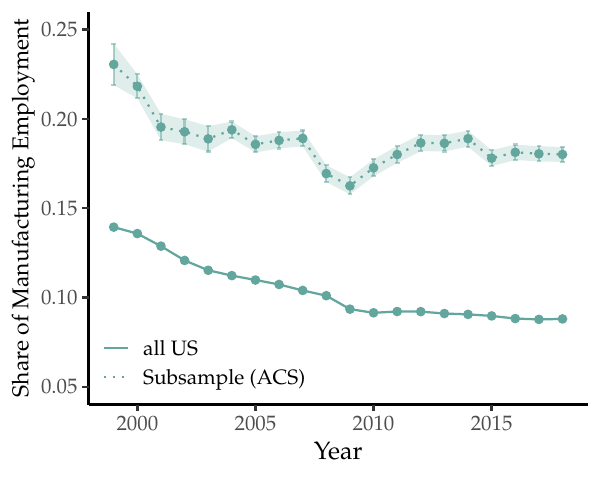}
            \subcaption{Manufacturing employment share}
            \label{fig:emp_US}
        \end{subfigure}
        \hfill
        \begin{subfigure}[b]{0.5\textwidth}
            \centering
            \includegraphics[width=0.85\textwidth]{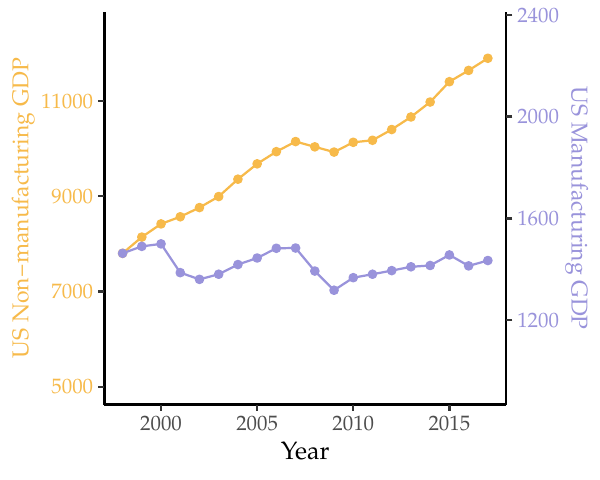}
            \subcaption{GDP by industry}
            \label{fig:gdp_US}
        \end{subfigure}
    \end{adjustbox}

    \vspace{0.5em}
    \captionsetup{font=footnotesize, justification=justified} 
    \caption*{\textit{Notes:} Left panel shows evolution of the manufacturing sector's employment share. The "all US" curve is computed directly from the Quarterly Census of Employment and Wages (QCEW) data of the Bureau of Labor Statistics, by dividing employment in the US manufacturing supersector (NAICS) by the overall US employment (excluding only the military). The "ACS subsample" curve represents the estimated probability of being employed in manufacturing for young white men (ages 18-30) with only a high school degree, obtained from the American Community Survey (ACS) by regressing a manufacturing dummy on year fixed effects. 
    Right panel displays the evolution of real US GDP by sector (manufacturing vs. nonmanufacturing) in billions of $1999$ dollars, using data from the Bureau of Economic Analysis (BEA).}
    \vspace{-1\baselineskip}
\end{figure}

\noindent The manufacturing sector's employment share has steadily declined in the US over the past decades \citep{baily2014us, pierce2016surprisingly, fort2018new}. As shown in Figure \ref{fig:emp_US}, from $1999$ to $2018$, its share fell from $13.9\%$ to $8.8\%$ of total US employment: a $36.7\%$ relative decline over $20$ years. Even among young white men with only a high school education, a group more likely to work in manufacturing, the sector's employment share dropped from $23\%$ to $18\%$ in the same period, a $21.7\%$ relative decline. \\
\indent This decline in manufacturing coincides with the rise of the nonmanufacturing sector, particularly services \citep{lee2006intersectoral, buera2012rise, autor2013growth}. 
As Figure \ref{fig:gdp_US} shows, US manufacturing GDP (in base-1999 dollars) barely changed from $1,462$ billion in $1998$ to $1,433$ billion in $2017$ (a $2\%$ decline), while nonmanufacturing grew by $52.5\%$, from $7,800$ to $11,895$ billion over the same period. \\
%
\begin{figure}[t]
	\centering
		\captionsetup{justification=centering} 
	\caption{Evolution of the difference of average (log) yearly earnings \\ in manufacturing ($D=1$) and nonmanufacturing sectors ($D=0$).}

	\includegraphics[width=0.50\linewidth]{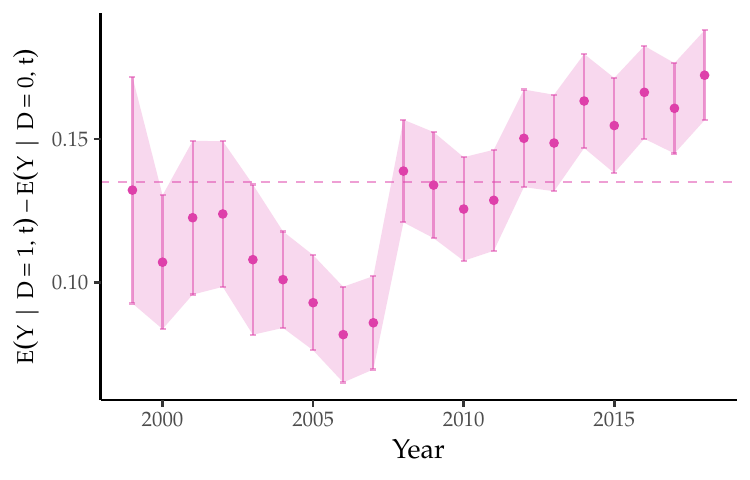}
	\label{fig:return_ols}
	\captionsetup{font=footnotesize, justification=justified} 
    \caption*{\textit{Notes:} $\mathbb{E}[$log(earnings)$|D=1, $ year$] - \mathbb{E}[$log(earnings)$|D=0, $ year$]$ is estimated by OLS, regressing log(earnings) on year fixed effects separately for the manufacturing and nonmanufacturing sectors. The horizontal line represents the $13.5\%$ average earnings premium for manufacturing jobs in the sample. }
    \vspace{-1\baselineskip}
\end{figure}
\indent A naive look at earnings by sector (Figure \ref{fig:return_ols}) suggests that (i) manufacturing workers earn $13.5\%$ more on average than their nonmanufacturing counterparts throughout the period and, perhaps surprisingly, (ii) this gap \textit{increased} from $13.2\%$ in $1999$ to $17.2\%$ in $2018$. \\
\indent However, this naive comparison is likely biased due to selection on unobservables: even after controlling for demographics (age, education, race), manufacturing workers systematically differ from those in other sectors \citep{heckmansedlacek1985}. 
In principle, an IV could address the endogeneity of sector choice, but finding one that affects sector choice without also affecting subsequent earnings is difficult. Fortunately, measures of local sector-specific strength can serve as valid semi-IVs instead, allowing for unbiased estimates of the returns to manufacturing (net of selection) and their evolution over time.\footnote{I remain agnostic about the reasons behind the evolution of returns to working in manufacturing. The literature highlights two candidate explanations: (i) technological change, which affects the task content of manufacturing jobs \citep{autor2003skill, goos2014explaining}, and (ii) trade, with rising import competition \citep{autor2013china}. \cite{autor2015untangling} propose methods to disentangle these two explanations.}

\subsection{Model specification and semi-IVs}

I estimate the effect of working in manufacturing ($D=1$) on the logarithm of the yearly earnings of workers ($Y$), measured in thousands of real $1999$ US dollars.\footnote{The manufacturing sector corresponds to the NAICS supersectors "31-33". The nonmanufacturing sector includes all other industries, private or public, except military jobs. }
I focus on young white male workers aged 18 to 30 with only a high school diploma and no college education, a demographic more likely to work in manufacturing and thus more affected by the sector's decline. To address the endogeneity of sector choice, I use the one-year lagged state-level manufacturing GDP ($Z_1$) and nonmanufacturing GDP ($Z_0$), measured in millions of $1999$ US dollars, as semi-IVs. These should be valid as discussed in Example 1 of Section \ref{subsec:example}. \\ %
\indent I estimate the following semi-parametric model for the potential earnings:
\begin{align}\label{eq:POappli}
	Y_d = \underbrace{\mu_{d}^{year} + \mu_{d}^{\text{state}} + \text{age} \beta^{age}_{d} + \text{age}^2 \beta^{age^2}_d}_{:=\  X \beta_d} +  \ log(Z_d) \delta_d + U_d,  
\end{align}
which corresponds to Model \eqref{eq:POsepa} with covariates $X$ including age, age$^2$, and state and year fixed effects. Sector-specific ($d$-specific) effects are estimated for all covariates, including fixed effects. The sector-specific state and year fixed effects control for intrinsic differences in state-level sectoral markets and national trends, capturing sector-specific costs, technological changes, and trade shocks. The sector-specific time fixed effects are particularly interesting as they track the evolution of earnings in both sectors and, consequently, the returns to working in manufacturing over time. With these fixed effects, the semi-IVs' identifying variations do not come from their levels but from deviations relative to sector-specific permanent state levels and the global time trend.\footnote{The semi-IV effects $\delta_d$ are assumed to be homogenous across time and place, allowing identification using local variations in the semi-IVs over time. Identification would not be possible with interacted year-state fixed effects or if the semi-IVs had state- and year-specific effects, such as $\delta_{d, state, year}$.} I assume time fixed effects in the outcome and selection equations absorb all temporal variation in $U_d$ and $V$. I therefore omit the $t$-index and treat these shocks as time-invariant, implying time-invariant $k_d(v)$ in the MTRs/MTE. This simplification permits pooled estimation across periods rather than separate period-specific models that rely only on within-period, cross-sectional (location-specific) variation. 
\noindent Since  $Y_d$ are not directly observed with only data on $(Y, D, Z, X)$, Model \eqref{eq:POappli} cannot be estimated directly. Instead, it is estimated following the semi-parametric procedure in Section \ref{sec:estimation_semi}.

\subsection{Data}
The individual-level data on earnings, employment, and sector choice for young white men come from the ACS $1\%$ yearly survey waves (2000-2019), effectively covering earnings from 1999 to 2018.\footnote{As the name suggests, each survey wave includes a random, representative sample of $1\%$ of the US population. However, from 2000 to 2004, the coverage was smaller, with the 2000-2004 waves containing only $0.13\%$, $0.43\%$, $0.38\%$, $0.42\%$, and $0.42\%$ of the population, respectively. } 
I include all US states except Hawaii, Alaska, and Washington D.C., as these are outliers in manufacturing employment, with some years in the sample having no observed young white male manufacturing workers there. State-level sector-specific GDP data come from the Bureau of Economic Analysis (BEA). All monetary variables are adjusted to real 1999 US dollars for consistency and comparability. 

\subsection{Results}

\begin{table}[h!] \centering 
  \caption{Direct effect of the semi-IVs on $P$ (first stage) and $Y_d$} 
  \label{tab:semiiv} 
  \resizebox{0.7\textwidth}{!}{ 
  \small
\begin{tabular}{@{\extracolsep{5pt}}lcccc} 
\\[-1.8ex]\hline 
\hline \\[-1.8ex] 
\\[-1.8ex] & $1^{st}$ stage & \multicolumn{3}{c}{Outcome, log(earnings)} \\ 
\cline{2-2} \cline{3-5} 
\\[-1.8ex] & $D$ & $Y_0$ & $Y_1$ & $Y_1 - Y_0$\\ 
\hline \\[-1.8ex] 

log($Z_0$): & & & & \\
log(Non-manufacturing GDP)
 & $-$0.388$^{***}$ & 0.438$^{***}$ &  & $-$0.438$^{***}$ \\ 
  & (0.057) & (0.028) &  & (0.028) \\ 
log($Z_1$): & & & & \\
 log(Manufacturing GDP) & 0.163$^{***}$ &  & 0.146$^{***}$ & 0.146$^{***}$ \\ 
  & (0.026) &  & (0.026) & (0.026) \\ 
  & & & & \\ 
\hline \\[-1.8ex] 
Observations & 476,117 & 350,146 & 79,783 & \\ 
\hline 
\hline \\[-1.8ex] 
\end{tabular} 
}
\captionsetup{font=footnotesize, justification=justified}
\caption*{\textit{Notes:} \hfill $^{*}$p$<$0.1; $^{**}$p$<$0.05; $^{***}$p$<$0.01. \\
Covariates $X$ include a quadratic term in age, along with state and year fixed effects. The first-stage model is a probit. The semi-IVs are measured in millions of $1999$ US dollars, and earnings in the log of thousands of $1999$ US dollars. 
I report the number of observations in the main (trimmed) sample. This number varies across bootstrap replications, as trimming in each replication is based on the re-estimated propensity score.} 
\vspace{-1\baselineskip}
\end{table}

\subsubsection{Propensity score estimation ($1^{\text{st}}$ stage)}

\begin{figure}[!h]
    \centering
    \caption{Propensity score support and distribution}
    \label{fig:semiiv_proba}


	\begin{adjustbox}{max width=1\textwidth,center}
    \begin{subfigure}[b]{0.5\textwidth}
         \centering
        \includegraphics[width=1\textwidth]{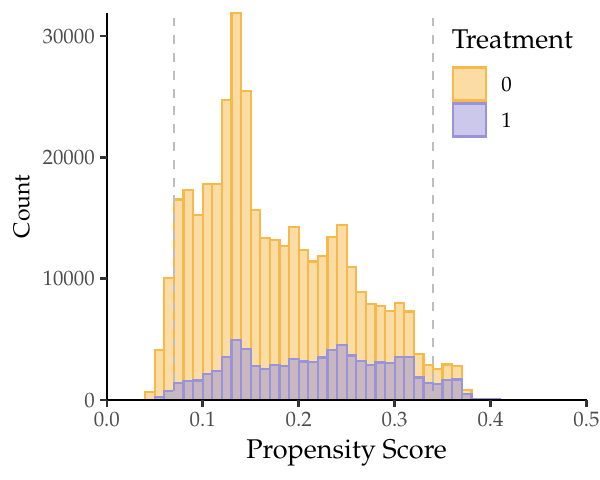}
        \caption{Histogram of $\hat{P}$ for $D=0$ and $D=1$}
        \label{fig:supp}
    \end{subfigure}
    \hfill 
    \begin{subfigure}[b]{0.5\textwidth}
        \centering
        \includegraphics[width=1\textwidth]{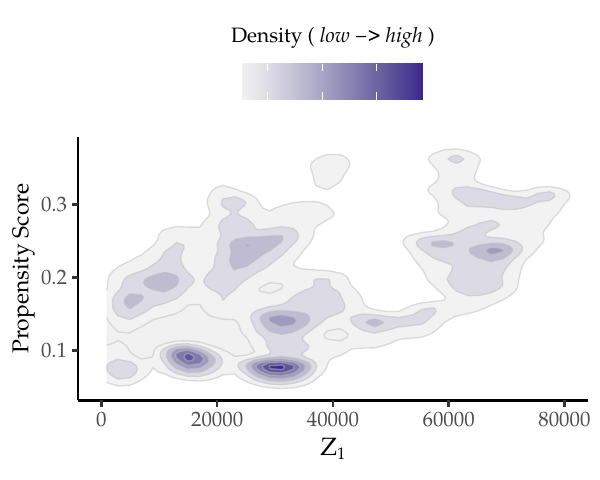}
        \caption{Density of $Z_1$ and $\hat{P}$}
        \label{fig:w1_proba}
    \end{subfigure}
    \end{adjustbox}

    \vspace{1em} 

    \captionsetup{font=footnotesize, justification=justified}
    \caption*{\textit{Notes:} The propensity score $P$ is estimated using a probit regression of $D$ on log($Z_0$), log($Z_1$), age, age$^2$, and state and year fixed effects. The semi-IVs are measured in millions of $1999$ US dollars. The vertical dashed grey lines in Figure \ref{fig:supp} represent the common support $\mathcal{P} = [0.079, 0.339]$. Figure \ref{fig:w1_proba} shows the joint distribution of $Z_1$ and $P$. The x-axis is truncated for readability, as some US states have larger $Z_1$. }
\end{figure}

\noindent \textbf{Propensity score.} I estimate the propensity score using a probit regression of $D$ on log($Z_0$), log($Z_1$), age, age$^2$, and state and year fixed effects. The results are reported in Column $1$ of Table \ref{tab:semiiv}. Both semi-IVs are relevant for sector choice $D$. 
As expected, all else equal, a larger (lagged) manufacturing GDP increases the probability that young workers select manufacturing jobs, while a larger nonmanufacturing GDP reduces it. Several factors may explain this. For instance, a larger sector is more attractive to young workers. Additionally, higher past GDP also allows the sector to recruit more in subsequent periods, leading to a passthrough effect from lagged GDP to future hires. \\ 
\indent Figure \ref{fig:w1_proba} provides a more detailed view of the distribution of $P$ conditional on $Z_1$. It is visible that $Z_0$ is relevant because, given any fixed $Z_1 = z_1$, $P$ still exhibits substantial variation. This is the variation used for identification (while further controlling for covariates). This shows that, even nonparametrically, we could identify large sets of MTE and LATE parameters at any fixed $Z_1=z_1$. \\

\noindent \textbf{Common Support.} Figure \ref{fig:supp} reports the distribution of the estimated propensity score in both the manufacturing (treated) and nonmanufacturing (untreated) subsamples. Overall, the propensity score ranges from $3.19\%$ to $42.12\%$, but the lowest and highest values do not appear in both subsamples. Since the estimation is noisier at the tails, I further trim the data by dropping observations with estimated propensity score below $7.9\%$ or above $33.9\%$. This corresponds to the common support after trimming the $2.5\%$ lowest and highest values of $P$ in each subsample.\footnote{All results are robust to less trimming, e.g., $1\%$, or even $0\%$, i.e., restricting to the common support.} As a result, MTE and MTRs are only estimated for unobserved resistance to treatment $V \in \mathcal{P} = [0.079, 0.339]$.\footnote{Following the applied literature with IVs \citep[e.g.][]{carneiroheckmanvytlacil2011}, the common support could be expanded by including additional semi-IVs or modeling interacted effects of the semi-IVs and covariates on $D$. Including more covariates would also widen the support of $P$, but this would rely on the separability assumption. Instead, we prefer to use variation induced by the semi-IVs for identification. } 

\begin{figure}[p]
    
    \vspace{-2cm}
    \centering
     {\renewcommand{\arraystretch}{0.8}
  \captionof{table}{(Some) Local Average Treatment Effects and Responses}
  \label{tab:late}
  \resizebox{0.85\textwidth}{!}{ 
  \small
    \begin{tabular}{@{\extracolsep{5pt}}lccc} 
\\[-1.8ex]\hline 
\hline \\[-1.8ex] 
\\[-1.8ex] & \color{mtr0}{\textbf{LATR$_0$}} & \color{mtr1}{\textbf{LATR$_1$}} & \color{mte}{\textbf{LATE}} \\ 
\\[-1.8ex] & $\mathbb{E}[Y_0|V \in \mathcal{S}_V]$ & $\mathbb{E}[Y_1|V \in \mathcal{S}_V]$ & $\mathbb{E}[Y_1 - Y_0|V \in \mathcal{S}_V]$\\ 
\\[-1.8ex] & & & \\
\hline \\[-1.8ex] 
\textbf{Set of $V$, $\mathcal{S}_V$:} & & & \\
& & & \\
Observable common support: & & & \\
$\mathcal{S}_V = [7.9\%, 33.9\%]$ & 2.499$^{***}$ & 2.573$^{***}$ & 0.075 \\
 & (0.207) & (0.128) & (0.241) \\
  & & & \\ 
 Low $V$: & & & \\
 $\mathcal{S}_V = [10\%, 15\%]$ & 2.184$^{***}$ & 2.918$^{***}$ & 0.734$^{**}$ \\
 & (0.313) & (0.061) & (0.316) \\
 & & & \\
 High $V$: & & & \\
 $\mathcal{S}_V = [25\%, 30\%]$ & 2.747$^{***}$ & 2.319$^{***}$ & $-$0.428$^{*}$ \\
 & (0.136) & (0.183) & (0.229) \\
 & & & \\
Specific value $V=v$: & \color{mtr0}{MTR$_0$} & \color{mtr1}{MTR$_1$} & \color{mte}{MTE} \\
$\rightarrow$ See Figure \ref{fig:est_effects} &  & & \\
& & & \\
\hline 
\hline \\[-1.8ex] 
\textit{Note:}  & \multicolumn{3}{r}{$^{*}$p$<$0.1; $^{**}$p$<$0.05; $^{***}$p$<$0.01} 
\end{tabular}
}
}
\vspace{1em}
\centering
\captionof{figure}{Estimated Marginal Treatment Effects and Responses}
\label{fig:est_effects}
    \begin{adjustbox}{max width=1\textwidth,center}
    \begin{subfigure}[b]{0.5\textwidth}
        \centering
        \includegraphics[width=1\textwidth]{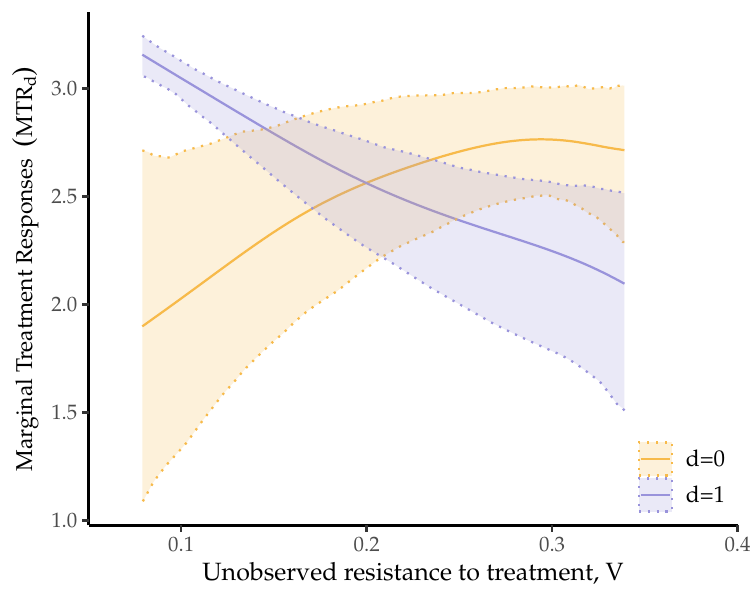}
    \end{subfigure}
    \hfill 
    \begin{subfigure}[b]{0.5\textwidth}
        \centering
        \includegraphics[width=1\textwidth]{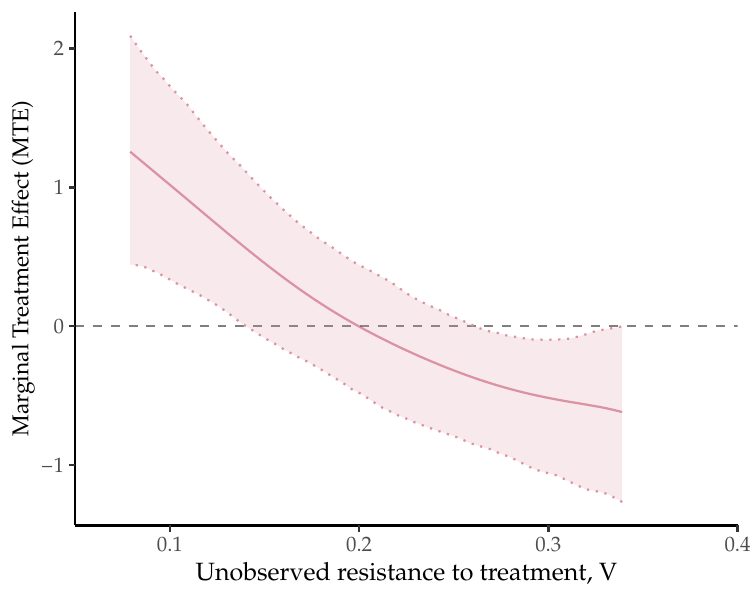}
    \end{subfigure}
    \end{adjustbox}
\captionsetup{font=footnotesize, justification=justified}
\caption*{\textit{Notes:} All estimates are evaluated at the mean covariates ($\bar{x}$) and semi-IVs ($\bar{z}_0, \bar{z}_1$) values. The covariates include age, age$^2$, and state and year fixed effects. The average age is $25.05$. The "average fixed effects" are based on the sample's average year and state distribution and correspond to averaging the state and year dummies. The semi-IVs are measured in millions of $1999$ US dollars, and earnings in the log of thousands of $1999$ US dollars. 
The first stage is estimated using a probit model (see Table \ref{tab:semiiv}). The partially linear model is estimated via double residual regression \citep{robinson1988root}, separately for each subsample $d=0$ and $d=1$. The first residual regression is performed via local linear regression with a Gaussian kernel and bandwidth $0.02$. The second residual regression is performed via local quadratic regression with a Gaussian kernel and bandwidth $0.0482$, chosen using the mean-squared error direct plug-in criteria of \cite{fangijbels1996}, as computed by \cite{calonico2019nprobust} (\texttt{nprobust} package). 
Results are almost insensitive to bandwidth choices between $0.02$ and $0.10$. Standard errors are block bootstrapped ($500$ replications) at the state level. In Table \ref{tab:late}, the LATE/LATR estimates are obtained by integrating the estimated MTE/MTR.}

\end{figure}

\subsubsection{Effects on earnings ($2^{\text{nd}}$ stage)} 
Given the semi-parametric model \eqref{eq:POappli}, the MTR take the form \eqref{eq:kd} as in Model \eqref{eq:POsepa}, i.e., 
\begin{align*}
	m_d(v, z_d, x) = \mathbb{E}[Y_d | V=v, Z_d=z_d, X=x ] = \underbrace{x' \beta_d + log(z_d) \delta_d}_{\text{effect of observables}} \quad + \ \underbrace{k_d(v).}_{\text{selection effect}}
\end{align*} 
First, I discuss the estimated effects of the semi-IVs (Table \ref{tab:semiiv}), then the estimated MTRs and MTE evaluated at average covariates and semi-IVs (Figure \ref{fig:est_effects} and Table \ref{tab:late}). \\

\noindent \textbf{Direct effects of the semi-IVs.} 
Table \ref{tab:semiiv} (Columns $2$-$4$) reports the estimated effects of the semi-IVs on $Y_0$, $Y_1$, and $Y_1-Y_0$, respectively. 
Each semi-IV significantly affects earnings in its corresponding sector. A $1\%$ increase in nonmanufacturing GDP raises the nonmanufacturing workers' earnings by $0.438\%$ in the next period. The pass-through from lagged manufacturing GDP to manufacturing workers' earnings is significantly smaller: a $1\%$ increase raises next-period manufacturing earnings by $0.146\%$. 
Note that the ratio of the effects of the semi-IVs on their respective potential outcomes is close to the ratio of their effects on the first stage. This suggests selection on gains, where the semi-IVs primarily influence the first stage through their effect on $Y_d$. Thus, selection is mainly driven by the comparison of the semi-IVs' effects rather than by each semi-IV individually. \\

\noindent \textbf{Heterogenous returns to working in manufacturing.}  
Figure \ref{fig:est_effects} reports the estimated MTR$_d$ (left panel) and MTE (right panel) curves at $V=v$ for an average individual with $X=\bar{x}, Z_0=\bar{z}_0, Z_1=\bar{z}_1$.\footnote{Due to the partially linear model, the MTR and MTE \textit{at the average} are equivalent to the \textit{average MTR and MTE} at $V=v$ in the sample. For the fixed effects, the "average fixed effects" already correspond to the sample's average distribution of state and year. }
These can only be estimated within the common support, i.e., for $V \in [0.079, 0.339]$. Over the entire common support, the estimated average earnings return to working in manufacturing (LATE) is $7.5\%$ and not significantly different from zero (Table \ref{tab:late}, first row). 
\noindent However, I find significant heterogeneity in the average potential outcomes (MTR) and treatment effects (MTE) across $V$ (Figure \ref{fig:est_effects}). Individuals with low resistance to treatment (small $V$) have, on average, high manufacturing earnings and low nonmanufacturing earnings. This is consistent with selection on gains: these individuals are the first to select into manufacturing because they benefit the most from it. For them, the returns to working in manufacturing are high: they earn more than twice as much in the manufacturing sector as in the nonmanufacturing sector (MTE $>1$). 
As $V$ increases, average manufacturing earnings decrease while nonmanufacturing earnings increase. Around $V=20\%$, the returns to working in manufacturing become negative. \\
\indent This heterogeneity is also reflected in the LATR/LATE estimates (Table \ref{tab:late}), obtained by integrating the MTR/MTE. Among individuals with relatively low resistance to treatment ($V \in [10\%, 15\%]$), average log earnings are $2.918$ in manufacturing (LATR$_1$) versus $2.184$ in nonmanufacturing (LATR$_0$).  This results in a large positive LATE of $0.734$, meaning that they earn, on average, $108\%$ more in manufacturing. Conversely, workers with relatively high $V$ ($V \in [25\%, 30\%]$) earn about $34.8\%$ less in manufacturing (LATE$=-0.428$). \\

\begin{figure}[p]

	\vspace{-1.5cm}
    \centering
    \captionsetup{justification=centering}
    \caption{Yearly evolution of the MTR, MTE, and manufacturing employment probability}
    \label{fig:semiiv_evol}

    \begin{subfigure}[b]{0.6\textwidth}
		\centering
        \includegraphics[width=0.9\textwidth]{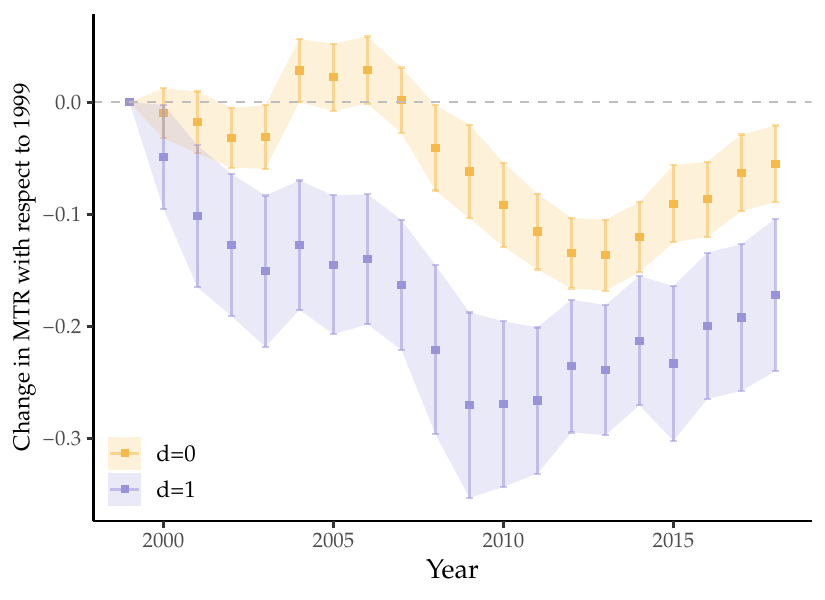}
        \caption{$\Delta_t$MTR$_0$ and $\Delta_t$MTR$_1$}
        \label{fig:evol_mtr}	
    \end{subfigure}

    \vspace{1em} 

	\begin{adjustbox}{max width=1\textwidth,center}
    \begin{subfigure}[b]{0.5\textwidth}
        \centering
        \includegraphics[width=0.9\textwidth]{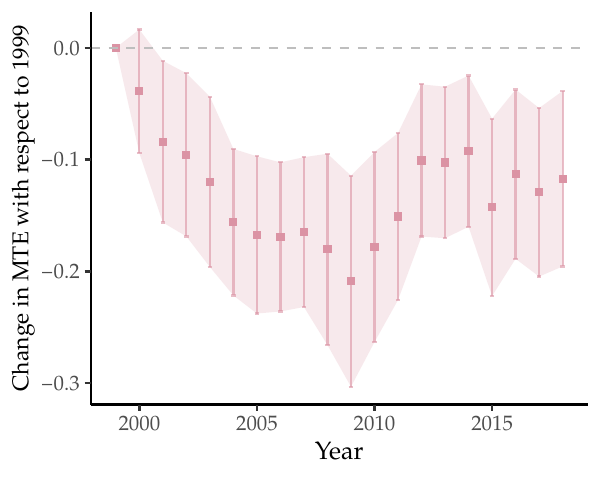}
        \caption{$\Delta_t$MTE }
        \label{fig:evol_mte}
    \end{subfigure}
    \hfill 
    \begin{subfigure}[b]{0.5\textwidth}
        \centering
        \includegraphics[width=0.9\textwidth]{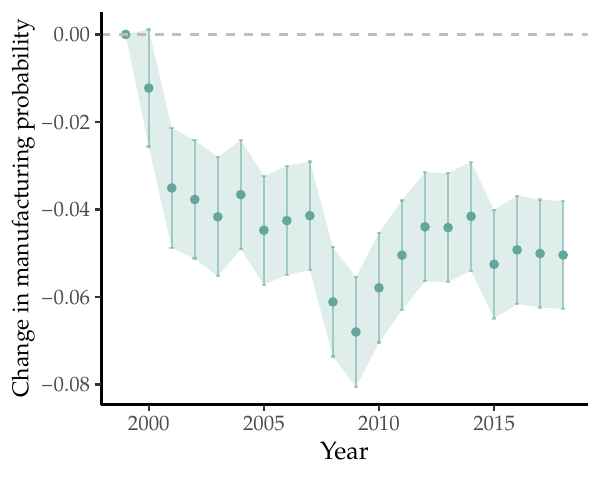}
        \caption{$\textrm{Pr}(D=1|$Year$=t)$.}
        \label{fig:evol_p}
    \end{subfigure}
    \end{adjustbox}

    \vspace{1em} 

   \captionsetup{font=footnotesize, justification=justified}
	\caption*{\textit{Notes:} Evolution of the MTR and MTE over time, relative to the baseline year $1999$. The time evolution accounts for the evolution of the semi-IVs over time. To ensure their effects are differenced out and do not influence the computations, state and age are fixed at arbitrary values, and comparisons are made at a fixed $V=v$ to remove the influence of time-invariant $k_d(v)$. For each year $t$, I compute the average evolution of sector GDP (semi-IVs), $log(\bar{z}_{0, t})$ and $log(\bar{z}_{1, t})$, along with the estimated year fixed effect $\mu_d^t$, for each $d$. The top panel shows the evolution of the MTR$_d$, $\Delta^{MTR_d}_{t - 1999} = \mu_d^t + ( log(\bar{z}_{d, t}) - log(\bar{z}_{d, 1999})) \delta_d$ for each $d=0, 1$. 
	The bottom left panel shows the evolution of MTE given by $\Delta^{MTE}_{t - 1999} = \Delta^{MTR_1}_{t - 1999} - \Delta^{MTR_0}_{t - 1999}$. The bottom right panel shows the evolution of the manufacturing employment probability, estimated separately using an OLS regression of sector choice $D$ on year fixed effects.}
\end{figure}


\subsubsection{Evolution of the returns to working in manufacturing over time}

Figure \ref{fig:semiiv_evol} shows the evolution of returns to working in manufacturing over time, relative to the $1999$ baseline. This evolution accounts for time fixed effects, as well as the changes in the semi-IVs over time and their resulting effects on earnings. 
As shown in Figure \ref{fig:evol_mte}, the returns to working in manufacturing declined until $2009$, down to $-20\%$, and have slightly rebounded since then, although they remain about $10\%$ lower than in $1999$.\footnote{I interpret these estimates as changes in the MTE, though under the semi-parametric specification, they also correspond to changes in the global average treatment effects. } 
I also report the evolution of the manufacturing employment probability in Figure \ref{fig:evol_p} and observe that it closely follows the evolution of the MTE. 
This suggests selection on earnings gains: workers primarily choose to enter manufacturing or not based on their expected earnings in each sector. The manufacturing employment share remains lower than in $1999$, largely because returns to working in manufacturing have not fully recovered. 
This pattern was undetectable in the naive (OLS) estimation of the returns to working manufacturing (Figure \ref{fig:return_ols}), which misleadingly suggests higher returns in $2018$ than in $1999$. Using the semi-IV approach to correct for sector choice endogeneity provides a clearer picture and helps explain the persistent decline in manufacturing employment. \\
\indent Perhaps surprisingly, the evolution of both MTRs (Figure \ref{fig:evol_mtr}) reveals a harsher reality for young, uneducated white men: while their manufacturing earnings ($Y_1$) have declined, their nonmanufacturing earnings ($Y_0$) have also fallen, albeit less sharply, since 1999. This highlights a marked deterioration in the economic prospects for this population, regardless of their sector choice. \\
\noindent \textbf{A practical remark.} As highlighted by this last result, the MTRs are often informative on their own. 
In general, they provide valuable information about the selection process, about the dependence between potential outcomes and resistance to treatment. 
In this application, we observe a standard scenario: $m_1$ is high when $V$ is low and decreases with $V$, while the reverse holds for $m_0$. However, many other scenarios could yield the same observed MTE. For instance, the MTE could be entirely driven by $m_1$ while $m_0$ would remain flat. Reporting only the MTE thus results in a loss of information. Since MTRs can also be estimated with IVs at no additional cost, I recommend always reporting them alongside the MTE.


\section{Conclusion}
This paper shows that semi-IVs provide a viable alternative to IVs for causal inference, particularly for estimating LATE and MTE. 
This novel approach is especially useful since semi-IVs are often more readily available, as illustrated by the examples and the returns to working in manufacturing application. Expanding the set of valid identifying variations broadens the empirical researchers' toolkit. This expansion should enable researchers to address important economic questions for which finding a credible IV was previously infeasible. 

%




\bibliographystyle{aer}
{\begin{spacing}{1}
\setlength{\bibsep}{4pt} 
\setlength{\baselineskip}{0pt} 
\footnotesize
\scriptsize
\bibliography{references}
\end{spacing}
}

\newpage
\appendix

\pagenumbering{arabic}

\newpage

\noindent {\LARGE \textbf{Supplementary materials: online appendices}} 

\section{Additional Literature Review}\label{app:literature} 

\noindent \textbf{Relaxing the exclusion restriction.} Given the importance of IVs for causal inference and the restrictiveness of the standard assumptions imposed on them, numerous papers explore identification while relaxing some of these assumptions \citep{dhaultfoeuillefevrier2015, torgovitsky2015, dechaisemartin2017, lee2018identifying, neweystouli2021, mogstadetal2021, kitagawa2021, goff2023vector} or provide tests of their validity \citep{kitagawa2015, caetanoetal2016, andresenhuber2021, dhaultfoeuilleetal2021, kwon2024testingmechanisms}. 
This paper focuses on relaxing the exclusion restriction and relates to the literature showing that identification can also be achieved without instruments or with included instruments, provided additional conditions hold, e.g., conditional partial independence \citep{mastenpoirier2018}, identification-at-infinity \citep{dhaultfoeuilleetal2018}, identification based on heteroskedasticity \citep{rigobon2003identification, kleinvella2010, lewbel2012, lewbel2018, lewbelschennachzhang2023}, or specific functional form or relevance conditions \citep{dhaultfoeuilleetal2021, gaowang2023, tsyawo2023feasible}. See \cite{lewbel2019identification} for an extensive discussion. 
Unlike these approaches, I use the \textit{same} general nonparametric nonseparable model -- i.e., generalized Roy models with weakly separable selection rule \citep{imbensangrist1994, vytlacil2002, heckmanvytlacil2005} -- as in the standard IV literature (extended with included semi-IVs). Crucially, I do not need to compensate for relaxing the exclusion restriction by adding further structure to the model (e.g., specifying the distribution of the error terms or introducing other assumptions). Instead, as with standard IVs, my identification strategy is "variable-based", relying only on the properties of the semi-IVs. Thus, under very general conditions, identification with semi-IVs follows the same logic as with IVs. \\
\indent Being able to obtain IV-like identification with "complementary invalid IVs" in nonseparable models is novel and is one of the main contributions of this paper and of \cite{bruneelbeyhum2024}. In the latter, we show that identification can be achieved by decomposing the exclusion and exogeneity properties of standard IVs into two other types of complementary invalid IVs, named "quasi-IVs": one that is excluded but possibly endogenous, and another that is exogenous but possibly included.  A similar idea appears in \cite{wang2023}, who proves that the LATE can be identified using two other (complementary) imperfect instruments: one violating the exclusion restriction and another violating monotonicity. A related idea is also found in \cite{kolesaretal2013}, who show that in linear models, identification can be achieved with many invalid (included) instruments, provided that their direct effects on outcomes are uncorrelated with their effects on the endogenous (continuous) regressor (which could be interpreted as a specific form of complementarity). 
Semi-IVs are another class of "complementary invalid IVs", which violate the full exclusion restriction but still satisfy some targeted exclusion restrictions from specific potential outcomes. 

\section{\cite{heckmanvytlacil2005} with semi-IVs}\label{app:heckmanvyt}

In this Appendix, I show how the Framework of Section \ref{sec:framework} can alternatively be written as a direct adaptation of the IV Framework of \cite{heckmanvytlacil2005}.  The main expositional difference is that I explicitly write the potential outcome equations here. 

\indent The potential outcomes and the selection are modeled as follows 
\begin{align}
	Y_d &= q_d(X, Z_d, U_d) \quad \quad \quad \quad \text{ for } d \in \mathcal{D}, \label{eq:hvoutcome} \\
	D &= \mathds{1}\{ g(X, Z) \geq V\}, \label{eq:hvselection}
\end{align}
where the semi-IVs, $Z=(Z_0, Z_1)$, and the covariates, $X$, are observed, while the "shocks", $(U_0, U_1, V)$, are unobserved. The observables are $Y, D, X$, and $Z$, meaning that all the semi-IVs are observed, regardless of the selected alternative. The scalar random variable $V$ may be correlated with $U=(U_0, U_1)$, yielding the endogeneity of $D$ with respect to the potential outcomes. The potential outcomes are thus determined as general nonseparable functions ($q_d$) of unrestricted shocks $U_d$ (which may contain several unobservables), which allows for rich forms of observed and unobserved heterogeneity, with (possibly) heterogenous treatment effects, general selection bias, and possible selection on gains. With the nonseparable potential outcome function, the semi-IVs can affect their corresponding potential outcomes in a completely unrestricted manner, also allowing for heterogenous effects of the semi-IVs on the potential outcomes from which they are not excluded.  \\
\indent The model can be described by the following set of assumptions. 

\begin{assumption}[Valid semi-IV model]\label{ass:heckmanvyt} The following conditions hold jointly
\begin{enumerate}[label=A\theassumption.\arabic*, itemsep=0pt, parsep=0pt] 
	\item \label{hvass:u1v} The random vectors $(U_1, V)$ are independent of $Z_0$ conditional on $Z_1$ and $X$. 
	\item \label{hvass:u0v} The random vectors $(U_0, V)$ are independent of $Z_1$ conditional on $Z_0$ and $X$. 
	\item \label{hvass:relevancew0} The term $g(Z_0, Z_1, X)$ is a nondegenerate random variable conditional on $Z_1$ and $X$. 
	\item \label{hvass:relevancew1} The term $g(Z_0, Z_1, X)$ is a nondegenerate random variable conditional on $Z_0$ and $X$. 
	\item \label{hvass:distv} The distribution of $V$ is absolutely continuous with respect to Lebesgue measure. 
	\item \label{hvass:regul1} $E|Y_1|$ and $E|Y_0|$ are finite. 
	\item \label{hvass:regul2} $1 > \textrm{Pr}(D=1 | X) > 0$. 
\end{enumerate}
\end{assumption}

\indent These assumptions are a direct adaptation of the IV assumptions of \cite{heckmanvytlacil2005} to the semi-IVs. As with IVs, they are equivalent to the main assumptions in Section \ref{sec:framework} \citep{vytlacil2002}. 
\noindent Indeed, the exclusion restriction holds because $Z_0$ is excluded from $Y_1$ in Equation \eqref{eq:hvoutcome}, and the exogeneity holds because of Assumption \ref{hvass:u1v}. Both of these imply that $Z_0$ is "excluded" from $Y_1$ conditional on $Z_1$ and $X$ (Assumption \ref{ass:exogeneity}): because it does not enter directly through the functional form, nor through the error term $U_1$ thanks to the independence. 
Similarly, assumption \ref{hvass:u0v} combined with Equation \eqref{eq:hvoutcome} imply that $Z_1$ is excluded from $Y_0$ conditional on $Z_0$ and $X$. \\
\indent A stronger, but often natural assumption, is to assume that 
\begin{align*}
	(U_0, U_1, V) \indep (Z_0, Z_1) | X.
\end{align*}
This is equivalent to assuming that $(Y_0(z_0,x), Y_1(z_1,x), D(z)) \indep (Z_0, Z_1) | X$, as in Footnote \ref{footnote:exogeneity}. This assumption is slightly stronger than the previous one, as it restricts how the semi-IVs affect their respective potential outcomes: they can only affect them through direct effects and not through their correlation with the unobserved $U_d$. \\
\indent The index model \eqref{eq:hvselection} for the selection is the same as the index model \eqref{eq:selection} in Section \ref{sec:framework}, and is equivalent to the Monotonicity assumption \ref{ass:monot} of \cite{imbensangrist1994}. Assumptions \ref{hvass:relevancew0} and \ref{hvass:relevancew1} are the relevance assumptions of $Z_0$ (resp. $Z_1$) conditional on $Z_1$ (resp. $Z_0$) and $X$. 
Assumption \ref{hvass:distv} is also assumed in the main text (with a normalization). \\
\indent Finally, Assumptions \ref{hvass:regul1} and \ref{hvass:regul2} are merely regularity conditions that ensure that the treatment parameters are well defined and that one observes a control and a treatment group for all $X$. 

%


\section{LATE \citep{imbensangrist1994} with semi-IVs}\label{app:late}

\noindent \textbf{Local Average Treatment Effect (LATE).} For any $X=x$, for any $z=(z_0, z_1)$ and $z'=(z_0', z_1') \in \mathcal{Z}$, with $P(z) = p < P(z') = p'$, define the counterpart of \cite{imbensangrist1994}'s LATE with semi-IVs as 
\begin{align*}
	\Delta_{LATE}(z, z', x) &= \mathbb{E}[ \ Y_1(z_1', x) - Y_0(z_0', x) \ | \ D(z', x)=1,  D(z, x) = 0, X=x \ ]. 
\end{align*}
In other words, this LATE is the mean gain for individuals (compliers) induced to switch from $D=0$ to $D=1$ by an exogenous change in $Z$ from $z$ to $z'$.\footnote{One could arbitrarily define the LATE differently, by comparing $Y_1(z_1) - Y_0(z_0)$, $Y_1(z_1') - Y_0(z_0)$ or $Y_1(z_1) - Y_0(z_0')$ instead. I choose this definition, which represents the treatment effect on the compliers in the new environment after the change in semi-IVs. Under this definition, the initial environment, $Z=z$, only serves to determine the initial propensity score, $P(z)$. This choice is arbitrary and not impactful since all these alternative parameters are identified as well. } The LATE is naturally decomposed into two local average treatment responses (LATR$_d$) defined as 
\begin{align*}
	\Delta_{LATR_d}(z, z', x) = \mathbb{E}[ \ Y_d(z_d', x) \ | \ D(z', x)=1,  D(z, x) = 0, X=x \ ] \quad \text{ for each } d=0, 1, 
\end{align*} 
such that $\Delta_{LATE}(z, z', x) = \Delta_{LATR_1}(z, z', x) - \Delta_{LATR_0}(z, z', x)$. \\
\indent Note that in terms of the general LATE and LATR$_d$ definitions of Section \ref{sec:identification}, we have 
\begin{align*}
	\Delta_{LATE}(z, z', x) = \text{LATE}(p, p', z_0', z_1', x), \quad \text{ and } \quad \Delta_{LATR}(z, z', x) = \text{LATR}(p, p', z_d', x), 
\end{align*}
since individuals with $D(z', x)=1; D(z, x)=0$ are the compliers with $p \leq V < p'$. Thus, $\Delta_{LATE}(z, z', x)$ can be identified as we identify $\text{LATE}(p, p', z_0', z_1', x)$, by comparing four key points.  
Let us see how to achieve this result alternatively, by adapting the reasoning of \cite{imbensangrist1994} to the presence of semi-IVs instead. \\

\noindent \textbf{Identification.} 
We want to identify $\Delta_{LATE}(z, z')$ using data on $(D, Y, Z_0, Z_1)$.  For any $z=(z_0, z_1)$ and $z'=(z_0', z_1') \in \mathcal{Z}$, with $P(z) = p < P(z') = p'$, we have 
\begin{align}\label{eq_late}
	&\mathbb{E}\big[ Y | Z=z' \big] - \mathbb{E}\big[ Y | Z=z \big] \nonumber \\
	&= \mathbb{E}[DY | Z=z'] - \mathbb{E}[DY|Z=z] \ + \ \mathbb{E}[(1-D)Y | Z=z'] - \mathbb{E}[(1-D)Y|Z=z]. \\ \nonumber 
\end{align}

\begin{figure}[!h]
\begin{center}
\begin{tikzpicture}[x=150pt,y=140pt, line width=0.25mm] 

\def\deltac{0.3}
\def\deltawzero{-1.5}
\def\deltawone{3.5}
\def\deltatwo{-2}
\def\Probazero{0.20} 
\def\Probaone{0.90} 

\def\wone{0.6}
\def\wzero{0.3}

\draw[<->](0,1)--(0,0)--(1,0); 
\draw (0, 1) node[above left=0pt, black]{$Z_1$};
\draw (1, 0) node[below right=0pt, black]{$Z_0$};
\begin{scope} 
\clip (0, 0) rectangle (1, 1);
\draw[scale=1, domain=0:1, smooth, variable=\x, black] plot ({\x}, {(\Probazero-\deltac-\deltawzero*\x)/(\deltawone+\deltatwo*\x)});
\draw[scale=1, domain=0:1, smooth, variable=\x, gray] plot ({\x}, {(\Probaone-\deltac-\deltawzero*\x)/(\deltawone+\deltatwo*\x)});

\end{scope}
\draw (1, 0.9) node[above right=0pt, black]{$P=p$};
\draw (0.7, 1) node[above right=0pt, gray]{$P=p'$};

\draw (0.01, \wone) -- (-0.01, \wone) node[left=0pt]{$z_1'$};
\draw (\wzero, 0.01) -- (\wzero, -0.01) node[below=0pt]{$z_0$};

\pgfmathsetmacro{\wzerop}{(\Probaone-\deltac-\deltawone*\wone)/(\deltawzero+\deltatwo*\wone)}

\coordinate (z0) at (\wzero, {(\Probazero-\deltac-\deltawzero*\wzero)/(\deltawone+\deltatwo*\wzero)}); 
\coordinate (z0') at (\wzerop, {(\Probazero-\deltac-\deltawzero*\wzerop)/(\deltawone+\deltatwo*\wzerop)}); 
\filldraw[black] (z0) circle (2pt) node[below right=0pt] {$z$};
\filldraw[mtr0color2] (z0') circle (2pt) node[above left=-3pt] {$\bar{z}^0$};
\draw[dotted, black](\wzero, 0)--(z0)--(0, {(\Probazero-\deltac-\deltawzero*\wzero)/(\deltawone+\deltatwo*\wzero)}) node[left=0pt, black]{$z_1$};
\draw[dotted, mtr0color2](\wzerop, 0)--(z0')--(0, {(\Probazero-\deltac-\deltawzero*\wzerop)/(\deltawone+\deltatwo*\wzerop)}) node[left=0pt, mtr0color2]{$\tilde{z}_1$};
\draw[mtr0color2] (0.01, {(\Probazero-\deltac-\deltawzero*\wzerop)/(\deltawone+\deltatwo*\wzerop)}) -- (-0.01, {(\Probazero-\deltac-\deltawzero*\wzerop)/(\deltawone+\deltatwo*\wzerop)});
\draw[black] (0.01, {(\Probazero-\deltac-\deltawzero*\wzero)/(\deltawone+\deltatwo*\wzero)}) -- (-0.01, {(\Probazero-\deltac-\deltawzero*\wzero)/(\deltawone+\deltatwo*\wzero)});

\coordinate (z1) at ({(\Probazero-\deltac-\deltawone*\wone)/(\deltawzero+\deltatwo*\wone)}, \wone); 
\coordinate (z1') at ({(\Probaone-\deltac-\deltawone*\wone)/(\deltawzero+\deltatwo*\wone)}, \wone); 
\filldraw[mtr1color2] (z1) circle (2pt) node[below right=0pt] {$\bar{z}^1$};
\filldraw[black] (z1') circle (2pt) node[above left=0pt] {$z'$};
\draw[dotted, mtr1color2] (0, \wone)--(z1)--({(\Probazero-\deltac-\deltawone*\wone)/(\deltawzero+\deltatwo*\wone)}, 0) node[below, mtr1color2]{$\tilde{z}_0$};
\draw[dotted, black] (0, \wone)--(z1')--({(\Probaone-\deltac-\deltawone*\wone)/(\deltawzero+\deltatwo*\wone)}, 0) node[below, black]{$z_0'$};
\draw[mtr1color2] ({(\Probazero-\deltac-\deltawone*\wone)/(\deltawzero+\deltatwo*\wone)}, 0.01) -- ({(\Probazero-\deltac-\deltawone*\wone)/(\deltawzero+\deltatwo*\wone)}, -0.01);
\draw[black] ({(\Probaone-\deltac-\deltawone*\wone)/(\deltawzero+\deltatwo*\wone)}, 0.01) -- ({(\Probaone-\deltac-\deltawone*\wone)/(\deltawzero+\deltatwo*\wone)}, -0.01);

\draw[->, ultra thick, red, line width=2pt, opacity=0.8] ($(z0) + (0.02, 0.03) $) -- ($(z1') - (0.02,0.03)$) node[midway, above] {};

\draw[<->, ultra thick, mtr1color2, line width=2pt, opacity=0.8] ($(z0) - (-0.02, 0) $) to[bend right=30] ($(z1) - (0,0.03)$) node[midway, above] {};

\draw[<->, ultra thick, mtr0color2, line width=2pt, opacity=0.8] ($(z0) - (-0.01, 0) $) to[bend right=20] ($(z0') - (0,0.03)$) node[midway, above] {};

\end{tikzpicture} 
\centering
\captionsetup{justification=centering}
\caption{Identification of $\Delta_{LATE}(z, z')$} 
\label{fig:late_identification2}
\end{center}
\end{figure}
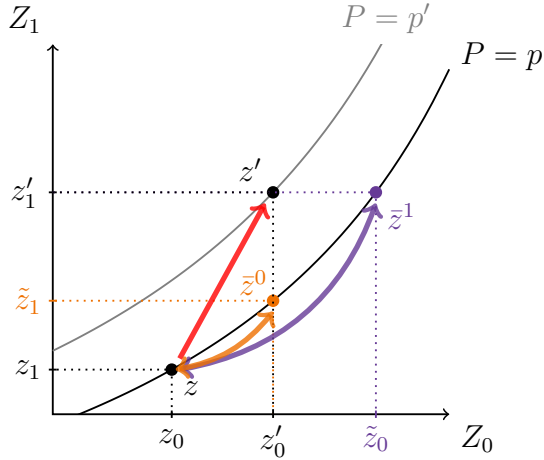

\noindent \textbf{Treated $\Delta_{LATR_1}(z, z')$.} First, focus on the treated sample (terms in $DY$). We have: 
\begin{align}\label{eq:latr1_app}
	&\mathbb{E}[DY|Z=z']-\mathbb{E}[DY|Z=z]  \nonumber \\
	&= \mathbb{E}[Y_1D(z') | Z=z'] - \mathbb{E}[Y_1D(z) | Z=z] \nonumber \\
	&= \mathbb{E}[Y_1 D(z') | Z_1=z_1'] - \mathbb{E}[Y_1 D(z)|Z_1=z_1] \nonumber \\
	&= \underbrace{\mathbb{E}[Y_1 (D(z') - D(z)) | Z_1=z_1']}_{= \ \text{selection effect given $Z_1=z_1'$}} \quad  + \quad \underbrace{\mathbb{E}[Y_1D(z) | Z_1=z_1'] - \mathbb{E}[Y_1D(z) | Z_1=z_1]}_{= \ \text{direct effect of } Z_1 \text{ from } z_1 \text{ to } z_1' \text{ when } D(z)=1}.
\end{align}
The first equality holds because when $D=1$, $Y = Y_1$, and by rewriting $D$ as the potential treatment $D(z)$. The second equality comes from the exclusion of $Z_0$ from $Y_1$ and $D(z)$ given $Z_1$, and the last equality is simply a rewriting where I added and subtracted $\mathbb{E}[YD(z)|Z_1=z_1']$. In words, an exogenous change in $Z$ from $z$ to $z'$ affects the average $DY$ for two reasons. First, as it would be the case with IVs, the shift in the semi-IVs yields a selection/composition effect because the set of individuals who select $D=1$ increases (since $p' > p$). There are compliers who choose $D(z')=1$ while they were choosing $D(z) = 0$ before the change. Second, because $Z_1$ changed from $z_1$ to $z_1'$, we observe a direct effect of the semi-IV change on the outcomes, since $Z_1$ is not excluded from $Y_1$. In this case, we need to isolate the effect of $Z_1$ on the always takers, who choose $D(z)=1$ even before the semi-IV change.  This is the main difference with standard IVs: to isolate the selection effect, we need to be able to identify this direct effect separately. \\
\indent To do so, we need to find a point $\bar{z}^1=(\tilde{z}_0, z_1')$ such that $P(\bar{z}^1) = P(z) = p$. Then, the direct effect is identified by the comparison of
\begin{align}\label{eq:latr1_app_compar}
	\mathbb{E}[DY | Z=\bar{z}^1] - \mathbb{E}[DY|Z=z] &= \mathbb{E}[Y_1D(z) | Z=\bar{z}^1 ] - \mathbb{E}[Y_1D(z) | Z=z] \nonumber \\
	&= \mathbb{E}[Y_1D(z) | Z_1=z_1' ] - \mathbb{E}[Y_1D(z) | Z_1=z_1], 
\end{align}
where we can drop the conditioning on $Z_0$ because of its exclusion from $Y_1$ and $D(z)$ given $Z_1$. This corresponds to the \textcolor{mtr1color2}{purple arrow} in Figure \ref{fig:late_identification2}. \\
\indent Plugging \eqref{eq:latr1_app_compar} in \eqref{eq:latr1_app}, we can isolate the selection effect as
\begin{align}\label{eq:latr1_selection_app}
	&\mathbb{E}[Y_1 (D(z') - D(z)) | Z_1=z_1'] \nonumber \\
	&= \Big( \mathbb{E}[DY|Z=z']-\mathbb{E}[DY|Z=z] \Big) - \Big( \mathbb{E}[DY | Z=\bar{z}^1] - \mathbb{E}[DY|Z=z] \Big) \nonumber \\
	&= \ \mathbb{E}[DY|Z=z'] - \mathbb{E}[DY | Z=\bar{z}^1]. 
\end{align}
In other words, the selection effect for $Y_1$ (close to $\Delta_{LATR_1}(z, z')$) is obtained by comparing the outcomes in the treated sample given $Z=z'$ to the ones given $Z=\bar{z}^1$. 
Note that it corresponds exactly to the two points comparison that would be used to identify LATR$_1(p, p', z_1')$ for the generalized LATR/LATE: a shift in the propensity score $P$ from $p$ to $p'$ by shifting the excluded semi-IV $Z_0$ from $\tilde{z}_0$ to $z_0$ while holding $Z_1=z_1'$ fixed. \\

\vspace{-1\baselineskip}

\noindent \textbf{Untreated $\Delta_{LATR_0}(z, z')$.}  Similarly, on the untreated sample (terms in $(1-D)Y$),  we have
\begin{align}\label{eq:latr0_app}
	&\mathbb{E}[(1-D)Y|Z=z']-\mathbb{E}[(1-D)Y|Z=z]   \\
	&= - \underbrace{\mathbb{E}[Y_0 (D(z') - D(z)) | Z_0=z_0']}_{:= \ \text{selection effect given $Z_0=z_0'$}} + \underbrace{\mathbb{E}[Y_0(1-D(z)) | Z_0=z_0'] - \mathbb{E}[Y_0(1-D(z)) | Z_0=z_0]}_{:= \ \text{direct effect of } Z_0 \text{ from } z_0 \text{ to } z_0' \text{ when } D(z)=0}. \nonumber
\end{align}
To isolate the direct effect of $Z_0$, we need to find a point $\bar{z}_0 = (z_0', \tilde{z}_1)$ such that $P(\bar{z}_0) = P(z) = p$ (cf the \textcolor{mtr0color2}{orange arrow} in Figure \ref{fig:late_identification2}), such that, 
\begin{align}\label{eq:latr0_app_compar}
	&\mathbb{E}[(1-D)Y | Z=\bar{z}_0] - \mathbb{E}[(1-D)Y | Z=z] \nonumber \\
	&= \mathbb{E}[Y_0(1-D(z)) | Z_0=z_0'] - \mathbb{E}[Y_0(1-D(z)) | Z_0=z_0]. 
\end{align}
Plugging \eqref{eq:latr0_app_compar} into \eqref{eq:latr0_app}, we can isolate the selection effect for the untreated as 
\begin{align}\label{eq:latr0_selection_app}
	&- \mathbb{E}[Y_0 (D(z') - D(z)) | Z_0=z_0] \nonumber \\
	&= \Big( \mathbb{E}[(1-D)Y|Z=z']-\mathbb{E}[(1-D)Y|Z=z] \Big) - \Big( \mathbb{E}[(1-D)Y | Z=\bar{z}^0] - \mathbb{E}[(1-D)Y|Z=z] \Big) \nonumber \\
	&=  \quad \mathbb{E}[(1-D)Y|Z=z'] - \mathbb{E}[(1-D)Y | Z=\bar{z}^0]. 
\end{align}
In other words, the selection effect for $Y_0$ (close to minus $\Delta_{LATR_0}(z, z')$) is obtained by comparing the outcomes in the untreated sample given $Z=z'$ to the ones given $Z=\bar{z}^0$. 
Note that it corresponds exactly to the two points comparison that would be used to identify LATR$_0(p, p', w_0)$ for the generalized LATR/LATE: a shift in the propensity score $P$ from $p$ to $p'$ by shifting the excluded semi-IV $Z_1$ from $\tilde{z}_1$ to $z_1'$ while holding $Z_0=z_0'$ fixed. \\

\noindent \textbf{LATE.} 
To obtain the $\Delta_{LATE}(z, z')$, notice that, because of the exclusion of the semi-IVs, 
\begin{align*}
	&\mathbb{E}[Y_1 (D(z') - D(z)) | Z_1=z_1'] = \mathbb{E}[Y_1 (D(z') - D(z)) | Z_1=z_1', Z_0=z_0'] \quad \text{ for any } z_0', \\
	\text{ and }\quad  &\mathbb{E}[Y_0 (D(z') - D(z)) | Z_0=z_0'] = \mathbb{E}[Y_0 (D(z') - D(z)) | Z_0=z_0', Z_1=z_1'] \quad \text{ for any } z_1'.
\end{align*}
Thus, we can rewrite and add \eqref{eq:latr1_selection_app} with \eqref{eq:latr0_selection_app} to obtain:
\begin{align*}
	&\mathbb{E}[(Y_1 - Y_0) (D(z') - D(z)) | Z_0 = z_0', Z_1=z_1'] \\
	&= \mathbb{E}[DY | Z=z'] - \mathbb{E}[DY | Z=\bar{z}^1] + \mathbb{E}[(1-D)Y | Z=z'] - \mathbb{E}[(1-D)Y | Z=\bar{z}^0]. 
\end{align*}
Since the first term is equal to $\Delta_{LATE}(z, z')  (p'-p)$, the LATE is identified by 
\begin{align}
	\Delta_{LATE}(z, z') &= \frac{\mathbb{E}[DY | Z=z'] - \mathbb{E}[DY | Z=\bar{z}^1] + \mathbb{E}[(1-D)Y | Z=z'] - \mathbb{E}[(1-D)Y | Z=\bar{z}^0]}{p'-p} \nonumber \\
	&= \frac{\mathbb{E}[Y | Z=z'] - \mathbb{E}[DY | Z=\bar{z}^1] - \mathbb{E}[(1-D)Y | Z=\bar{z}^0]}{p'-p}, \nonumber 
\end{align}
if both $\bar{z}^1=(\tilde{z}_0, z_1')$ with $P(z_1) = P(z)$ and $\bar{z}^0=(z_0', \tilde{z}_1)$ with $P(\bar{z}_0) = P(z)$ exist. This formula corresponds exactly to the same two by two points comparisons that would be used to identify the generalized LATE$(p, p', z_0', z_1')$. Additionally here, given the specific parameter definition, we compare two different points, $\bar{z}^0$ and $\bar{z}^1$, to the same $z'$, requiring only $3$ different points/environments in total. The definition of the LATE and its identification are strongly tied to $Z=z'$. On the contrary, the original point, $z$, only pins down $p=P(z)$, but its underlying values, $(z_0, z_1)$ are not important otherwise.

\section{Direct effect of the semi-IVs (targeted-policy evaluation)}\label{app:directeffect}

\indent In general, for each $d=0,1$, an exogenous change in $Z_d$ from $z_d$ to $z_d'$ has two effects on the observed average outcome $Y$: (i) a \textit{selection effect} since it alters the composition of the treated and untreated population by shifting the treatment incentives ($P$), and (ii) a \textit{direct effect} since, holding the composition fixed, the change in $Z_d$ directly affects the potential outcomes $Y_d$.\footnote{To discuss the effects of \textit{exogenous changes} in the semi-IVs, one must adopt the stronger version of the exogeneity assumption \ref{ass:exogeneity} outlined in footnote \ref{footnote:exogeneity}, i.e., $(Y_0(z_0), Y_1(z_1), D(z)) \indep (Z_0, Z_1) | X$. This ensures that the semi-IVs are exogenous, influencing the potential outcomes only through general direct effects (because they are not excluded), without being correlated with $d$-specific unobservables. This stronger assumption is perfectly reasonable in most applications. }  
The MTR and LATR capture these selection effects, which are identified by varying $P$ at fixed $Z_d$ using the excluded $Z_{1-d}$. 
In many cases, the direct effects of the semi-IVs on their respective potential outcomes are also of interest. For instance, when $Z_d$ is a $d$-specific targeted policy -- such as a subsidy or a tax cut/increase for a given industry -- evaluating the direct effect of $Z_d$ on $Y_d$ is itself valuable.\footnote{This is akin to a mediation analysis for targeted policies, decomposing the overall effect of an exogenous change in $Z_d$ on $Y$ into the direct effect of $Z_d$ on $Y_d$ and the indirect effect mediated by selection $D$.} In the manufacturing example, the direct effect corresponds to the pass-through from manufacturing and nonmanufacturing characteristics to the earnings of the workers in the corresponding sectors. \\
\indent To identify these direct effects, net of the selection effects, the general intuition is to vary $Z_d$ while holding $P$ fixed. Holding $P$ fixed requires using the other semi-IV, $Z_{1-d}$.
\noindent For example, conditional on $X=x$ left implicit as before, for each $d$, define the \textit{average direct effect} of $Z_d$ on the $D=d$ group (i.e., the \textit{treated} or \textit{untreated} population) when $P=p$ by
\begin{align}\label{eq:wdatteffect}
	\Delta_{Z_d}^{D=d}(z_d, z_d', p) = &\quad \mathbb{E}[Y_d | D=d, Z_d=z_d', P=p] 
	- \mathbb{E}[Y_d | D=d, Z_d=z_d, P=p]. 
\end{align}
This is the average effect of exogenously varying $Z_d$ from $z_d$ to $z_d'$ on the individuals who choose $D=d$ when $P=p$, i.e., those with unobserved resistance to treatment $V \leq p$ if $D=1$ and $V > p$ if $D=0$. These average targeted policy effects of $Z_d$ on $Y_d$, conditional on $D=d$, may depend on the underlying selected population $V$ and thus vary with $P$. 
They are also identified using the excluded semi-IV as an IV. Specifically, for any policy change from $Z_d=z_d$ to $z_d'$, if there exists two distinct points, $\tilde{z} = (z_0, z_1)$ and $\tilde{z}' = (z_0', z_1')$, such that $P(\tilde{z}) = P(\tilde{z}') = p$, then the direct policy effect $\Delta_{Z_d}^{D=d}(z_d, z_d', p)$ is identified by
\begin{align*}
		\Delta_{Z_d}^{D=d}(z_d, z_d', p) = \mathbb{E}[Y | D=d, Z=\tilde{z}'] - \mathbb{E}[Y | D=d, Z=\tilde{z}] \quad \text{ for each } d=0,1.
\end{align*}
In words, the policy effect of changing $Z_d$ from $z_d$ to $z_d'$ is identified if implemented in environments with various $Z_{1-d}$, allowing $P=p$ to be observed with both values of $Z_d$. \\  
\indent The direct effects of $Z_d$ for individuals with $V=v$ are also identified. Define 
\begin{align}\label{eq:wdeffect}
	\Delta_{Z_d}(z_d, z_d', v) = \mathbb{E}[Y_d | Z_d=z_d', V=v] - \mathbb{E}[Y_d | Z_d=z_d, V=v],  
\end{align}
the average effect of changing $Z_d$ from $z_d$ to $z_d'$ on $Y_d$ for individuals with unobserved resistance to treatment $V=v$. This is directly identified by 
 \begin{align*}
	\Delta_{Z_d}(z_d, z_d', v) = m_d(v, z_d') - m_d(v, z_d), 
\end{align*}
if the corresponding MTR$_d$ are identified at $(v, z_d)$ and $(v, z_d')$.\footnote{Similarly, the policy effect on a set of compliers with $V \in [v, v')$ is identified using the LATR$_d$ as
\vspace{-0.5\baselineskip}
\begin{align*}
	\Delta_{Z_d}(z_d, z_d', v, v') &= \mathbb{E}[Y_d | Z_d=z_d', v \leq V < v'] - \mathbb{E}[Y_d | Z_d=z_d, v \leq V < v'] 
	= \text{LATR}_d(v, v', z_d') - \text{LATR}_d(v, v', z_d). 
\end{align*}
\vspace{-2\baselineskip}} \\
\indent The direct effects of the semi-IVs are generally heterogenous with respect to $V$. Since these effects can be identified at different values of the unobserved $V$, one can test their homogeneity. Moreover, testing whether they are homogenous and null -- i.e., if $\Delta_{Z_d}(z_d, z_d', v) = 0$ for all $z_d, z_d' \in \mathcal{Z}_d$ and $v \in [0, 1]$ -- provides a test of whether $Z_d$ is a valid IV. Indeed, if $Z_d$ has no effect on $Y_d$, it is excluded from both potential outcomes, making it a valid IV. 

\section{Estimation: Robinson double residual regression}\label{app:estimation}

This appendix details \cite{robinson1988root}'s double residual regression procedure estimating the partially linear model of Equation \eqref{eq:plm} on the treated and untreated subsamples, i.e., 
\begin{align*}
	\mathbb{E}[Y | D=d, Z_d, P, X] = X \beta_d + Z_d \delta_d + \kappa_d(P), \quad \text{ for } d =0,1. 
\end{align*}
As the counterpart estimation of MTE with IVs, the estimation consists of three steps. \\

\noindent \textbf{1$^{st}$ stage: propensity score.} Estimate the propensity score $P(Z_0, Z_1, X)$ by regressing $D$ on $(Z_0, Z_1, X)$. I use a probit model for this stage, but linear probability models, logit, or nonparametric alternatives are applicable. One can check here whether the semi-IVs are relevant or not. If they are, one can proceed using the fitted propensity score, $\hat{P}$, as an observed variable in the following stages. \\

\noindent \textbf{2$^{nd}$ stage: direct effect of the semi-IVs (and covariates).} On the subsample with $D=d$, estimate $\widehat{\mathbb{E}}[Y|D=d, \hat{P}]$, $\widehat{\mathbb{E}}[X|D=d, \hat{P}]$ and $\widehat{\mathbb{E}}[Z_d|D=d, \hat{P}]$ for all $X$ and $Z_d$ variables, using local (linear) regressions of $Y, X, $ and $Z_d$ on $\hat{P}$. 
Then, create the residuals, $r_{variable}^d = variable - \widehat{\mathbb{E}}[variable|D=d, \hat{P}]$ for each variable $Y, Z_d,$ and $X$. Notice that, according to \eqref{eq:POsepa}, we have
\begin{align*}
	&Y_d - \mathbb{E}[Y_d | D=d, P]  \\
	&= (X - \mathbb{E}[X | D=d, P]) \beta_d + (Z_d - \mathbb{E}[Z_d | D=d, P])\delta_d + \smash{\overbrace{U_d - \mathbb{E}[U_d | D=d, P]}^{:= \ \Epsilon}},
\end{align*}
with $\mathbb{E}[ \Epsilon | r_X^d, r_{Z_d}^d] = 0$ by the separability and independence assumptions. So, on the subsample $D=d$, the linear regression (without intercept) of 
\begin{align*}
	r_Y^d = r_X^d \beta_d + r_{Z_d}^d \delta_d + \Epsilon,
\end{align*}
provides consistent estimates of the effect of the covariates, $\hat{\beta}_d$, and of the semi-IVs $Z_d$, $\hat{\delta}_d$, on $Y_d$. The role of the excluded semi-IV, $Z_{1-d}$ is to avoid perfect colinearity between $Z_d$ and $P$ here, i.e., to have variations in $P$ at fixed $Z_d$. \\

\noindent \textbf{3$^{rd}$ stage: $k_d(v)$.} 
Once $\beta_d$ and $\delta_d$ are estimated for both $d=0,1$, construct
\begin{align*}
\tilde{Y}_i = Y_i - ( X_i\hat{\beta}_d -	 Z_{di} \hat{\delta}_d))\mathds{1}\{D_i = d\},  
\end{align*}
i.e., the outcome net of the effects of the covariates and semi-IVs. If $\hat{\beta}_d$ and $\hat{\delta}_d$ are consistently estimated, by analogy, using \eqref{eq:POsepa}, we have that 
\begin{align*}
	\tilde{Y}_i = D_i \widehat{U}_{1i} + (1-D_i) \widehat{U}_{0i}, 
\end{align*}
i.e., $\tilde{Y}$ provides consistent estimates of the unobservables $U_d$ when $D=d$. Then, for the treated individuals we have
\begin{align*}
	\mathbb{E}[D\tilde{Y} | \hat{P}=p]= p  \ \mathbb{E}[U_1|D=1, \hat{P}=p] = p\  \mathbb{E}[U_1|V \leq p] = p \int_0^p \mathbb{E}[U_1 | V=v] \frac{1}{p} dv = \int_0^p k_1(v) dv. 
\end{align*}
Consequently, for any $v$ in the interior of the support of $\hat{P}$ given $D=1$,
\begin{align*}
	\frac{\partial\mathbb{E}[D\tilde{Y}_i | \hat{P}=p]}{\partial p} \Big|_{p=v}= \ \widehat{k_1}(v). 
\end{align*}
\indent Similarly, for any $v$ in the interior of the support of $\widehat{P}$ given $D=0$, 
\begin{align*}
	- \ \frac{\partial\mathbb{E}[(1-D)\tilde{Y}_i | \hat{P}=p]}{\partial p} \Big|_{p=v}= \ \widehat{k_0}(v). 
\end{align*}
So to estimate $k_d(v)$, estimate $\mathbb{E}[\tilde{Y}_i\mathds{1}\{D=d\} | \hat{P}=p]$ and its derivative using local quadratic regressions of $\tilde{Y}D$ and $\tilde{Y}(1-D)$ on $\widehat{P}$.\footnote{Following \cite{fangijbels1996}, in order to estimate the first order derivative I specify a local polynomial of the order of the derivative $+ 1$, hence a local quadratic regression. The \texttt{semiIVreg} package includes several options for the kernel and bandwidth choices. The default option is to use the Gaussian kernel and to select the bandwidth according to the mean-squared error direct plug-in criteria of \cite{fangijbels1996}, as computed by \cite{calonico2019nprobust} (\texttt{nprobust} package).} \\
\indent Notice that the identification and estimation in the separable model \eqref{eq:POsepa} is very close to the general identification proof. The main advantage of separability is that $k_d(v)$ does not depend on $z_d$ (or $x$). As a result, one can identify $k_d(v)$ for any $v$ in the interior of the support of $\widehat{P}$ given $D=d$, instead of only on the support of $\widehat{P}$ given $D=d$ and $Z_d=z_d$ and $X=x$ as in the general proof.\footnote{In the general identification proof, I never specified that the support of $P$ was taken conditional on $D=d$. This is because, in terms of identification, if $0 < P < 1$, we should always observe some individuals with $D=1$ and some with $D=0$. With small samples, this is not necessarily the case, hence the additional precision for the estimation. } 


\section{Monte Carlo Simulations}\label{app:montecarlo}
\subsection{Heterogenous treatment effect}\label{app:mc_heterogenous}
I simulate a simple generalized Roy model, close to the simulations of \cite{heckmanurzuavytlacil2006}, but including semi-IVs. 

\subsubsection{Generalized Roy Model}
\noindent\textbf{Outcome and selection.} 
I simulate the following generalized Roy model with heterogenous treatment effects: 
\begin{align*}
	Y_d &= \mu_d + Z_d \delta_d + U_d, \\
	D &= \mathds{1}\{ \underbrace{\alpha + Z_0 \alpha_0 + Z_1 \alpha_1}_{:= \ g(Z)} - \tilde{V} \geq 0 \}, 
\end{align*}
with $\tilde{V} \indep (Z_0, Z_1)$, and $\mathbb{E}[U_d | \tilde{V}, Z_0, Z_1] = \mathbb{E}[U_d | \tilde{V}]$ such that the semi-IVs are valid and even satisfy the stronger Separability assumption \ref{ass:separability}. \\ 

\noindent \textbf{Semi-IVs.} The semi-IVs are correlated and simulated as
\[
\begin{pmatrix}
Z_0 \\
Z_1 
\end{pmatrix} \sim \mathcal{N} \left( \begin{pmatrix} 
\mu_{Z_0} \\ 
\mu_{Z_1} 
\end{pmatrix}, 
\begin{pmatrix} 
\sigma^2_{Z_0} & \sigma_{Z_0Z_1} \\ 
\sigma_{Z_0Z_1} & \sigma^2_{Z_1} 
\end{pmatrix} \right).
\]
If $\delta_0 \neq 0$, $Z_0$ is not excluded from $Y_0$ and is thus not a valid IV. However, it is still excluded from $Y_1$ conditional on $Z_1$. Because of the correlation between the semi-IVs, it is important to condition on $Z_1$ for the exogeneity to hold. 
Conversely, $Z_1$ is not a valid IV if $\delta_1 \neq 0$; it is, however, well excluded from $Y_0$ conditional on $Z_0$. 
Despite its nonexclusion from $Y_0$, if $\alpha_0 \neq 0$, $Z_0$ is relevant and thus a valid semi-IV excluded from $Y_1$. Similarly, if $\alpha_1 \neq 0$, $Z_1$ is relevant and thus a valid semi-IV excluded from $Y_0$.  \\

\noindent \textbf{Error distribution.} The joint distribution of $U_0$ and $U_1$ is given by:
\begin{align*}
    \begin{pmatrix} U_0 \\ U_1 \end{pmatrix} \sim \mathcal{N}\left(\begin{pmatrix} 0 \\ 0 \end{pmatrix}, \begin{pmatrix} \sigma^2_{U_0} & \sigma_{U_0U_1} \\ \sigma_{U_0U_1} & \sigma^2_{U_1} \end{pmatrix}\right)
\end{align*}
Then, the (unnormalized) resistance to treatment, $\tilde{V}$, is defined as a function of $U_1$, $U_0$ and some other unobserved utility cost of treatment, $C$, 
\begin{equation*}
    \tilde{V} = -(U_1 - U_0) + C,
\end{equation*}
where $C \sim \mathcal{N}(0, \sigma^2_{\text{cost}})$ and is independent of $(U_0, U_1)$. 
The dependence of $\tilde{V}$ on $U_1$ and $U_0$ is the reason why the treatment $D$ is endogenous in this model. Here $\tilde{V}$ directly includes the comparison between $U_1$ and $U_0$, implying some \textit{selection on gains}. Fixing $U_0$ and $C$, if $U_1$ is very large, then $Y_1$ is also large and the resistance to treatment $\tilde{V}$ is very low, meaning that the individual would be one of the first to select into treatment: thus $D=1$ is very likely. A similar reasoning holds when $U_0$ and thus $Y_0$ is very low. Reciprocally, if $U_1$ is very low or if $U_0$ is very high, then $\tilde{V}$ is high, and the individual is not likely to select into treatment. If the utility cost of treatment, $C$, is high, $\tilde{V}$ is also high, and the individual is less likely to select into treatment. \\ 
\indent By simple computations, we have that 
\begin{align*}
	\tilde{V} \sim \mathcal{N}(0, \underbrace{\sigma^2_{U_0} + \sigma^2_{U_1} - 2 \sigma_{U_0 U_1} + \sigma^2_C}_{:= \ \sigma^2_{\tilde{V}}}). 
\end{align*}
Since $\tilde{V}$ is normally distributed, the first stage is a probit model. As in the main text, normalize $\tilde{V}$ to $V=F_{\tilde{V}}(\tilde{V}) \sim \text{Unif}(0, 1)$, where $V$ represents the ranks of the unobserved resistance to treatment, with a low $V$ meaning a low resistance, and thus higher chances to select $D=1$. \\

\noindent \textbf{Theoretical MTR and MTE.} Given the model specification, we can derive a closed form for the MTR and MTE. I do not use the knowledge of the true parametric distribution of the errors for the semi-parametric estimation, but it will be helpful to compare the estimation results to the truth. 
\noindent We have
\begin{align*}
	k_d(v) = \mathbb{E}[ U_d | V=v ] = \mathbb{E}[U_d | \tilde{V} = F_{\tilde{V}}^{-1}(v)]
\end{align*}
Since $(\tilde{V}, U_0, U_1)$ are trivariate normal it yields the closed-form for $k_d(v)$:
\begin{align*}
	k_0(v) &= \frac{\sigma^2_{U_0} - \sigma_{U_0U_1}}{\sigma^2_{\tilde{V}}} \Big( F_{\tilde{V}}^{-1}(v) - \mu_{\tilde{V}} \Big),\\
	\text{ and } \quad  k_1(v) &= \frac{-\sigma^2_{U_1} + \sigma_{U_0U_1}}{\sigma^2_{\tilde{V}}} \Big( F_{\tilde{V}}^{-1}(v) - \mu_{\tilde{V}} \Big), 
\end{align*}
where $\mu_{\tilde{V}} = 0$. Then, the MTR$_d$ and MTE are given by:
\begin{align*}
	m_0(v, z_0) &= \mu_0 + z_0 \delta_0 + k_0(v), \\
	m_1(v, z_1) &= \mu_1 + z_1 \delta_1 + k_1(v), \\
	\text{ and } \quad MTE(v, z_0, z_1) &= \mu_1 - \mu_0 + z_1 \delta_1 - z_0 \delta_0 + k_1(v) - k_0(v). 
\end{align*}


\begin{figure}[t!]
\centering
\captionof{figure}{Monte-Carlo simulation with heterogenous treatment effects.}
\captionsetup{font=small, justification=raggedright, singlelinecheck=false}
\caption*{%
\centering
Specification with $N = 10,000$ observations.  \\
\begin{tabular}{lc} 
\textit{Parameters} & \\
Outcome: & $\mu_0 = 3.2$, $\mu_1 = 3.6$, $\delta_0 = 1$, $\delta_1 = 1.3$. \\
Selection: & $\alpha = -0.2$, $\alpha_0 = -1.2$, $\alpha_1 = 1$. \\
Semi-IVs: & $\mu_{Z_0} = 0$, $\mu_{Z_1} = 0$, $\sigma^2_{Z_0} = 1$, $\sigma^2_{Z_1} = 0.8$, $\sigma_{Z_0Z_1} = 0.3$. \\
Errors: & $\sigma^2_{U_0} = 1$, $\sigma^2_{U_1} = 1.5$, $\sigma_{U_0U_1} = 0.5$, $\sigma^2_C = 1.5$. \\
\end{tabular}%
}

\captionsetup{font=normalsize, justification=centering}
 \centering
  {\small
    \begin{tabular}{@{\extracolsep{5pt}}lcccc} 
\\[-1.8ex]\hline 
\hline 
\\[-1.8ex]  & \multicolumn{2}{c}{semi-IV} & \multicolumn{2}{c}{OLS} \\ 
\cline{2-3} \cline{4-5} 
\\[-1.8ex] Coefficients & Mean & (SD) & Mean & (SD) \\
\hline \\[-1.8ex] 
\textit{Effects of the semi-IVs.} & & & &\\
$\quad$ $Z_0$: $\delta_0 = 1$ & 1.000 & (0.022) & 0.932 & (0.015) \\
$\quad$ $Z_1$: $\delta_1 = 1.3$ & 1.300 & (0.022) & 1.217 & (0.020) \\ 
& & & & \\
\textit{Treatment (OLS).} & & & & \\
$\quad $ Intercept: $\mu_0 = 3.2$ & & & 3.395 & (0.013) \\
$\quad $ $D$: \ \ $\mu_1 - \mu_0 = 0.4$ & & & 0.625 & (0.022)\\
\\
\hline 
\hline \\[-1.8ex] 
\end{tabular}
}
\captionsetup{font=footnotesize, justification=justified}
\caption*{
\noindent \textit{Notes:} The OLS corresponds to the naive a regression of $Y$ on a constant, $D$, $DZ_1$ and $(1-D)Z_0$. We do not report estimates of $\mu_0$ nor $\mu_1$ with semi-IVs since we do not estimate it directly. 
}


\label{fig:MC_heter}
    \begin{subfigure}[b]{0.48\textwidth}
        \centering
        \includegraphics[width=1\textwidth]{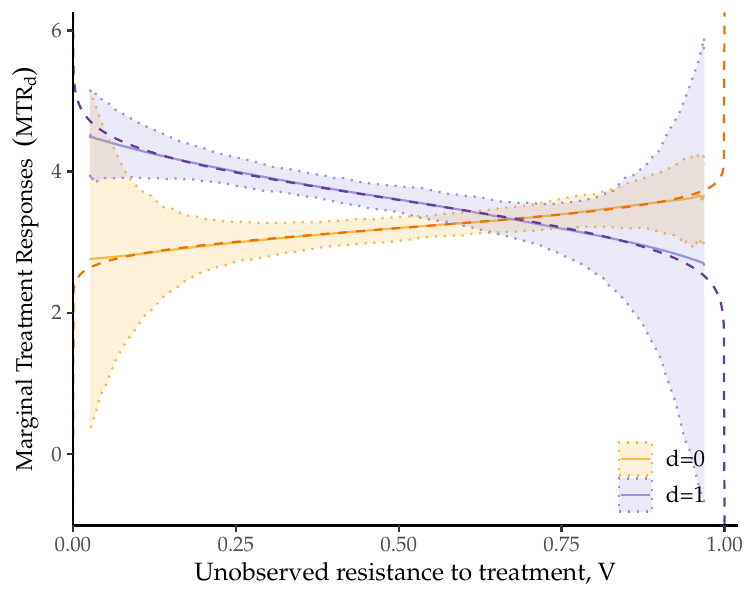}
    \end{subfigure}
    \hfill 
    \begin{subfigure}[b]{0.48\textwidth}
        \centering
        \includegraphics[width=1\textwidth]{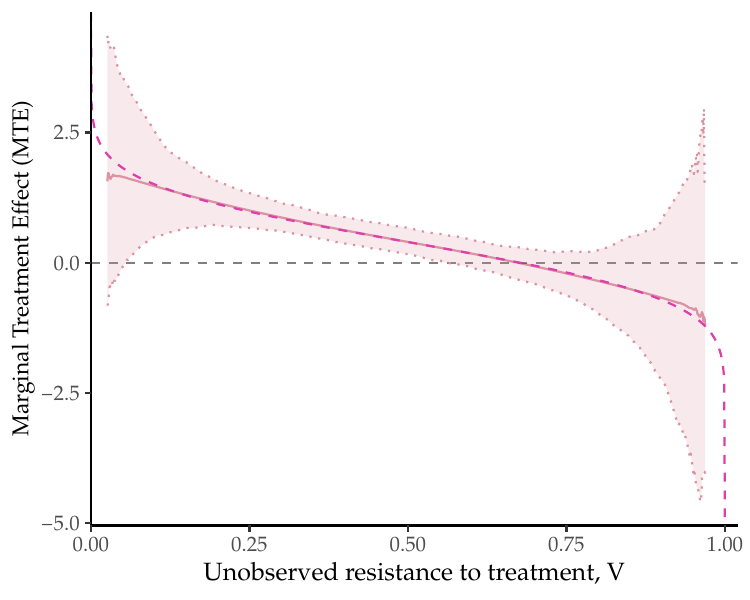}
    \end{subfigure}
    \vspace{0.5em} 
\captionsetup{font=footnotesize, justification=justified}
\caption*{
\noindent \textit{Notes:} Results of the estimation from $500$ Monte Carlo simulations. The table provides the estimated coefficients for the linear part of the model (the direct effects of the semi-IVs), while the straight lines in the figures provide the estimated MTRs and MTE, taken at a reference individual with $(Z_0, Z_1)=(0, 0)$ (i.e., their averages). In this special case, $m_d(v, z_d=0) = \mu_d + k_d(v)$. The colored areas represent the intervals where $95\%$ of the estimates lie. The dashed lines represent the true theoretical MTRs and MTE. We use arbitrary bandwidths of $0.10$ for both stages of the double residual regression. The results are robust to any bandwidth choice between $0.05$ and $0.20$. We trim the support of $V$ to $[0.025, 0.970]$, such that the common support of $\hat{P}$ includes these in almost all simulations. 
}
\end{figure}

\setstretch{1.25} 
\setlength{\abovedisplayskip}{6pt} 
\setlength{\belowdisplayskip}{6pt}

\subsubsection{Monte Carlo simulations}

Since the semi-IVs are valid, I can estimate the MTRs and MTE applying the semi-parametric estimation procedure described in Section \ref{sec:estimation_semi}. Figure \ref{fig:MC_heter} reports the estimation results from $500$ Monte Carlo simulations. \\
\indent First, for the linear part of the semi-parametric model, the effects of the semi-IVs ($\delta_d$) are precisely estimated. Not surprisingly, running a naive OLS assuming that the error term is independent of $D$ yields biased estimates of both the effects of the semi-IVs and the difference in intercepts (treatment effects). \\
\indent Second, for the nonparametric part of the model, I report the estimated MTRs and MTE at the mean semi-IVs, i.e., $(Z_0, Z_1) = (0,0)$. The estimates are reported over the common support of the propensity given $D=1$ and $D=0$, which I trim to $[0.025, 0.97]$ such that almost all simulations have observations with $\hat{P}$ in this common support. The estimated MTRs and MTEs are unbiased compared to their theoretical counterpart. They are also precisely estimated with $95\%$ of the estimates being in tight boundaries. This precision is remarkable because, with "only" $10,000$ observations, there are not so many observations in the neighborhood of any given $V=v$ in the treated or untreated subsamples. The semi-parametric estimation works well, even without making any assumption about the shape of the MTRs (especially the $k_d(v)$ functions) or about the distribution of errors. The estimates are noisier at the tails by construction of the estimator. 

\subsection{Homogenous treatment effect}\label{app:mc_homogenous}
\subsubsection{Constant treatment effect model}

Let us now simulate simpler models with homogenous treatment effects. The outcome and selection are given by 
\begin{align*}
	Y_d &= \mu_d + Z_d \delta_d + U, \\
	D &= \mathds{1}\{ \alpha + Z_0 \alpha_0 + Z_1 \alpha_1 - \tilde{V} \geq 0 \}, 
\end{align*}
where $(U, V) \indep (Z_0, Z_1)$ such that the semi-IVs are well excluded and satisfy a stronger form of separability than Assumption \ref{ass:separability}. 
Compared to the heterogenous treatment effect model, the main difference is that it is the same $U$ that affects both potential outcomes. Think about it as a single unobservable, e.g., the unobserved ability. 
As a consequence, the individual treatment effects 
\begin{align*}
	Y_1 - Y_0 = \mu_1 - \mu_0 + Z_1\delta_1 - Z_0 \delta_0, 
\end{align*}
are constant given $(Z_0, Z_1)$, and independent of $U$. 
In this case, one can follow \eqref{eq:tsls} and write a single outcome equation to be estimated: 
\begin{align}\label{eq:tsls_app}
	Y = \mu_0 + D(\mu_1 - \mu_0)  + D Z_1 \delta_1 + (1-D) Z_0 \delta_0 + U.
\end{align}
Then, the MTE are also independent of $V$, and equal to the ATE for all $v$:
\begin{align*}
	\text{ATE}(z_0, z_1) = \mathbb{E}[Y_1 - Y_0 | Z_0=z_0, Z_1=z_1] =\mu_1 - \mu_0 + z_1\delta_1 - z_0 \delta_0. 
\end{align*}
In particular, $ATE(0, 0) = \mu_1 - \mu_0$, i.e., the effect of $D$ in Equation \eqref{eq:tsls_app}.  \\
\indent The semi-IV follow a bivariate normal distribution as in Section \ref{app:mc_heterogenous}:  
\[
\begin{pmatrix}
Z_0 \\
Z_1 
\end{pmatrix} \sim \mathcal{N} \left( \begin{pmatrix} 
\mu_{Z_0} \\ 
\mu_{Z_1} 
\end{pmatrix}, 
\begin{pmatrix} 
\sigma^2_{Z_0} & \sigma_{Z_0Z_1} \\ 
\sigma_{Z_0Z_1} & \sigma^2_{Z_1} 
\end{pmatrix} \right).
\]

\indent About the shocks, there is endogeneity because the outcome shock, $U$, and the unobservable resistance to treatment, $\tilde{V}$, are correlated as follows:
\begin{align*}
    \begin{pmatrix} U \\ \tilde{V} \end{pmatrix} \sim \mathcal{N}\left(\begin{pmatrix} 0 \\ 0 \end{pmatrix}, \begin{pmatrix} \sigma^2_{U} & \sigma_{U\tilde{V}} \\ \sigma_{U\tilde{V}} & \sigma^2_{\tilde{V}} \end{pmatrix}\right). 
\end{align*}
As usual, we normalize $V=F_{\tilde{V}}(\tilde{V})$. 

\subsubsection{Monte Carlo simulations} 

This model can be estimated using several approaches with semi-IVs. 
\noindent First, one can use the simple 2SLS approach described in Section \ref{sec:estimation_2sls}. 
\noindent One can also estimate the model with the general semi-parametric (SP) approach of Section \ref{sec:estimation_semi}: in this case, we estimate heterogenous MTEs and MTRs varying with $V$ (at the reference individual with $(Z_0, Z_1)=(0,0)$), and then compute their averages by integrating MTE$(v, 0, 0)$ over the support of $P$ (the range of the support should not matter since the true treatment effects are assumed to be homogenous here). The averages of $m_1(v, 0), m_0(v, 0)$ and MTE$(v, 0, 0)$ provide an estimate of $\mu_1$, $\mu_0$ and $\mu_1 - \mu_0 = $ATE$(0, 0)$, respectively. As for the effects of the semi-IVs, they are already estimated as a single coefficient in the semi-parametric approach. This procedure has the advantage of being robust when treatment effects are heterogenous. Unfortunately, this comes at the cost of some efficiency loss. \\
\indent A better approach is to impose the homogeneity constraint on the treatment effects in the semi-parametric estimation (SP-H). To do so, note that, since $(U, V) \indep (Z_0, Z_1)$, we have $(U, V) \indep P$ since $P=P(Z)$, and we can write
\begin{align*}
	\mathbb{E}[U] = \mathbb{E}[U | P ] = \mathbb{E}[U | D=1, P] P + \mathbb{E}[U | D=0, P] (1-P). 
\end{align*}
Denote $\kappa_d(P) = \mathbb{E}[U | D=d, P]$. Under constant treatment effects we have\footnote{One can easily show that in this case, $k_1(p) = \kappa_1(p) + p \kappa_1'(p)$ and $k_0(p) = \kappa_0(p) - (1-p) \kappa_0'(p)$ are equal by construction, which is expected since they both represent the same $\mathbb{E}[U|V=p]$.} 
\begin{align}\label{eq:relationkappa}
	\kappa_0(P) = - \kappa_1(P)  \frac{P}{1-P}. 
\end{align}
As a consequence, Equation \eqref{eq:plm} can be written as  
\begin{align*}
	\mathbb{E}[Y | D=0, Z_0, P] &= \mu_0 + \delta_0 Z_0 + \kappa_0(P) = \mu_0 + \delta_0 Z_0 - \kappa_1(P) \frac{P}{1-P}, \\
	\text{ and } \quad \mathbb{E}[Y | D=1, Z_1, P] &= \mu_1 + \delta_1 Z_1 + \kappa_1(P). 
\end{align*}
Pooling these two, one can estimate a single model for the outcome variable $Y$, i.e., 
\begin{align*}
	&\mathbb{E}[ Y | D, Z_0, Z_1, P] = D \mathbb{E}[Y | D=1, Z_1, P] + (1-D) \mathbb{E}[Y | D=0, Z_0, P] \\
	&= \mu_0 + D(\mu_1-\mu_0) + DZ_1\delta_1 - (1-D)Z_0\delta_0 + \Big( D - (1-D) P/(1-P)\Big) \kappa_1(P). 
\end{align*}
So, by regressing $Y$ on a constant, $D$, $DZ_1$, $(1-D)Z_0$ and a flexible function of $P$ multiplied by $(D - (1-D)P/(1-P))$, we estimate an homogenous treatment effect model. The \textit{homogeneity constraint} is that the same $\kappa_1(P)$ is estimated for all $D$ (treated or not). I specify this flexible function $\kappa_1(P)$ as a polynomial function, in the spirit of nonparametric sieve estimation. The advantage of the pooling is that the MTRs are estimated using all the information and not just the separate subsamples. Thus, I take full advantage of the knowledge that $k_0(v)$ must be equal to $k_1(v)$ under homogeneity.  \\

\begin{table}[!p]
\captionsetup{font=normalsize, justification=centering}
\caption{Monte Carlo simulations with homogenous treatment effects}\label{tab:homo1}
\captionsetup{font=small, justification=raggedright, singlelinecheck=false}
\caption*{%
\centering
Specification: with $N = 10,000$ observations.  \\
\begin{tabular}{lc} 
\textit{Parameters} & \\
Outcome: & $\mu_0 = 3.2$, $\mu_1 = 3.6$, \\
& Specification 1: $\delta_0 = 0.8$, $\delta_1 = 0.5$; Specification 2: $\delta_0 = \delta_1 =0$. \\
Selection: & $\alpha = 0$, $\alpha_0 = -0.7$, $\alpha_1 = 0.7$. \\
Semi-IVs: & $\mu_{Z_0} = 0$, $\mu_{Z_1} = 0$, $\sigma^2_{Z_0} = 1$, $\sigma^2_{Z_1} = 1$, $\sigma_{Z_0Z_1} = 0.5$. \\
Errors: & $\sigma^2_{U} = 1$, $\sigma^2_{\tilde{V}} = 1.5$, $\sigma_{U\tilde{V}} = 0.6$. \\
\end{tabular}%
}
 \centering
     {\renewcommand{\arraystretch}{0.9}
  \resizebox{0.9\textwidth}{!}{ 
  {\small
    \begin{tabular}{@{\extracolsep{5pt}}lcccccc} 
\\[-1.8ex]\hline 
\hline 
\\[-1.8ex] & True & OLS & IV & \multicolumn{3}{c}{semi-IV} \\ 
\cline{3-3} \cline{4-4} \cline{5-7} 
\\[-1.8ex]  & & & & (2SLS) & (SP) & (SP-H) \\
\hline \\[-1.8ex] 
\textbf{Specification 1:} & & & & \\
$\quad $ Intercept: $\mu_0$ & 3.2 & 3.614 & 3.515 & 3.202 & 3.196 & 3.199 \\
& & (0.013) & (0.041) & (0.037) & (0.091) & (0.027) \\
\\
$\quad $ $D$: $\mu_1 - \mu_0 $ & 0.4 & -0.428 & 0.068  & 0.396 & 0.405 & 0.401\\
& & (0.019) & (0.074) & (0.071) & (0.123) & (0.049) \\
\\
$\quad$ $Z_0(1-D)$: $\delta_0$ & 0.8 & 0.706 &  & 0.800  & 0.800  & 0.801 \\
& & (0.013) & & (0.028)  & (0.016)  & (0.015) \\
\\
$\quad$ $Z_1D$: $\delta_1$ &  0.5 & 0.595 & & 0.503  & 0.501 & 0.500 \\ 
& & (0.013) & & (0.029) & (0.016) & (0.015) \\
\\
\textbf{Specification 2:} $\delta_0 = \delta_1 = 0$ \\
\textit{$\rightarrow$ semi-IVs are valid IVs.} \\
\\
$\quad $ Intercept: $\mu_0$ & 3.2 & 3.613 & 3.200 & 3.202 & 3.205 & 3.199 \\
& & (0.013) & (0.023) & (0.036) & (0.090) & (0.026) \\
\\
$\quad $ $D$: $\mu_1 - \mu_0 $ & 0.4 & -0.427 & 0.400  & 0.397 & 0.393 & 0.401\\
& & (0.019) & (0.042) & (0.067) & (0.129) & (0.048) \\
\\
$\quad$ $Z_0(1-D)$: $\delta_0$ & 0 & -0.094 &  & -0.001  & 0.000  & 0.000 \\
& & (0.012) & & (0.025)  & (0.016)  & (0.014) \\
\\
$\quad$ $Z_1D$: $\delta_1$ &  0 & 0.095 & & -0.001  & 0.000 & 0.000 \\ 
& & (0.014) & & (0.026) & (0.017) & (0.016) \\
\hline  
\multicolumn{7}{l}{ $N \quad = \quad 10,000$ observations for every simulations} \\
\hline 
\hline \\[-1.8ex] 
\end{tabular}
}
}
}
\vspace{1em}
\captionsetup{font=footnotesize, justification=justified}
\caption*{
\noindent \textit{Notes:} The OLS column corresponds to the estimation of \eqref{eq:tsls_app} by OLS, i.e., wrongly assuming that $D$ is exogenous such that $\mathbb{E}[U | D, Z_0, Z_1] = 0$. The IV column corresponds to the 2SLS of $Y$ on $D$, using $Z_0$ and $Z_1$ as standard IVs. Then, I implement three different semi-IV estimation procedures. The semi-IV 2SLS corresponds to the simple 2SLS method described in Section \ref{sec:estimation_2sls}, with a linear probability model of $D$ on $Z_0$ and $Z_1$ in the first stage, and using the estimated $\widehat{P}, \widehat{P}W_1, (1-\widehat{P})W_0$ as (optimal) instruments in a 2SLS afterwards. The semi-IV "SP" column corresponds to the semi-parametric estimation of Section \ref{sec:estimation_semi} and of Appendix \ref{app:mc_heterogenous}, i.e., I estimate the model with heterogenous treatment effects (bandwidth arbitrarily set to $0.10$, and trimming the $1\%$ top and bottom of the support of $P$ for treated and untreated), and then compute the ATE as the average of the $\widehat{MTE}(v, 0, 0)$ and the "intercept" as the average of $\widehat{m}_0(v, 0)$. Finally, the last semi-IV "SP-H" corresponds to the "semi-parametric approach with homogeneity constraint" estimation approach, which imposes that the treatment effects must be homogenous by imposing the relation \eqref{eq:relationkappa} between $\kappa_0$ and $\kappa_1$. 
}

\end{table}
\setstretch{1.25} 
\setlength{\abovedisplayskip}{6pt} 
\setlength{\belowdisplayskip}{6pt}

\begin{figure}[!t]
\centering
	\caption{Semi-parametric estimation with homogeneity constraint (SP-H) of the MTRs}\label{fig:MC_homo1}
    \includegraphics[width=0.55\textwidth]{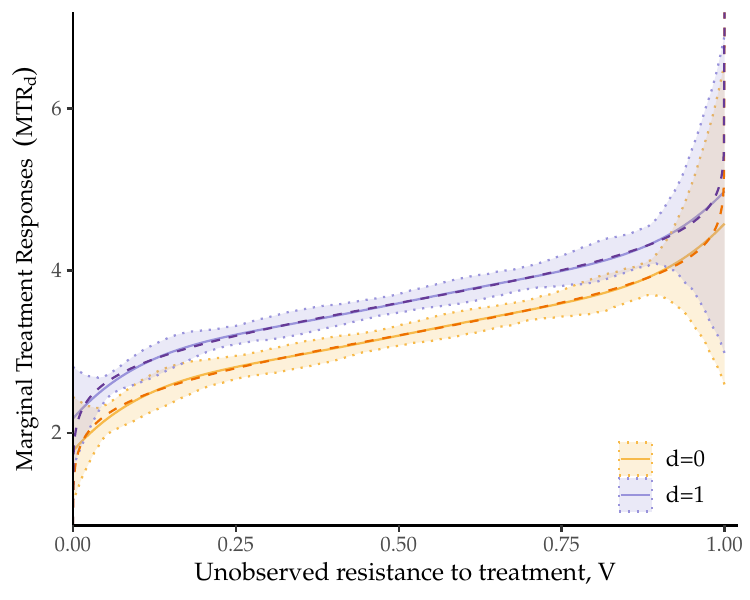}
    \vspace{0.5em} 
\captionsetup{font=footnotesize, justification=justified}
\caption*{
\noindent \textit{Notes:} Results of the estimation of the MTRs, at $(Z_0, Z_1) = (0,0)$, from $500$ Monte Carlo simulations of the first specification (Table \ref{tab:homo1}), using the semi-parametric estimation with homogeneity constraint on the treatment effect. By construction $m_1(v, 0) = m_0(v, 0) + \mu_1 - \mu_0$. We specify the control function $\kappa_1(P)$ as a polynomial of order $5$ function of $P$. Results are almost insensitive to polynomial orders between $2$ and $5$.
}
\end{figure}

\noindent \textbf{Results.} Table \ref{tab:homo1} reports the coefficients from the model estimated using the three different semi-IV methods discussed above, and compares them to the estimates of a naive OLS and of an IV regression using the semi-IVs as valid IVs. 
I estimate two specifications: the first one is a general semi-IV model with homogenous treatment effect, and the second one is a special case where the semi-IVs are in fact excluded ($\delta_0 = \delta_1 = 0$) and could thus be used as valid IVs. This specification provides an idea of the loss of efficiency one faces by being too cautious and considering a valid IV is only partially excluded (by considering it as a semi-IV). \\
\indent Overall, all the estimation methods yield unbiased estimates of the parameters. However, using the general approach (SP), which is robust to heterogenous treatment effects, comes at the cost of reduced efficiency. The simple semi-IV 2SLS is more efficient than the general approach. The semi-parametric approach with homogeneity constraint (SP-H) outperforms the two other alternatives. The efficiency gains stem from the fact that we pool all the observations to estimate the control function $\kappa_1(P)$. Perhaps surprisingly, in the case where the semi-IVs are, in fact, valid IVs, using the SP-H approach causes almost no efficiency loss with respect to the standard IV-2SLS using $Z_0$ and $Z_1$ as IVs. The minor difference in standard errors is due to the fact that the semi-IV approach has to estimate two additional parameters for each semi-IV. Given how large the bias of the IV regression can be if the semi-IVs are invalid IVs (cf Specification 1), being cautious and considering them as semi-IVs in case of doubt may be recommended. \\
\indent In the event where the semi-IVs are valid IVs, note that the semi-IV approaches very precisely estimate that the effects of the semi-IVs on the outcome is null. This suggests that semi-IVs can be used to test the validity of some IVs (although it requires having another complementary valid semi-IV). \\
\indent Finally, in Figure \ref{fig:MC_homo1}, I report the estimation of the MTRs, at $(Z_0, Z_1) = (0, 0)$, using the semi-parametric estimation approach with homogeneity constraint (SP-H). These estimates are unbiased and very precise. Even with homogenous treatment effects, one should report the estimates of the MTRs. Indeed, it provides crucial information about the correlation between $V$ and $U$, i.e., about the relation between the unobservables in the outcome and the selection (positively related in the specification of Figure \ref{fig:MC_homo1}).

\subsection{Replicability} 
All the simulations of this section have been implemented using the companion \texttt{semiIVreg} package \citep{semiivreg} and its \texttt{simul$\_$data()} function to simulate generalized Roy models. See the package \href{https://cbruneelzupanc.github.io/semiIVreg}{webpage} for more details on how to proceed and replicate these simulations. 


\section{(Unordered) Discrete Treatment $D$}\label{app:discrete}

\noindent The model with binary treatment can be extended to problems with (unordered) discrete  treatment, $D \in \mathcal{D}=\{0, ..., J-1\}$ with $J > 2$ mutually exclusive alternatives. As in the binary case, one can nonparametrically identify similar treatment effects (pooled LATE, margin-specific MTE) with semi-IVs as with IVs. The underlying logic remains the same as in the binary case: by shifting the excluded semi-IVs, one can vary the set of compliers while keeping the included semi-IV fixed.

\subsection{The model with discrete $D$} 

\noindent \textbf{Potential outcomes.} As in the binary case, there is one potential outcome per alternative, $Y_d$, but one only observes the outcome, 
\begin{align*}
	Y = Y_D = \sum_{d=0}^{J-1} Y_d \mathds{1}\{D=d\}. 
\end{align*}

\noindent \textbf{Alternative-specific semi-IVs.} 
We have $J$ different (one per distinct value of $\mathcal{D}$) continuously distributed \textit{alternative $d$-specific semi-IVs}, $Z_d$. Denote $Z=(Z_0, Z_1, ..., Z_{J-1})$ the observed set of all semi-IVs. For all $j \in \mathcal{D}$, define $D_j$ a dummy $=1$ if $D=j$, $0$ otherwise. Define potential treatments as $D_j(Z)$. Similarly, denote $P_d(Z,X) = \mathbb{E}[D_j|Z,X]=\textrm{Pr}(D=j|Z,X)$, and $\mathbf{P}(Z,X)=(P_0(Z,X), P_1(Z,X), ..., P_{J-1}(Z,X))$ the generalized propensity score, i.e., the set of all $d$-specific selection probabilities, which sum to $1$. 
The semi-IVs are valid if they satisfy the following generalization of Assumption \ref{ass:semiiv} to $J$ alternatives:\footnote{As in Section \ref{sec:framework}, I do not write the exclusion and exogeneity separately. The exclusion holds because $Y_d(z,x) = Y_d(z_d,x)$ and does not depend on any $z_{-d}$, for all $z \in \mathcal{Z}$.} 

\begin{assumption}[Valid semi-IVs, discrete $D$]\label{ass:semiiv_discrete} 
The semi-IVs satisfy the following conditions:
\begin{enumerate}[label=A\theassumption.\arabic*]
	\item (Exogeneity and Exclusion)\label{ass:exogeneity_discrete} For all $z \in \mathcal{Z}$, (i) $D(z) \indep Z | X$, (ii) $(Y_d, D(z)) \indep Z_{-d} | (Z_d, X)$ where $Z_{-d}$ is the set of all the semi-IVs, except $Z_d$. 
	\item (Relevance)\label{ass:relevance_discrete} The generalized propensity score $\mathbf{P}(Z, X)$ is a nondegenerate random variable conditional on $(Z_{-d}, X)$ for all $d \in \mathcal{D}$. \end{enumerate}
\end{assumption}
\noindent In other words, each semi-IV $Z_d$ may, at most, not be excluded from one of the potential outcome, $Y_d$. Hence the name \textit{$d$-specific semi-IVs}. In addition, they must also be relevant, i.e., have an effect on the selection probabilities. Now, we add some further restriction on the selection process. \\

\noindent \textbf{Discrete choice index model.} 
Following the identification of unordered discrete treatment with IVs \citep{heckmanurzuavytlacil2006, heckmanurzuavytlacil2008, heckmanvytlacil2007b}, I extend the monotonicity to the discrete case by imposing a discrete choice index model.\footnote{This discrete index formulation is not necessary to achieve identification with semi-IVs. For example, see how I identify margin-specific treatment effects as in \cite{mountjoy2022community} using semi-IVs under only partial monotonicity with a comparable compliers assumption (Section \ref{subsec:mountjoy}). However, the discrete index model \eqref{eq:selection_late_multi} has nice properties and simplifies the exposition. } 
The value of choosing $D=d$ for the individual is 
\begin{align}\label{eq:selection_late_multi} 
	R_d &= u_d(Z_d, X) + \Epsilon_d, 
\end{align} 
where $u_d$ is the latent utility of choosing $D=d$, common to every individuals, and $\Epsilon_d$ are the $d$-specific continuously distributed unobserved shocks. The $\Epsilon_d$ can be correlated across alternatives. 
 Note that the semi-IVs are also $d$-specific in the selection process here: $Z_d$ should only affect $u_d$ and not any other $u_k$ (conditional on $Z_k$). Note that this $d$-specificity is also required to identify discrete treatment with $d$-specific IVs. The advantage is that, with semi-IVs, this $d$-specificity is a natural extension of the outcome exclusion to the selection and should be seamlessly satisfied by most semi-IVs. \\
\indent The individual chooses the alternative that yields the highest utility, i.e., 
\begin{align*}
	D = \underset{k\in \mathcal{D}}{\text{argmax }} R_k.
\end{align*}
\indent Since only $J-1$ comparison of utilities matter for the discrete decision, $D$, we typically normalize the shocks $\Epsilon$ with respect to $\Epsilon_0$, i.e., we define $V_d = \Epsilon_d - \Epsilon_0$ for $d  \in \{1, ..., J-1 \}$. Denote $\mathbf{V}=(V_1, ..., V_{J-1}) \in \pmb{\mathcal{V}}$. 
\noindent Under this model, one can re-express Assumption \ref{ass:semiiv_discrete} as:

{\theoremstyle{plain}
\newtheorem*{specialassumption2}{Assumption \ref{ass:semiiv_discrete}'}  
\begin{specialassumption2} The semi-IVs $Z$ satisfy the following properties
\begin{enumerate}[label=A\theassumption'.\arabic*]
	\item (Exogeneity and Exclusion) (i) $\mathbf{V} \indep Z | X$. (ii) $(Y_d, \mathbf{V}) \indep Z_{-d} | (Z_d, X)$ where $Z_{-d}$ is the set of all the semi-IVs, except $Z_d$. 	
	\item (Relevance) For all $d \in \mathcal{D}$, $\partial u_d(z_d, x)/\partial Z_d \neq 0$ for all $z_d, x$. 
\end{enumerate}
\end{specialassumption2} 
For expositional simplicity, I further assume that $\partial u_d(z_d, x)/\partial Z_d > 0$ for all $d, z_d, x$, i.e., that the semi-IVs have a positive effect on their corresponding alternative utilities. In other words, think of $Z_d$ as an incentive, a "push variable" in favour of $D$.\footnote{This is an expositional normalization. If one has a semi-IV that deters participation into $D$, just focus on $-Z_d$ instead of $Z_d$. If a semi-IV has a relevant but ambiguous effect (sometimes positive, sometimes negative), the analysis still holds, but it complicates the exposition. } \\

\subsection{Example and comparison with discrete $D$ using IVs} 

Most examples of semi-IVs discussed in Section \ref{subsec:example} can naturally be extended to discrete choice models. For example, if one were to study occupation/sector choice with multiple sectors, the local market size (e.g., GDP) of each sector could serve as valid $d$-specific semi-IVs. It is likely that the characteristics of a specific sector only affect the utility of workers working in this sector, so the discrete choice structure is naturally satisfied. A similar reasoning applies more generally to many other discrete models, for example, discrete location choice models using local amenities as semi-IVs. \\
\indent Note that the counterpart identification of discrete treatment with IVs \citep{heckmanurzuavytlacil2006, heckmanurzuavytlacil2008, heckmanvytlacil2007b, kirkeboenleuvenmogstad2016, mountjoy2022community} generally uses several ($J-1$) "$d$-specific IVs" or IVs with a larger support (of size $J$).\footnote{\cite{klinewalters2016} being an exception using a single binary assignment to the head start program as an IV in a discrete treatment setup.}  
This limits the range of applications that identify the effect of discrete treatment with IVs because, even more so than in the binary case, it is difficult to find $J-1$ $d$-specific IVs, that affect the utility of only one alternative without affecting its corresponding potential outcome or the latent utility of the other alternatives.
Fortunately, finding $d$-specific semi-IVs, which can also affect the $d$-specific outcome, is considerably easier.
To identify the same treatment effects, researchers face a trade-off between finding $J-1$ $d$-specific IVs or $J$ $d$-specific semi-IVs. As in the binary case, the semi-IVs are usually the same variable adapted across alternatives (e.g., sector-specific GDP) so finding one more semi-IV is mostly costless, making the semi-IV approach more naturally suited to the discrete treatment case.\footnote{As in the binary case, it is also better if the semi-IVs are continuously distributed, while it is not always necessary for the IVs. However, the identification of marginal treatment effects with IVs also requires continuous IVs anyway \citep{mountjoy2022community}. Moreover, most examples of semi-IVs are, in fact, continuous. } 

\subsection{Identification}
For expositional simplicity, omit $X$ in the notation and proceed conditional on $X=x$. 
\noindent We have data on $(Y, D, Z)$ and on the generalized propensity score variable $\mathbf{P} = \mathbf{P}(Z)$, because the function $\mathbf{P}(z)$ is nonparametrically identified for all $z \in \mathcal{Z}$. 
I first discuss the properties of the discrete choice model, then define the relevant treatment effects of interest with semi-IVs -- the counterpart to those with IVs --  and finally address their identification.

\begin{figure}[h!]
	\centering
	\caption{Treatment choice as a partition}\label{fig:partitiondiscrete}
	\begin{minipage}{0.45\textwidth}
	\begin{tikzpicture}[x=80pt,y=80pt, line width=0.25mm] 
    	\def\uzero{0.4}
    	\def\uone{0.2}
    	\def\utwo{0.4}

		\draw[<->](-1,1)--(-1,-1)--(1,-1); 
		\draw (-1, 1) node[above left=0pt, black]{$V_2$};
		\draw (1, -1) node[below right=0pt, black]{$V_1$};
		
		\node[below=0pt] at (\uzero-\uone, -1) {$v_1=u_0(z_0) - u_1(z_1)$}; 
		\node[above=1pt, rotate=90 ] at (-1, \uzero-\utwo) {$v_2=u_0(z_0) - u_2(z_2)$}; 
		\draw[thick] (\uzero-\uone, -1.03) -- (\uzero-\uone, -0.97); 
		\draw[thick] (-1.03, \uzero - \utwo) -- (-0.97, \uzero - \utwo); 

		\begin{scope}
		\clip (-1, -1) rectangle (1, 1);	

		\draw[thick] (-1, \uzero-\utwo) -- (\uzero-\uone, \uzero-\utwo) -- (\uzero-\uone, -1);
		\draw[domain=(\uzero - \uone):1,smooth,variable=\x,black,thick] plot ({\x},{\uone - \utwo +\x});
		
		\fill[mtr0color, opacity=0.3] (-1,-1) rectangle (\uzero-\uone, \uzero-\utwo); 
		\fill[mtr1color, opacity=0.3] (\uzero-\uone, -1) -- (\uzero-\uone, \uzero-\utwo) -- (1, \uone - \utwo + 1) -- (1, -1) -- cycle;
		\fill[mtecolor, opacity=0.3] (-1, \uzero-\utwo) -- (\uzero-\uone, \uzero-\utwo) -- (1, \uone - \utwo + 1) -- (1, 1) -- (-1, 1) -- cycle;
		\end{scope}
		
		\node[] at (-1+\uzero*0.5-\uone*0.5+0.5, -1 + \uzero*0.5-\utwo*0.5 + 0.5) {$D(z)=0$}; 
		\node[] at (-1+\uzero*0.5-\uone*0.5+0.5, \uzero-\utwo + 0.5-\uzero*0.5+\utwo*0.5) {$D(z)=2$}; 
		\node[] at (\uzero - \uone +0.5 - \uzero*0.5+\uone*0.5, -1 + \uzero*0.5-\utwo*0.5 + 0.5) {$D(z)=1$}; 
		
    
    \end{tikzpicture}
    \begin{center}
	$\quad \quad \quad  $(a) Partition of $D$ at $Z=z$
	\end{center}
	\end{minipage} \hspace{0.5cm}
	\begin{minipage}{0.45\textwidth}
		\begin{tikzpicture}[x=80pt,y=80pt, line width=0.25mm] 
    	\def\uzero{0.4}
    	\def\uone{0.2}
    	\def\utwo{0.4}
    	
		\draw[<->](-1,1)--(-1,-1)--(1,-1); 
		\draw (-1, 1) node[above left=0pt, black]{$V_2$};
		\draw (1, -1) node[below right=0pt, black]{$V_1$};

		\node[below=0pt] at (\uzero-\uone, -1) {$v_1(z)$}; 
		\node[left=0pt] at (-1, \uzero-\utwo) {$v_2(z)$}; 
		\draw[thick] (\uzero-\uone, -1.03) -- (\uzero-\uone, -0.97); 
		\draw[thick] (-1.03, \uzero - \utwo) -- (-0.97, \uzero - \utwo); 

		\begin{scope}
		\clip (-1, -1) rectangle (1, 1);	
		\draw[thick] (-1, \uzero-\utwo) -- (\uzero-\uone, \uzero-\utwo) -- (\uzero-\uone, -1);
		\draw[domain=(\uzero - \uone):1,smooth,variable=\x,black,thick] plot ({\x},{\uone - \utwo +\x});
		
		
		\end{scope}

		\def\uzerop{0.4}
    	\def\uonep{0.7}
    	\def\utwop{-0.2}

		\node[below=0pt] at (\uzerop-\uonep, -1) {$v_1'(z')$}; 
		\node[left=0pt] at (-1, \uzerop-\utwop) {$v_2'(z')$}; 
		\draw[thick] (\uzerop-\uonep, -1.03) -- (\uzerop-\uonep, -0.97); 
		\draw[thick] (-1.03, \uzerop - \utwop) -- (-0.97, \uzerop - \utwop); 

		\begin{scope}
		\clip (-1, -1) rectangle (1, 1);
		\draw[thick, dashed] (-1, \uzerop-\utwop) -- (\uzerop-\uonep, \uzerop-\utwop) -- (\uzerop-\uonep, -1);
		\draw[domain=(\uzerop - \uonep):1,smooth,variable=\x,black,thick, dashed] plot ({\x},{\uonep - \utwop +\x});
		
		\fill[mtr0color, opacity=0.3] (-1,-1) rectangle (\uzerop-\uonep, \uzerop-\utwop); 
		\fill[mtr1color, opacity=0.3] (\uzerop-\uonep, -1) -- (\uzerop-\uonep, \uzerop-\utwop) -- (1, \uonep - \utwop + 1) -- (1, -1) -- cycle;
		\fill[mtecolor, opacity=0.3] (-1, \uzerop-\utwop) -- (\uzerop-\uonep, \uzerop-\utwop) -- (1, \uonep - \utwop + 1) -- (1, 1) -- (-1, 1) -- cycle;
		\end{scope}

		\node (p1) at (\uzerop-\uonep, \uzerop-\utwop) {}; 
		\node (p) at (\uzero-\uone, \uzero-\utwo) {}; 
		\draw[->, thick, line width=0.5mm] (p) -- (p1); 
		

		\node[rotate=45, above] at (0.5, 0.5) {{\footnotesize \begin{tabular}{c} $1 \leftarrow 2$ \\ compliers \end{tabular}}};
		\node[rotate=0, above] at (-0.05, -0.7) {{\footnotesize\begin{tabular}{c} $1 \leftarrow 0$ \\ compliers \end{tabular}}};
		\node[rotate=0, above] at (-0.65, 0.1) {{\footnotesize\begin{tabular}{c} $0 \leftarrow 2$ \\ compliers \end{tabular}}};
	\end{tikzpicture}
	
	\begin{center}
	$\quad \quad \   $(b) Change in $Z$ from $z$ to $z'$
	\end{center}
	\end{minipage}
\end{figure}
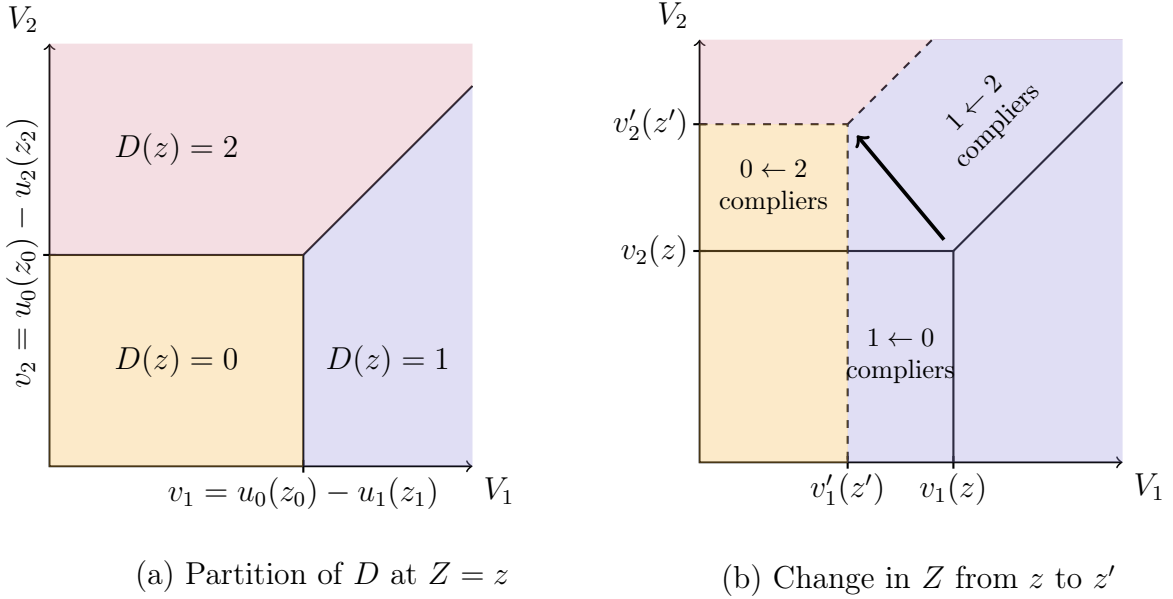

\subsubsection{Identifying properties of the model}\label{subsec:discreteproperty}

%
%
\indent The multinomial choice model for the selection leads to a clear partitioning of the $J-1$ dimensional space of $\mathbf{V}$ into $D(z)=0, 1, ..., J-1$, as illustrated in Figure \ref{fig:partitiondiscrete} for $J=3$.\footnote{The partitioning comes from the individuals comparing the different alternatives of the treatment and picking the one yielding the highest utility. Assuming ties break in favour of the lowest $d$, with the discrete index model, we have:	
\vspace{-1\baselineskip}
\begin{center}
\resizebox{\linewidth}{!}{
\begin{tabular}{c|c|c}
	$D(z) = 0$ & $D(z) = 1$ & $D(z)=2$  \\
	$\iff$ & $\iff$ & $\iff$ \\
	$\left\{\begin{array}{l l} 
	u_0(z_0) + \Epsilon_0 &\geq u_1(z_1) + \Epsilon_1, \\ 
	u_0(z_0) + \Epsilon_0 &\geq u_2(z_2) + \Epsilon_2.\end{array}\right.$ & 
	$\left\{\begin{array}{l l} 
	u_1(z_1) + \Epsilon_1 &> u_0(z_0) + \Epsilon_0, \\ 
	u_1(z_1) + \Epsilon_1 &\geq u_2(z_2) + \Epsilon_2.\end{array}\right.$ & 
	$\left\{\begin{array}{l l} 
	u_1(z_1) + \Epsilon_1 &> u_0(z_0) + \Epsilon_0, \\ 
	u_1(z_1) + \Epsilon_1 &\geq u_2(z_2) + \Epsilon_2.\end{array}\right.$ \\
	$\iff$ & $\iff$ \\
	$\left\{\begin{array}{l l} 
	V_1 &\leq u_0(z_0) - u_1(z_1), \\ 
	V_2 &\leq u_0(z_0) - u_2(z_2). \end{array}\right.$ & 
	$\left\{\begin{array}{l l} 
	V_1 &> u_0(z_0) - u_1(z_1), \\ 
	V_2 &\leq u_1(z_1) - u_2(z_2) + V_1.\end{array}\right.$ & 
	$\left\{\begin{array}{l l} 
	V_2 &> u_0(z_0) - u_2(z_2), \\ 
	V_2 &> u_1(z_1) - u_2(z_2) + V_1. \end{array}\right.$
	\end{tabular} 
	}
	\end{center}
	This yields the partitioning observed in Figure \ref{fig:partitiondiscrete}.
} 
From this partition, we have that $P_d(z) = \textrm{Pr}(D(z) = d)$ which is the total mass of all $V$ in the area with $D(z) = d$ (normalizing the total mass of $\pmb{\mathcal{V}}$ to 1). 
This partitioning yields several key properties, similar to the ones we had in the binary treatment case, that can be leveraged for identification (even without identifying $u_d(z_d)$).  I discuss these properties in the context of $J=3$, but they hold more generally for any $J \geq 3$. \\

\noindent \textit{1. Generalized propensity score sufficient statistics.} For any $z, z' \in \mathcal{Z}$ with $z'\neq z$, if $\mathbf{P}(z) = \mathbf{P}(z')$ then $D(z) = D(z')$ for all individuals.\footnote{`\textit{Proof}': as visible in Figure \ref{fig:partitiondiscrete}, the partitioning of $\pmb{\mathcal{V}}$ into $D$ is fully characterized by the two values, $v_1 = u_0(z_0) - u_1(z_1)$ and $v_2 = u_0(z_0) - u_2(z_2)$. These two values give the coordinates of the "indifferent individual", i.e., the individual indifferent between every treatment alternative. By construction, there is only one point of indifference between the three alternatives in this model. So, the coordinates of the indifferent individual pin down the partitioning of $\mathcal{V}$, and, thus, the generalized propensity score $\mathbf{P}$. In fact, there is a one-to-one mapping between the indifferent individual coordinates and $\mathbf{P}$. Indeed, since all $V \in \pmb{\mathcal{V}}$ have a non-negative mass, going from any $(v_1, v_2)$ to $(v_1', v_2')$ always leads to a different $\mathbf{P}$. This is straightforward graphically when $J=3$: for any $V$ in the interior of $\pmb{\mathcal{V}}$, a shift to the top right necessarily increases $P_0$, a shift to the bottom right increases $P_2$, a shift to the top left increases $P_1$, and a shift to the bottom left reduces $P_0$. 
\noindent Now, the effects of $Z$ on the partition is fully characterized by their effects on the latent $d$-specific utilities, and thus, on the coordinates of the indifferent individual. Since there is a one-to-one mapping between this and $\mathbf{P}$, the effect of any $Z$ on the partitioning of $D$ is fully characterized by the observable $\mathbf{P}$ it generates. 
} \\
\indent In other words, as in the binary case, the generalized propensity score serves a sufficient statistic to summarize the effect of the semi-IVs on the selection, i.e., on the partition of $\pmb{\mathcal{V}}$ into $D$. One could denote $D(\mathbf{P})$ instead of $D(Z)$, representing the choices of $D$ given that the propensity score is $\mathbf{P}$. Then the property means that if $\mathbf{P}(z)=\mathbf{p}$, then $D(z) = D(\mathbf{p})$ for all individuals. 
Importantly, this discrete version of the index sufficiency property requires \textit{all} the probabilities to be equal. Only observing that for some $j$, $P_j(z) = P_j(z')$, is not sufficient and does not bring useful information about the partitioning.
This property is useful in order to isolate the direct effects of the semi-IVs net of the selection, by moving $Z$ while holding $\mathbf{P}$ fixed. As in the binary case
\begin{align*}
	&\Delta^{D=d}_{Z_d}(z_d, z_d', \mathbf{p}) = \mathbb{E}[Y_d | D=d, Z_d=z_d', \mathbf{P}=\mathbf{p}] - \mathbb{E}[Y_d | D=d, Z_d=z_d, \mathbf{P}=\mathbf{p}], 
\end{align*}
is directly identified from the data provided one can move $Z_d$ while holding $\mathbf{P}$ fixed, which is achievable by moving all the other semi-IVs as well. \\


\noindent \textit{2. Reachability without shifting $Z_d$.} For any $d \in \mathcal{D}$, provided that all the semi-IVs are sufficiently relevant, for any $z=(z_0, ..., z_{J-1})$ in the interior of $\mathcal{Z}$ with $\mathbf{P}(z) = \mathbf{p}$, one can reach any $\mathbf{p'}$ in the neighbourhood of $\mathbf{p}$ without shifting $Z_d=z_d$, by shifting only the $J-1$ other semi-IVs.\footnote{`\textit{Proof}': this is visible in Figure \ref{fig:partitiondiscrete}(b). A shift in $Z$ from $z$ to $z'$ leads to a shift from $v_1(z) = u_0(z_0) - u_1(z_1)$ to $v_1(z') = u_0(z_0')-u_1(z_1')$ and from $v_2(z)=u_0(z_0) - u_2(z_2)$ to $v_2(z') = u_0(z_0') - u_2(z_2')$. Any $\tilde{z}'$ with $v_1(\tilde{z}')=v_1(z')$ and $v_2(\tilde{z}')=v_2(z')$ would yield the same $\mathbf{P}=\mathbf{p}'$. Imagine one wants to fix $Z_0=z_0$, i.e., $\tilde{z}'=(z_0, \tilde{z}_1', \tilde{z}_2')$. At fixed $Z_0=z_0$, $u_0(z_0)$ is fixed. But one can still adjust $v_1$ and $v_2$ by moving only $u_1$ and $u_2$ through moves of $Z_1$ and $Z_2$. Typically, picking $\tilde{z}_1'$ such that $u_1(\tilde{z}_1') = u_0(z_0) - u_0(z_0') + u_1(z_1')$ yields $v_1(\tilde{z}') = v_1(z')$, and picking $\tilde{z}_2'$ such that $u_2(\tilde{z}_2') = u_0(z_0) - u_0(z_0') + u_1(z_2')$ yields $v_2(\tilde{z}') = v_2(z')$. Similarly, if one wants to fix $Z_1$ or $Z_2$, one can shift only $(Z_0, Z_2)$ or $(Z_0, Z_1)$, respectively. This is feasible because the marginal individual is determined by $\mathbf{V}$, which is of dimension $J-1$, while $Z$ is of dimension $J$ and impacts $J$ different $u_d$ functions. Fixing a specific $Z_d=z_d$ is like normalizing all the $u_j$ with respect to a specific alternative $u_d(z_d)$ baseline. This is feasible with this discrete choice structure with $d$-specific semi-IVs. } 
\noindent In other words, as in the binary case, if the semi-IVs are relevant, one can shift $\mathbf{P}$ while holding a specific $Z_d$ fixed. This permits the use of the excluded semi-IVs, $Z_{-d}$, as IVs for a specific $Y_d$, without having to shift $Z_d$. 
In general, one needs to shift all the other semi-IVs to mimic the change in a single semi-IV on the propensity score, hence the need for $J$ semi-IVs in the discrete case. 


\subsubsection{Margin-specific compliers}


A fundamental challenge with multivalued treatment is that, as with IVs, a shift in the semi-IVs generates more complex selection patterns than in the binary case, resulting in multiple flows of \textit{margin-specific compliers}. Let us define the "\textit{$d'\leftarrow d$ compliers}" as 
\begin{align*}
	C_{d'\leftarrow d}(z, z') := Compliers_{d'\leftarrow d}(z, z') = \Big\{ \mathbf{V} \in \pmb{\mathcal{V}}: D(z) = d, D(z')=d' \Big\},
\end{align*}
i.e., the individuals who shift from $D=d$ when $Z=z$ to $D=d'$ when $Z=z'$. Using the sufficiency property of the propensity score, one can define these margin-specific compliers in terms of the generalized propensity score, $\mathbf{P}$, instead, i.e., 
\begin{align*}
	C_{d'\leftarrow d}(\mathbf{p}, \mathbf{p'}) = \Big\{ \mathbf{V} \in \pmb{\mathcal{V}}: D(\mathbf{p}) = d, D(\mathbf{p'})=d' \Big\}. 
\end{align*}
Notice that, for any $z, z'$ such that $\mathbf{P}(z) = \mathbf{p}$ and $\mathbf{P}(z') = \mathbf{p'}$, we have 
\begin{align*}
	C_{d'\leftarrow d}(z, z') = C_{d'\leftarrow d}(\mathbf{P}(z), \mathbf{P}(z')) = C_{d'\leftarrow d}(\mathbf{p}, \mathbf{p'}).\footnotemark 
\end{align*}
\footnotetext{Notice that, in the binary treatment case, one had $Compliers_{1\leftarrow 0}(p, p') = \{ V: p \leq V < p'\}$, i.e., one could characterize the set of compliers in terms of the latent resistance to treatment $V$. In the discrete treatment case, this is not possible because the mass associated with each $\mathbf{V}=(V_1, V_2)$ point in $\pmb{\mathcal{V}}$ is unknown (because the correlation among the $\Epsilon_d$ is unrestricted). Therefore, one cannot identify the $d$-specific utilities ($u_d$) associated with each alternative $d$. Thus, one does not know the exact mapping between $\mathbf{P}$, $\mathbf{V}$, and all the $u_d$. One only knows that there is a one-to-one mapping, which allows us to define the complier areas using the observable shift of the generalized propensity score they correspond to. 
This complicates the identification of margin-specific treatment effects. Indeed, for instance, without further assumption, one cannot identify the shift in $\mathbf{P}$ that would yield only the $1\leftarrow 0$ compliers of Figure \ref{fig:partitiondiscrete}(b) (induced into $D=1$ by the shift in $Z$ from $z$ to $z'$) to shift back from $D=0$ into $D=1$. Fortunately, following \cite{mountjoy2022community}, one can proceed by using marginal shifts in $d$-specific semi-IVs to identify these margin-specific MTEs. 
}
\indent These multiple flows of compliers are visible in Figure \ref{fig:partitiondiscrete}(b) where the shift from $z$ to $z'$ leads to three different flows of compliers: \textit{$1\leftarrow 2$ compliers}, who switch from $D(z)=2$ to $D(z')=1$, \textit{$1\leftarrow 0$ compliers}, who switch from $D(z)=0$ to $D(z')=1$, and \textit{$0\leftarrow 2$ compliers} who switch from $D(z)=2$ to $D(z)=0$.  
\noindent This makes it harder to identify, and even to define, the treatment parameters of interest. For example, for the $1\leftarrow 0$ compliers, the treatment effect is $Y_1 - Y_0$, but for the $1\leftarrow 2$ compliers, the relevant margin of treatment is $Y_1 - Y_2$. 
\noindent The analysis is even harder if one focuses on the effect of the shift on $Y_0$: because the shift from $z$ to $z'$ lead to some individuals shifting into $D(z')=0$ from $D(z)=2$ (the $0\leftarrow 2$ compliers) but also some individuals shifting out of $D(z)=0$ to $D(z')=1$ (the $1\leftarrow 0$ compliers). \\
\indent An interesting property of the discrete index model is that, as visible in Figure \ref{fig:singlesemiivshift}, if one increases only one semi-IV, $Z_d$, then there will only be $J-1$ flows of compliers from the other alternatives into $D=d$. Thus, I analyse only treatment effects induced by shifts in a single semi-IV at a time, keeping in mind that any general shift in $Z$ can be decomposed into multiple single semi-IV shifts. Next, I define the treatment effects of interest with semi-IVs. \\ 

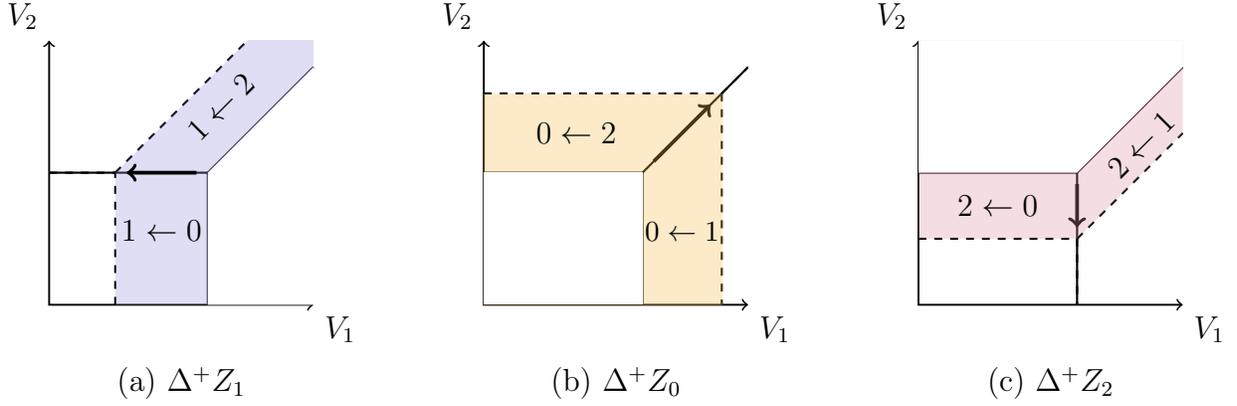
\begin{figure}[!h]

	\centering
	\caption{Complier flows with one semi-IV shift ($J=3$)}\label{fig:singlesemiivshift}
	\begin{minipage}{0.30\textwidth}
	\begin{tikzpicture}[x=50pt,y=50pt, line width=0.25mm] 
    	\def\uzero{0.4}
    	\def\uone{0.2}
    	\def\utwo{0.4}
    	
		\draw[<->](-1,1)--(-1,-1)--(1,-1); 
		\draw (-1, 1) node[above left=0pt, black]{$V_2$};
		\draw (1, -1) node[below right=0pt, black]{$V_1$};

		\begin{scope}
		\clip (-1, -1) rectangle (1, 1);	
		\draw[thick] (-1, \uzero-\utwo) -- (\uzero-\uone, \uzero-\utwo) -- (\uzero-\uone, -1);
		\draw[domain=(\uzero - \uone):1,smooth,variable=\x,black,thick] plot ({\x},{\uone - \utwo +\x});
		
		\fill[mtr1color, opacity=0.3] (\uzero-\uone, -1) -- (\uzero-\uone, \uzero-\utwo) -- (1, \uone - \utwo + 1) -- (1, -1) -- cycle;
		
		\end{scope}
		
		\def\uzerop{0.4}
    	\def\uonep{0.9}
    	\def\utwop{0.4}

		\begin{scope}
		\clip (-1, -1) rectangle (1, 1);
		\draw[thick, dashed] (-1, \uzerop-\utwop) -- (\uzerop-\uonep, \uzerop-\utwop) -- (\uzerop-\uonep, -1);
		\draw[domain=(\uzerop - \uonep):1,smooth,variable=\x,black,thick, dashed] plot ({\x},{\uonep - \utwop +\x});
		
		\fill[mtr1color, opacity=0.3] (\uzerop-\uonep, -1) -- (\uzerop-\uonep, \uzerop-\utwop) -- (1, \uonep - \utwop + 1) -- (1, -1) -- cycle;
		\end{scope}

		\node (p1) at (\uzerop-\uonep, \uzerop-\utwop) {}; 
		\node (p) at (\uzero-\uone, \uzero-\utwo) {}; 
		\draw[->, thick, line width=0.5mm] (p) -- (p1);

		\fill[white, opacity=1] (\uzero-\uone, -1) -- (\uzero-\uone, \uzero-\utwo) -- (1, \uone - \utwo + 1) -- (1, -1) -- cycle;

		\node[rotate=45] at (0.3, 0.5) {$1\leftarrow 2$ };
		\node[rotate=0, above] at (-0.15, -0.6) {$1\leftarrow 0$ };
	
	\end{tikzpicture}
	\centering (a) $\Delta^+ Z_1$
	\end{minipage} \hfill 
	\begin{minipage}{0.30\textwidth}
		\begin{tikzpicture}[x=50pt,y=50pt, line width=0.25mm] 
    	\def\uzero{0.4}
    	\def\uone{0.2}
    	\def\utwo{0.4}
    	
		\draw[<->](-1,1)--(-1,-1)--(1,-1); 
		\draw (-1, 1) node[above left=0pt, black]{$V_2$};
		\draw (1, -1) node[below right=0pt, black]{$V_1$};

		
		\begin{scope}
		\clip (-1, -1) rectangle (1, 1);	
		\draw[thick] (-1, \uzero-\utwo) -- (\uzero-\uone, \uzero-\utwo) -- (\uzero-\uone, -1);
		\draw[domain=(\uzero - \uone):1,smooth,variable=\x,black,thick] plot ({\x},{\uone - \utwo +\x});
		
		\fill[mtr0color, opacity=0.3] (-1,-1) rectangle (\uzero-\uone, \uzero-\utwo); 
		
		\end{scope}

		\def\uzerop{1.0}
    	\def\uonep{0.2}
    	\def\utwop{0.4}

		\begin{scope}
		\clip (-1, -1) rectangle (1, 1);
		\draw[thick, dashed] (-1, \uzerop-\utwop) -- (\uzerop-\uonep, \uzerop-\utwop) -- (\uzerop-\uonep, -1);
		\draw[domain=(\uzerop - \uonep):1,smooth,variable=\x,black,thick, dashed] plot ({\x},{\uonep - \utwop +\x});
		\end{scope}

		\node (p1) at (\uzerop-\uonep, \uzerop-\utwop) {}; 
		\node (p) at (\uzero-\uone, \uzero-\utwo) {}; 
		\draw[->, thick, line width=0.5mm] (p) -- (p1); 
	
		
		\fill[mtr0color, opacity=0.3] (-1,-1) rectangle (\uzerop-\uonep, \uzerop-\utwop);
		\fill[white, opacity=1] (-1,-1) rectangle (\uzero-\uone, \uzero-\utwo); 
		
		\node[rotate=0] at (-0.3, 0.3) {$0\leftarrow 2$ };
		\node[rotate=0, above] at (0.5, -0.6) {{\small $0\leftarrow 1$} };
		
		\end{tikzpicture}
		\centering (b) $\Delta^+ Z_0$
	\end{minipage} 
	\hfill\begin{minipage}{0.30\textwidth}
		\begin{tikzpicture}[x=50pt,y=50pt, line width=0.25mm] 
    	\def\uzero{0.4}
    	\def\uone{0.2}
    	\def\utwo{0.4}
    	
		\draw[<->](-1,1)--(-1,-1)--(1,-1); 
		\draw (-1, 1) node[above left=0pt, black]{$V_2$};
		\draw (1, -1) node[below right=0pt, black]{$V_1$};

		
		\begin{scope}
		\clip (-1, -1) rectangle (1, 1);	
		\draw[thick] (-1, \uzero-\utwo) -- (\uzero-\uone, \uzero-\utwo) -- (\uzero-\uone, -1);
		\draw[domain=(\uzero - \uone):1,smooth,variable=\x,black,thick] plot ({\x},{\uone - \utwo +\x});
		

		\end{scope}

		\def\uzerop{0.4}
    	\def\uonep{0.2}
    	\def\utwop{0.9}

		\begin{scope}
		\clip (-1, -1) rectangle (1, 1);
		\draw[thick, dashed] (-1, \uzerop-\utwop) -- (\uzerop-\uonep, \uzerop-\utwop) -- (\uzerop-\uonep, -1);
		\draw[domain=(\uzerop - \uonep):1,smooth,variable=\x,black,thick, dashed] plot ({\x},{\uonep - \utwop +\x});
		\end{scope}

		\node (p1) at (\uzerop-\uonep, \uzerop-\utwop) {}; 
		\node (p) at (\uzero-\uone, \uzero-\utwo) {}; 
		\draw[->, thick, line width=0.5mm] (p) -- (p1); 
	
		\fill[mtecolor, opacity=0.3] (-1, \uzerop-\utwop) -- (\uzerop-\uonep, \uzerop-\utwop) -- (1, \uonep - \utwop + 1) -- (1, 1) -- (-1, 1) -- cycle;
		\fill[white, opacity=1] (-1, \uzero-\utwo) -- (\uzero-\uone, \uzero-\utwo) -- (1, \uone - \utwo + 1) -- (1, 1) -- (-1, 1) -- cycle;
		
		\node[rotate=0] at (-0.4, -0.25) {$2\leftarrow 0$ };
		\node[rotate=45, above] at (0.8, 0.1) {$2\leftarrow 1$ };
		
		\end{tikzpicture}
		\centering (c) $\Delta^+ Z_2$
	\end{minipage} 
\centering
\end{figure}

\subsubsection{Margin-specific LATE}

\noindent Start with a given set of semi-IVs $Z = z \in \mathcal{Z}$. From here onwards, focus on a shift in only one semi-IV, $Z_d$, from $z_d$ to $z_d'$, leading from $z$ to $z'$ with associated propensity scores $\mathbf{P}(z)=\mathbf{p} \neq \mathbf{P}(z') = \mathbf{p'}$ with $p_d' > p_d$ and $p_{k}' \leq p_k$ for all $k \neq d$. For example, imagine that all else equal, the GDP of the manufacturing sector ($D=d$) increases, making it more attractive and diverting workers from the other sectors into manufacturing. What is the treatment effect of this change for these compliers/diverted workers? It depends on the choice they would have made without the change, i.e., $D(z)$. 
\noindent For complying workers diverted from a sector $k \neq d$, the \textit{margin-specific local average treatment effects} are given by: 
\begin{align}\label{eq:discretelatemargin}
	\text{LATE}_{d\leftarrow k}(z, z') = \mathbb{E}[Y_d - Y_k | \mathbf{V} \in C_{d \leftarrow k}(z, z'), Z_d=z_d', Z_k=z_k'], 
\end{align}
with  $z_k'=z_k$ since only $z_d$ was shifted. 
More generally, 
\begin{align}\label{eq:discretelatemargin2}
\text{LATE}_{d\leftarrow k}(z, z') &:=  \text{LATE}_{d\leftarrow k}(\mathbf{p}, \mathbf{p'}, z_d', z_k') \nonumber \\
	&= \mathbb{E}[Y_d - Y_k | \mathbf{V} \in C_{d \leftarrow k}(\mathbf{p}, \mathbf{p'}), Z_d=z_d', Z_k=z_k'].
\end{align}
One can split these LATE into two LATRs, for the outcomes $Y_j$ with $j \in \{d, k\}$, as 
\begin{align}\label{eq:discreteLATR}
\text{LATR}^j_{d\leftarrow k}(z, z') &= \mathbb{E}[Y_j | \mathbf{V} \in C_{d \leftarrow k}(z, z'), Z_j=z_j'], \\
	 &= \mathbb{E}[Y_j | \mathbf{V} \in C_{d \leftarrow k}(\mathbf{p}, \mathbf{p'}), Z_j=z_j']  := \text{LATR}^j_{d\leftarrow k}(\mathbf{p}, \mathbf{p'}, z_j'). \nonumber \\ \nonumber
\end{align}

\noindent \textbf{(non)-Identification.} For a shift in $Z_d$, the LATR$^k_{d\leftarrow k}$ are identified for any $k \neq d$, while for $Y_d$, the LATR$^d_{d\leftarrow k}$ are not directly identified separately for any $k$ of origin. \\
\indent First, for any potential outcome $Y_j$ with $j=k \neq d$, the $\text{LATR}^k_{d\leftarrow k}(\mathbf{p}, \mathbf{p'}, z_k)$ are directly identified from the data by 
\begin{align}\label{eq:latrjdiscreteidentification}
	\text{LATR}^k_{d\leftarrow k}(\mathbf{p}, \mathbf{p'}, z_k) = - \frac{\mathbb{E}[Y D_k | Z=z'] - \mathbb{E}[Y D_k | Z=z]}{P_k(z') - P_k(z)}. 
\end{align}
Indeed, following the reasoning of the binary case and using that $z_k'=z_k$, we have
\begin{align}\label{eq:latrjdiscreteidentification_details}
	&\mathbb{E}[YD_k | Z=z'] - \mathbb{E}[YD_k | Z=z] \nonumber \\
	&= \mathbb{E}[Y_kD_k(\mathbf{p'}) | Z_k=z_k'] - \mathbb{E}[Y_kD_k(\mathbf{p}) | Z_k=z_k] \nonumber \\
	&= \mathbb{E}[Y_k ( D_k(\mathbf{p'}) - D_k(\mathbf{p})) | Z_k=z_k ] \nonumber \\
	&= - \ \mathbb{E}[Y_k | \mathbf{V} \in C_{d \leftarrow k}(\mathbf{p}, \mathbf{p'}), Z_k = z_k] \times (P_k(\mathbf{p'}) - P_k(\mathbf{p})), 
\end{align}
where $P_k(\mathbf{p})$ is the $k$ element of $\mathbf{p}$ and is equal to $P_k(z)$ if $P(z) = \mathbf{p}$. The minus sign in the last equality comes from the fact that $D_k(\mathbf{p'}) - D_k(\mathbf{p})$ can only equals $0$ (for the stayers who keep $D=k$) or $-1$ (for the `compliers' who flow out of $D=k$).  
Since we only shifted $Z_d$ to reach $Z=z'$ from $Z=z$, $z_k'=z_k$, and there is no direct effect of the included semi-IV on $Y_k$. $Z_d$ serves as an IV for $Y_k$, as in the binary treatment case. More importantly, the reason why we can identify this treatment response if that we can map individuals with $\{ D_k(z')=0; D_k(z)=1\}$ to the compliers $C_{d \leftarrow k}(z, z')$ or $C_{d \leftarrow k}(\mathbf{p}, \mathbf{p'})$. Indeed, under the model assumptions, when one shifts $Z_d$, there is only an outflow of individuals from $D=k$ to $D=d$. There is no inflow into any $D=k$ with $k\neq d$. 
So the complying set $d\leftarrow k$ of mass $P_k(z) - P_k(z')$ is identified from the perspective of $Y_k$, and we can identify LATR$_{d\leftarrow k}^k$ as we would have with an IV. The intuition is that the observable differences in $YD_k$ are driven solely by individuals who moved from $D=k$ to $D=d$ due to the change in semi-IVs. \\
\indent However, for the potential outcome corresponding to the shifted semi-IV, $Z_d$, i.e., for $Y_j$ with $j=d$, one cannot identify LATR$^d_{d\leftarrow k }$ for any $k \neq d$. Indeed, even though we observe
\begin{align*}
	\mathbb{E}[Y D_d | Z=z'] - \mathbb{E}[Y D_d | Z=z],
\end{align*}
this comparison is not enough to identify the margin-specific treatment effects of interest. Because, as in its counterpart with IVs, when shifting $Z$ from $z$ to $z'$ by moving only $Z_d$, there are $J-1$ inflows of compliers into $D=d$. As a consequence, observing 
\begin{align*}
	D_d(z) = 0; D_d(z') = 1 \iff \mathbf{V} \in \underset{k \in \mathcal{D}; k \neq d}\bigcup C_{d \leftarrow k}(z, z').
\end{align*}
Without further assumption, we cannot determine the origin of the compliers selecting $D=d$, and we only observe the total inflow (union of all the $d\leftarrow k$ compliers) of all the compliers selecting $D=d$ due to the change in $Z$. 
This is why we can only directly identify a \textit{pooled LATE} across all the margins. In the terminology of \cite{heckmanurzuavytlacil2006}, this is the treatment effect of one choice ($d$) versus the  "\textit{next best alternative}". 

\subsubsection{Pooled LATE, or LATE versus the next best alternative}
Define the \textit{pooled LATE} as the total treatment effect on $Y$ for all the individuals who switch to $D=d$ due to an exogenous shift in $Z_d$ from $z_d$ to $z_d'$, holding all the other semi-IVs fixed. This pooled LATE is equal to
\begin{align}\label{eq:latepooled}
	\text{LATE}_d(z, z') = \sum_{k \in \mathcal{D}; k \neq d} \text{LATE}_{d\leftarrow k}(z, z') \times \frac{ P_k(z) - P_k(z')}{P_d(z') - P_d(z)}.  
\end{align}

\noindent Plugging \eqref{eq:discretelatemargin} and \eqref{eq:discreteLATR} into \eqref{eq:latepooled}, one can show that this LATE is equal to a sum of LATR$^j_{d\leftarrow k}$, 
\begin{align}\label{eq:latepooled2}
	\text{LATE}_d(z, z') =& \overbrace{\sum_{k \in \mathcal{D}; k \neq d} \text{LATR}^d_{d\leftarrow k}(z, z') \frac{ P_k(z) - P_k(z')}{P_d(z') - P_d(z)}}^{=: \ \text{LATR}_d(z, z') } \nonumber \\
	&- \sum_{k \in \mathcal{D}; k \neq d} \text{LATR}^k_{d\leftarrow k}(z, z') \frac{ P_k(z) - P_k(z')}{P_d(z') - P_d(z)},  
\end{align} 
where all the LATR$^k_{d\leftarrow k}$ are identified as described previously, so only the weighted sum of LATR$^d_{d\leftarrow k}$ over all $k\neq d$ remains to be identified. 
Since $P_k(z) - P_k(z')$ is simply the mass of the $d\leftarrow k$ compliers, and $P_d(z') - P_d(z)$ the total mass of all the compliers that come to $D=d$, this sum is equal to
\begin{align*}
	\text{LATR}_d(z, z') &= \mathbb{E}[ Y_d | \mathbf{V} \in \underset{k \in \mathcal{D}; k \neq d}\bigcup C_{d \leftarrow k}(z, z'), Z_d = z_d'] \\
	&= \mathbb{E}[Y_d | D_d(z) = 0, D_d(z') = 1, Z_d=z_d'] \\
	&=  \mathbb{E}[Y_d | D_d(\mathbf{p}) = 0, D_d(\mathbf{p'}) = 1, Z_d=z_d'] \\
	&=: \text{LATR}_d(\mathbf{p}, \mathbf{p'}, z_d').
\end{align*}
With this formulation, one can see that the total LATR$_d$ can be identified using a more standard approach, comparing $YD_d(z')$ and $YD_d(z)$. \\
\indent Because we have semi-IVs instead of IVs, one additional step remains: the shift from $Z=z$ to $z'$ induces a shift from $Z_d=z_d$ to $z_d'$, so, as in the binary case, one cannot directly compare $\mathbb{E}[Y D_d | Z=z'] - \mathbb{E}[Y D_d | Z=z]$ to obtain LATR$_d(z, z')$, because of the effect of the included semi-IV $Z_d$ on $Y_d$. 
Instead, we proceed as in the binary case, i.e., by identifying the LATR by shifting the (generalized) propensity score while holding $Z_d=z_d$ fixed.  
\noindent If there exists a $\tilde{z}=(\tilde{z}_0, ..., \tilde{z}_{J-1})$ with $\tilde{z}_d=z_d'$ and such that $\mathbf{P}(\tilde{z}) = \mathbf{p}$, then we have that\footnote{The proof is similar to the one developed in \eqref{eq:latrjdiscreteidentification_details}, except that individuals with $D_d(z') - D_d(\tilde{z}) = 1$ correspond the union of all $d\leftarrow k$ compliers here. }
\begin{align}\label{eq:latrdidentification}
	\text{LATR}_d(\mathbf{p}, \mathbf{p'}, z_d') = \frac{\mathbb{E}[YD_d | Z=z'] - \mathbb{E}[YD_d | Z=\tilde{z}]}{P_d(z') - P_d(\tilde{z})}.
\end{align}
Indeed, if there exists a $Z=\tilde{z}$ with the same propensity score as $z$, then a change from $Z=\tilde{z}$ to $z'$ induces the same flow of compliers as a change from $z$ to $z'$, i.e., $C_{d\leftarrow k}(z, z') = C_{d\leftarrow k}(\tilde{z}, z') = C_{d\leftarrow k} (\mathbf{p}, \mathbf{p'})$ for each specific margin $k \neq d$, and thus over all margins as well. So we use this change from $\tilde{z}$ to $z'$, which does not change $Z_d$, to isolate the selection effect at $Z_d=z_d'$. 
As in the binary case, a change from $\tilde{z}$ to $z'$ to shift $\mathbf{P}$ without changing $Z_d$ requires a shift in all the other semi-IVs. \\
\indent Provided all the semi-IVs are sufficiently relevant -- which is likely to be the case for small changes in $Z_d$ and in propensity scores -- $\tilde{z}$ exists (see the reachability property of Subsection \ref{subsec:discreteproperty}), and we can identify LATR$_d(z, z')$. In this case,  the pooled LATE$_d(z, z')$ is also identified by plugging \eqref{eq:latrdidentification} and \eqref{eq:latrjdiscreteidentification} into \eqref{eq:latepooled2}. Therefore, in the discrete treatment case, semi-IVs can identify the pooled LATE exactly as IVs. \\
\indent Similarly, one could identify the MTR and MTE in the discrete case, as the limit case where $Z_d$ varies marginally and so does $P_d(Z)$. The adjustments to identify the pooled MTE are straightforward, so I skip the details here.

\subsubsection{Margin-specific MTE}\label{subsec:mountjoy}
While identifying margin-specific LATE is difficult (with either IVs or semi-IVs), the identification of margin-specific MTE remains possible due to specific properties of the complier sets at the margin. In fact, with IVs, this identification is also achievable while relaxing the discrete choice index model into a partial monotonicity and a comparable complier assumption \citep{mountjoy2022community}.\footnote{It may also be possible to relax the monotonicity assumption in other directions with semi-IVs, as \cite{lee2018identifying} or \cite{heckmanpinto2018} do with IVs. However, these are less straightforward than the relaxation of \cite{mountjoy2022community}, which naturally applies to semi-IVs.} I show that these IV results also hold with semi-IVs. \\ 

\noindent \textbf{Application.} I focus on the same question as \cite{mountjoy2022community}, while noting that the methodology applies to many other previously mentioned examples. Students make an education choice between no college ($D=0$), two-year colleges ($D=1$), or four-year colleges ($D=2$). We want to identify the effect of these choices on subsequent earnings, at age $30$ for example. We have 3 $d$-specific semi-IVs capturing $d$-specific costs or benefits. The tuition fees of two and four-year colleges are natural candidates for $Z_1$ and $Z_2$, respectively: all else equal, for $d \in \{1,2\}$, students who do not pick $D=d$ do not pay these tuition fees and do not receive the education associated with these fees, so the fees should have no effect on their outcomes (conditional on the included $Z_d$). Moreover, the tuition fees should be relevant for the selection into the different types of colleges. Unlike the IV approach, I do not assume that the tuition fees are excluded. For instance, I do not rule out that higher tuition for two-year colleges implies a better quality of education in these two-year colleges and thus a better outcome for individuals who selected $D=1$. For $D=0$, $Z_0$ could be a measure of the local market conditions for the non-college educated individuals (e.g., their unemployment or wage rate). This captures the opportunity cost of going to college and may affect $Y_0$, but should not affect college attendees ($D=1$ or $2$). Note that the three semi-IVs should be taken at the time of the treatment decision, when the individuals were $17$ or $18$ years old, while the outcome is taken later (e.g., at age $30$). The main difference with \cite{mountjoy2022community} is that I need the third semi-IV here ($Z_0$). The advantage is that I do not need to find two fully excluded IVs, which is infeasible in many other applications.\footnote{\cite{mountjoy2022community} uses the distance to two and four-year colleges as $d$-specific IVs.} 
\noindent Because we normalize $Z_d$ to have a positive effect on $P_d$, we define $Z_1$ and $Z_2$ as "minus the tuition fees", so that a marginal \textit{increase} in $Z_1$ corresponds to a marginal \textit{decrease} in two-year college tuition. \\
\indent In this Framework, a decrease in two-year college tuition ($\Delta^+ Z_1$) directly affects the outcome of individuals with $D=1$ and, as in the IV case, induces two distinct flows of compliers for whom we aim to identify the margin-specific MTE: (i) the $1\leftarrow 0$ compliers, who switch from $D=0$ to $D=1$ due to the marginal change in $Z_1$ (democratization effect), and (ii) the $1 \leftarrow 2$ compliers who switch from $D=2$ to $D=1$ (diversion effect). Due to these two flows, the overall effect of $D_1$ on $Y$ due to the change in $Z_1$, measured by the pooled MTE, is ambiguous and less informative. Shifting individuals without college into two-year college generally improves their future earnings, whereas diverting students from four-year college may be detrimental. Ideally, we want to identify the margin-specific MTEs on both margins separately, as these are more informative. I show how to do this with semi-IVs. \\

\noindent \textbf{Margin-specific marginal compliers, MTR and MTE.} \\
Define the \textit{marginal complier} switching from $D=k$ to $D=d$ when $Z_d$ increases marginally at $Z=z$ by $C_{d\leftarrow k}^{Z_d+}(z)$, and the marginal complier switching from $D=k$ to $D=d$ when $Z_k$ decreases marginally at $Z=z$ by $C_{d\leftarrow k}^{Z_k-}(z)$. \\
\indent Without loss of generality, from here onwards, focus on a marginal increase in $Z_1$. 
For each potential outcome $Y_d$ with $d \in \mathcal{D}$, define the MTR$^d_{1\leftarrow k}$ at $Z=z$ as
\begin{align*}
	\text{MTR}^d_{1\leftarrow k}(z) = \mathbb{E}[Y_d | \mathbf{V} \in C^{Z_1+}_{1\leftarrow k}(z), Z_d=z_d] \quad \text{ for all } k \neq 1,   
\end{align*}
i.e., the average potential outcome $Y_d$ for marginal compliers who switch from $D=k$ to $D=1$ due to a marginal increase in $Z_1$ at $Z=z$. \\
\indent From there, the \textit{margin-specific MTE} can be defined as:
\begin{align*}
	\text{MTE}_{1 \leftarrow k}(z) &= \mathbb{E}[Y_1 - Y_k | \mathbf{V} \in C^{Z_1+}_{1\leftarrow k}(z), Z_k=z_k, Z_1=z_1] 
	= \text{MTR}^1_{1\leftarrow k}(z) - \text{MTR}^k_{1\leftarrow k}(z). 
\end{align*}
These are the treatment effects for the marginal compliers who are induced into $D=1$ from $D=k$ because of the marginal increase in $Z_1$ when $Z=z$. \\

\begin{figure}[!h]
	\centering
	\caption{\cite{mountjoy2022community}'s identification strategy for MTE$_{1\leftarrow 0}(z)$}\label{fig:mountjoy}

	\begin{minipage}{0.45\textwidth}
	\centering
	\begin{tikzpicture}[x=60pt,y=50pt, line width=0.25mm] 
    	\def\uzero{0.4}
    	\def\uone{0.2}
    	\def\utwo{0.4}
    	
		\draw[<->](-1,1)--(-1,-1)--(1,-1); 
		\draw (-1, 1) node[above left=0pt, black]{$V_2$};
		\draw (1, -1) node[below right=0pt, black]{$V_1$};

		\begin{scope}
		\clip (-1, -1) rectangle (1, 1);	
		\draw[thick] (-1, \uzero-\utwo) -- (\uzero-\uone, \uzero-\utwo) -- (\uzero-\uone, -1);
		\draw[domain=(\uzero - \uone):1,smooth,variable=\x,black,thick] plot ({\x},{\uone - \utwo +\x});
		
		\fill[mtr1color, opacity=0.3] (\uzero-\uone, -1) -- (\uzero-\uone, \uzero-\utwo) -- (1, \uone - \utwo + 1) -- (1, -1) -- cycle;
		
		\end{scope}
		
		\def\uzerop{0.4}
    	\def\uonep{0.7}
    	\def\utwop{0.4}

		\begin{scope}
		\clip (-1, -1) rectangle (1, 1);
		\draw[thick, dashed] (-1, \uzerop-\utwop) -- (\uzerop-\uonep, \uzerop-\utwop) -- (\uzerop-\uonep, -1);
		\draw[domain=(\uzerop - \uonep):1,smooth,variable=\x,black,thick, dashed] plot ({\x},{\uonep - \utwop +\x});
		
		\fill[mtr1color, opacity=0.3] (\uzerop-\uonep, -1) -- (\uzerop-\uonep, \uzerop-\utwop) -- (1, \uonep - \utwop + 1) -- (1, -1) -- cycle;
		\end{scope}

		\node (p1) at (\uzerop-\uonep, \uzerop-\utwop) {}; 
		\node (p) at (\uzero-\uone, \uzero-\utwo) {}; 
		\draw[->, thick, line width=0.5mm] (p) -- (p1);

		\fill[white, opacity=1] (\uzero-\uone, -1) -- (\uzero-\uone, \uzero-\utwo) -- (1, \uone - \utwo + 1) -- (1, -1) -- cycle;

		\node[rotate=45, font=\footnotesize] at (0.4, 0.5) {$1\leftarrow 2$ };
		\node[rotate=0, above, font=\footnotesize] at (-0.05, -0.6) {$1\leftarrow 0$ };
	
	\end{tikzpicture} 
	\par \vspace{0.1em} 
    \centering{\small (a) $\Delta^+ Z_1$} 
	\end{minipage} \hfill 
	\begin{minipage}{0.45\textwidth}
	\centering 
		\begin{tikzpicture}[x=60pt,y=50pt, line width=0.25mm] 
    	    	
    	\def\uzero{0.4}
    	\def\uone{0.2}
    	\def\utwo{0.4}
    	
		\draw[<->](-1,1)--(-1,-1)--(1,-1); 
		\draw (-1, 1) node[above left=0pt, black]{$V_2$};
		\draw (1, -1) node[below right=0pt, black]{$V_1$};

		
		\begin{scope}
		\clip (-1, -1) rectangle (1, 1);	
		\draw[thick] (-1, \uzero-\utwo) -- (\uzero-\uone, \uzero-\utwo) -- (\uzero-\uone, -1);
		\draw[domain=(\uzero - \uone):1,smooth,variable=\x,black,thick] plot ({\x},{\uone - \utwo +\x});
		
		\fill[mtr0color, opacity=0.3] (-1,-1) rectangle (\uzero-\uone, \uzero-\utwo); 
		
		\end{scope}

		\def\uzerop{-0.1}
    	\def\uonep{0.2}
    	\def\utwop{0.4}

		\begin{scope}
		\clip (-1, -1) rectangle (1, 1);
		\draw[thick, dashed] (-1, \uzerop-\utwop) -- (\uzerop-\uonep, \uzerop-\utwop) -- (\uzerop-\uonep, -1);
		\draw[domain=(\uzerop - \uonep):1,smooth,variable=\x,black,thick, dashed] plot ({\x},{\uonep - \utwop +\x});
		\end{scope}

		\node (p1) at (\uzerop-\uonep, \uzerop-\utwop) {}; 
		\node (p) at (\uzero-\uone, \uzero-\utwo) {}; 
		\draw[->, thick, line width=0.5mm] (p) -- (p1); 
	
		\fill[white, opacity=1] (-1,-1) rectangle (\uzerop-\uonep, \uzerop-\utwop); 

		\node[rotate=0, font=\footnotesize] at (-0.6, -0.25) {$2\leftarrow 0$ };
		\node[rotate=0, above, font=\footnotesize] at (-0.05, -0.8) {$1\leftarrow 0$ };
		
		\end{tikzpicture}
		\par \vspace{0.1em} 
    	\centering{\small (b) $\Delta^- Z_0$} 
	\end{minipage} 

	\vspace{1em} 
		
	
	\begin{minipage}{0.45\textwidth}
	\centering
	\begin{tikzpicture}[x=60pt,y=50pt, line width=0.25mm] 
    	\def\uzero{0.4}
    	\def\uone{0.2}
    	\def\utwo{0.4}
    	
    	\def\uzerop{0.4}
    	\def\uonep{0.7}
    	\def\utwop{0.4}
    	
    	\def\uzeropp{-0.1}
    	\def\uonepp{0.2}
    	\def\utwopp{0.4}

		\begin{scope}
		\clip (-1, -1) rectangle (1, 1);	
		\fill[mtr1color, opacity=0.3] (\uzero-\uone, -1) -- (\uzero-\uone, \uzero-\utwo) -- (1, \uone - \utwo + 1) -- (1, -1) -- cycle;
		\fill[mtr0color, opacity=0.3] (-1,-1) rectangle (\uzero-\uone, \uzero-\utwo); 
		
		
		\end{scope}
		

		\begin{scope}
		\clip (-1, -1) rectangle (1, 1);
		
		\fill[mtr1color, opacity=0.3] (\uzerop-\uonep, -1) -- (\uzerop-\uonep, \uzerop-\utwop) -- (1, \uonep - \utwop + 1) -- (1, -1) -- cycle;
		\end{scope}


		\fill[white, opacity=1] (\uzero-\uone, -1) -- (\uzero-\uone, \uzero-\utwo) -- (1, \uone - \utwo + 1) -- (1, -1) -- cycle;

		
		\begin{scope}
		\clip (-1, -1) rectangle (1, 1);
		\fill[white, opacity=1] (-1,-1) rectangle (\uzeropp-\uonepp, \uzeropp-\utwopp); 
		\end{scope}

				\begin{scope}
		\clip (-1, -1) rectangle (1, 1);
		\fill[yellow, opacity=0.9] (\uzeropp-\uonepp, -1) -- (\uzeropp-\uonepp, \uzeropp-\utwopp) -- (\uzero - \uone, \uzero - \utwo) -- (\uzero - \uone, -1) -- cycle;	
		\fill[green, opacity=0.9] (\uzeropp-\uonepp, \uzeropp-\utwopp) -- (\uzero - \uone, \uzero - \utwo) -- (\uzeropp - \uonepp, \uzero - \utwo) -- cycle;	
		\end{scope}

		\begin{scope}
		\clip (-1, -1) rectangle (1, 1);	
		\draw[thick] (-1, \uzero-\utwo) -- (\uzero-\uone, \uzero-\utwo) -- (\uzero-\uone, -1);
		\draw[domain=(\uzero - \uone):1,smooth,variable=\x,black,thick] plot ({\x},{\uone - \utwo +\x});
		\draw[thick, dashed] (-1, \uzeropp-\utwopp) -- (\uzeropp-\uonepp, \uzeropp-\utwopp) -- (\uzeropp-\uonepp, -1);
		\draw[domain=(\uzeropp - \uonepp):1,smooth,variable=\x,black,thick, dashed] plot ({\x},{\uonepp - \utwopp +\x});
		\draw[thick, dashed] (-1, \uzerop-\utwop) -- (\uzerop-\uonep, \uzerop-\utwop) -- (\uzerop-\uonep, -1);
		\draw[domain=(\uzerop - \uonep):1,smooth,variable=\x,black,thick, dashed] plot ({\x},{\uonep - \utwop +\x});
		\end{scope}

		\draw[<->](-1,1)--(-1,-1)--(1,-1); 
		\draw (-1, 1) node[above left=0pt, black]{$V_2$};
		\draw (1, -1) node[below right=0pt, black]{$V_1$};
			
	\end{tikzpicture}
	
		\par \vspace{0.1em} 
    	\centering{\small (c) Overlap (yellow)} 
	\end{minipage} \hfill 
	\begin{minipage}{0.45\textwidth}
	\centering
	\begin{tikzpicture}[x=60pt,y=50pt, line width=0.25mm] 
    	\def\uzero{0.4}
    	\def\uone{0.2}
    	\def\utwo{0.4}
    	
    	\def\uzerop{0.4}
    	\def\uonep{0.3}
    	\def\utwop{0.4}
    	
    	\def\uzeropp{0.3}
    	\def\uonepp{0.2}
    	\def\utwopp{0.4}

		\begin{scope}
		\clip (-1, -1) rectangle (1, 1);	
		\fill[mtr1color, opacity=0.3] (\uzero-\uone, -1) -- (\uzero-\uone, \uzero-\utwo) -- (1, \uone - \utwo + 1) -- (1, -1) -- cycle;
		\fill[mtr0color, opacity=0.3] (-1,-1) rectangle (\uzero-\uone, \uzero-\utwo); 
		
		
		\end{scope}
		

		\begin{scope}
		\clip (-1, -1) rectangle (1, 1);
		
		\fill[mtr1color, opacity=0.3] (\uzerop-\uonep, -1) -- (\uzerop-\uonep, \uzerop-\utwop) -- (1, \uonep - \utwop + 1) -- (1, -1) -- cycle;
		\end{scope}


		\fill[white, opacity=1] (\uzero-\uone, -1) -- (\uzero-\uone, \uzero-\utwo) -- (1, \uone - \utwo + 1) -- (1, -1) -- cycle;

		
		\begin{scope}
		\clip (-1, -1) rectangle (1, 1);
		\fill[white, opacity=1] (-1,-1) rectangle (\uzeropp-\uonepp, \uzeropp-\utwopp); 
		\end{scope}

				\begin{scope}
		\clip (-1, -1) rectangle (1, 1);
		\fill[yellow, opacity=0.9] (\uzeropp-\uonepp, -1) -- (\uzeropp-\uonepp, \uzeropp-\utwopp) -- (\uzero - \uone, \uzero - \utwo) -- (\uzero - \uone, -1) -- cycle;	
		\fill[green, opacity=0.9] (\uzeropp-\uonepp, \uzeropp-\utwopp) -- (\uzero - \uone, \uzero - \utwo) -- (\uzeropp - \uonepp, \uzero - \utwo) -- cycle;	
		\end{scope}

		\begin{scope}
		\clip (-1, -1) rectangle (1, 1);	
		\draw[thick] (-1, \uzero-\utwo) -- (\uzero-\uone, \uzero-\utwo) -- (\uzero-\uone, -1);
		\draw[domain=(\uzero - \uone):1,smooth,variable=\x,black,thick] plot ({\x},{\uone - \utwo +\x});
		\draw[thick, dashed] (-1, \uzeropp-\utwopp) -- (\uzeropp-\uonepp, \uzeropp-\utwopp) -- (\uzeropp-\uonepp, -1);
		\draw[domain=(\uzeropp - \uonepp):1,smooth,variable=\x,black,thick, dashed] plot ({\x},{\uonepp - \utwopp +\x});
		\draw[thick, dashed] (-1, \uzerop-\utwop) -- (\uzerop-\uonep, \uzerop-\utwop) -- (\uzerop-\uonep, -1);
		\draw[domain=(\uzerop - \uonep):1,smooth,variable=\x,black,thick, dashed] plot ({\x},{\uonep - \utwop +\x});
		\end{scope}

		\draw[<->](-1,1)--(-1,-1)--(1,-1); 
		\draw (-1, 1) node[above left=0pt, black]{$V_2$};
		\draw (1, -1) node[below right=0pt, black]{$V_1$};

	\end{tikzpicture}
	
		\par \vspace{0.1em} 
    	\centering{\small (d) Overlap for marginal changes} 
	\end{minipage} \hfill 

\centering
\vspace{1em}
\end{figure}
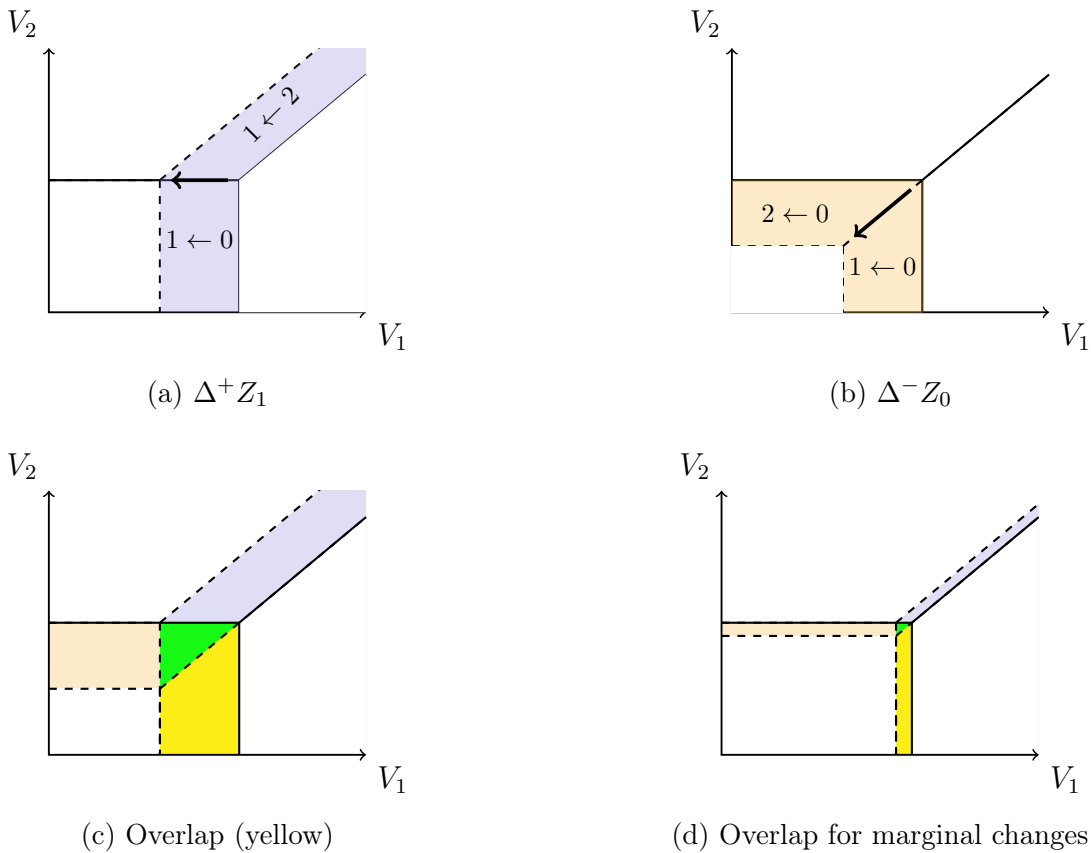

\noindent \textbf{Same marginal compliers property.} The key challenge is to disentangle both flows of margin-specific compliers for $D=1$. Fortunately, under the discrete choice model, if we focus only on marginal changes, one can isolate both margins as illustrated in Figure \ref{fig:mountjoy}. For any $Z=z$, an increase in $Z_1$ causes two flows of compliers into $D=1$: $1\leftarrow 0$ and $1\leftarrow 2$. Similarly, a decrease in $Z_0$ causes two flows of compliers out of $D=0$: $1\leftarrow 0$ and $2\leftarrow 0$. A key insight from \cite{mountjoy2022community} is that, for marginal increases in $Z_1$ and marginal decreases in $Z_0$ at $Z=z$, the marginal compliers are the \textit{same}, i.e., 
\begin{align*}
	C^{Z_1+}_{1\leftarrow 0}(z) = C^{Z_0-}_{1\leftarrow 0}(z),
\end{align*}
under the discrete choice index model. This is visible in the panel (d) of Figure \ref{fig:mountjoy}: the overlap (yellow area) becomes the entire $1\leftarrow 0$ set at the margin (the green discrepancy vanishes).  
Similarly, the marginal compliers are the same for the other margin $1 \leftarrow 2$, i.e.,
\begin{align*}
	C^{Z_1+}_{1\leftarrow 2}(z) = C^{Z_2-}_{1\leftarrow 2}(z). 
\end{align*}

\noindent \textbf{Identification.} With these "\textit{same marginal compliers}" properties, one can proceed to the identification. As in the binary case, we identify the marginal effect of $D$ on $Y_d$ without shifting $Z_d$, avoiding confounding direct effects of $Z_d$ on $Y_d$. This is possible by shifting only the other semi-IVs, $Z_{-d}$, which are excluded from $Y_d$ and can serve as IVs for $Y_d$. \\ 
\indent For $Y_0$ and $Y_2$, using the marginal version of \eqref{eq:latrjdiscreteidentification_details}, we directly identify the marginal treatment responses at $Z=z$ by marginally increasing the excluded $Z_1$, i.e., 
\begin{align*}
	\text{MTR}_{1\leftarrow 0}^0(z) &=  \mathbb{E}[Y_0 | \mathbf{V} \in C^{Z_1+}_{1\leftarrow 0}(z), Z_0=z_0] = \frac{\partial \mathbb{E}[Y_0 D_0 | Z=z]}{\partial Z_1}\Big/ \frac{\partial P_0(z)}{\partial Z_1},  \\
	\text{ and } \  \text{MTR}_{1\leftarrow 2}^2(z) &=  \mathbb{E}[Y_2 | \mathbf{V} \in C^{Z_1+}_{1\leftarrow 2}(z), Z_2=z_2] = \frac{\partial \mathbb{E}[Y_2 D_2 | Z=z]}{\partial Z_1}\Big/ \frac{\partial P_2(z)}{\partial Z_1}. 
\end{align*}
\indent Then, to identify separately the two margin-specific MTRs for $Y_1$, one needs to marginally decrease $Z_0$ and $Z_2$ and use the "same marginal compliers property" of the discrete index model. More precisely, for $1\leftarrow k$ compliers, with $k=0$ or $2$, we obtain:
\begin{align*}
	\text{MTR}_{1\leftarrow k}^1(z) &= \mathbb{E}[Y_1 | \mathbf{V} \in C^{Z_1+}_{1\leftarrow k}(z), Z_1=z_1] \\
	&= \mathbb{E}[Y_1 | \mathbf{V} \in C^{Z_k-}_{1\leftarrow k}(z), Z_1=z_1] = \frac{\partial \mathbb{E}[Y_1 D_1 | Z=z]}{\partial Z_k^-}\Big/ \frac{\partial P_1(z)}{\partial Z_k^-},  
\end{align*}
by decreasing $Z_k$ and holding $Z_1=z_1$ fixed. 
Notice that $\partial P_1(z)/\partial Z_k^- = -\partial P_k(z)/\partial Z_1^+$ because these are the same set of compliers with the same mass. \\
\indent Thus, for any $k=0, 2$, the margin-specific MTE are identified by 
\begin{align*}
	\text{MTE}_{1 \leftarrow k}(z) = \text{MTR}^1_{1\leftarrow k}(z) - \text{MTR}^k_{1\leftarrow k}(z) \quad \text{ for any } Z=z \text{ in the interior of } \mathcal{Z}. 
\end{align*}
\indent Incidentally, note that these derivations are very close to the one in \cite{mountjoy2022community}, who already uses a second IV to identify the two margins and who also already condition  the MTE on $Z=z$. The only slight difference is that I use three different semi-IVs to disentangle the two different margins, while \cite{mountjoy2022community} only uses a second IV to pin down one of the margins, and then uses the difference with the pooled MTE to obtain the remaining margin. Thus, the only minor distinction is that I identify the pooled MTE differently, without shifting $Z_1$ for MTR$^1$ in order to avoid the direct effects.   \\

\noindent \textbf{Relaxing the discrete index model.} As in \cite{mountjoy2022community}, it is straightforward that the previous proof still holds under weaker assumptions than the discrete index model. The necessary ingredients are an (unordered) partial monotonicity assumption combined with \textit{comparable compliers} properties (instead of strictly the `same' compliers). Both of these are implied by the discrete index model, which is stronger than necessary. Let us adapt these two assumptions with semi-IVs.

\begin{assumption}[Unordered Partial Monotonicity]\label{ass:upm} For all $z=(z_0, ..., z_{J-1}) \in \mathcal{Z}$, define $z'=(z_0', ..., z_{J-1}')$ with $z_d' > z_d$, and the other $z_k' = z_k$ for all $k\neq d$. We have
\begin{align*}
	D_d(z') \geq D_d(z) \quad \text{ and } \quad D_k(z') \leq D_k(z),  
\end{align*}
with each inequality holding strictly for at least some individuals (relevance). 
\end{assumption}

\noindent \textbf{Comparable Compliers ($d\leftarrow k$) property.} For $d, k \in \mathcal{D}$ with $d\neq k$, for all $z=(z_0, ..., z_{J-1}) \in \mathcal{Z}$, define $z'=(z_0', ..., z_{J-1}')$ with $z_d' > z_d$ and $z_j' = z_j$ for all $j \neq d$, and $\tilde{z}=(\tilde{z}_0, ..., \tilde{z}_{J-1})$ with $\tilde{z}_k < z_k$ and $\tilde{z}_j = z_j$ for all $j \neq k$. We say that the compliers from a marginal increase in $Z_d$ or decrease in $Z_k$ at $Z=z$ are comparable, if for $j \in \{d, k\}$ we have
\begin{align*}
	&\underset{\tilde{z}_k\uparrow z_k}{lim} \ \mathbb{E}[ Y_j | D(\tilde{z}) = d, D(z) = k, Z_j=z_j] = \underset{z_d'\downarrow z_d}{lim} \ \mathbb{E}[ Y_j | D(z') = d, D(z) = k, Z_j=z_j]. 
	\end{align*}

\begin{assumption}[Comparable Compliers, $J=3$, $Z_1$ increase]\label{ass:comparablecomplier} The comparable complier properties hold for both $1\leftarrow 0$, and $1 \leftarrow 2$ compliers. 
	
\end{assumption}


%

\indent The unordered partial monotonicity (Assumption \ref{ass:upm}) is sufficient to ensure that shifting only $Z_d$ only induces flows into $D=d$ from all the other alternatives. 
\noindent The comparable compliers property is a weaker form of the "same compliers" property implied by the discrete choice model. The previous identification proof only requires that this property holds for $1\leftarrow 0$ and $1\leftarrow 2$ compliers (Assumption \ref{ass:comparablecomplier}). Together, assumptions \ref{ass:upm} and \ref{ass:comparablecomplier} are sufficient to identify the margin-specific MTE following the previous reasoning, without imposing the stronger discrete index selection model. \\
\indent Both assumptions are direct adaptations of the ones in \cite{mountjoy2022community} with IVs, and I refer to the original paper for more discussion. The only distinction with semi-IVs is that one needs a comparable compliers assumption around both margins, $1\leftarrow 2$ and $1\leftarrow 0$, instead of only one of the two. Fortunately, this is a weak and natural additional assumption to make. \\ 

\noindent \textbf{Concluding remark.}
\noindent In summary, for nonparametric identification of margin-specific MTE, one can use $J$ $d$-specific semi-IVs instead of $J-1$ $d$-specific IVs. This is convenient for applied researchers since semi-IVs are easier to find. Moreover, one can also identify the direct effects of the semi-IVs, which may be of interest in their own right. For instance, one can identify the effect of subsidizing two-year college tuitions on the subsequent earnings of two-year college attendees at different margins of compliance. 

\end{document}